\documentclass[11pt,a4paper]{article}

\usepackage{epsfig}
\usepackage{amsmath}
\usepackage{amssymb}
\usepackage{verbatim}
\usepackage{bm}
\usepackage{dsfont}
\usepackage[footnotesize]{caption}
\usepackage{subfigure}
\usepackage{cancel}
\usepackage{cite}
\usepackage{textcomp}
\usepackage{calc}
\usepackage{geometry}
\usepackage{bbm}
\usepackage{color}

\newcommand{\modulo}[1]{\overset{\textrm{\tiny$\left(#1\right)$}}{=}}

\begin{document}


\noindent May 2015
\hfill IPMU 15-0075

\vskip 1.5cm

\begin{center}
{\LARGE\bf Peccei-Quinn Symmetry from\\\smallskip Dynamical Supersymmetry Breaking}

\vskip 2cm

{\large Keisuke~Harigaya$^{\,a,b}$, Masahiro~Ibe$^{\,a,b}$,
Kai~Schmitz$^{\,b}$, Tsutomu~T.~Yanagida$^{\,b}$}\\[3mm]
{\it{
a ICRR, The University of Tokyo, Kashiwa, Chiba 277-8582, Japan\\
b Kavli IPMU (WPI), UTIAS, The University of Tokyo, Kashiwa, Chiba 277-8583, Japan}}
\end{center}

\vskip 1cm


\begin{abstract}


\noindent The proximity of the Peccei-Quinn scale to the scale of supersymmetry
breaking in models of pure gravity mediation hints at a common dynamical
origin of these two scales.
To demonstrate how to make such a connection manifest, we embed the Peccei-Quinn
mechanism into the vector-like model of dynamical supersymmetry breaking \`a la IYIT.
Here, we rely on the anomaly-free discrete $Z_4^R$ symmetry
required in models of pure gravity mediation to solve the $\mu$ problem
to protect the Peccei-Quinn symmetry from the dangerous effect of
higher-dimensional operators.
This results in a rich phenomenology featuring
a QCD axion with a decay constant of $\mathcal{O}\left(10^{10}\right)\,\textrm{GeV}$ and
mixed WIMP/axion dark matter.
In addition, exactly five pairs of extra
$\mathbf{5}$ and $\mathbf{5}^*$ matter multiplets, directly coupled to
the supersymmetry breaking sector and with masses
close to the gravitino mass, $m_{3/2} \sim 100 \,\textrm{TeV}$,
are needed to cancel the $Z_4^R$ anomalies.


\end{abstract}


\thispagestyle{empty}

\newpage


\tableofcontents


\section{Introduction}
\label{sec:introduction}


\subsection{Common dynamical origin of the PQ and the SUSY breaking scale}
\label{subsec:commonorigin}


The Peccei-Quinn (PQ) mechanism~\cite{Peccei:1977hh} is widely regarded as the most promising
approach to solving the strong $CP$ problem~\cite{Kim:1986ax}
in quantum chromodynamics (QCD).
Not only does it offer a dynamical explanation for the absence of $CP$
violation in strong interactions, it also gives rise to a rich and testable
phenomenology.
In particular, it predicts the existence of a new pseudoscalar particle, the
axion~\cite{Weinberg:1977ma}, which could even form dark matter (DM) or at
least contribute a sizable fraction to it~\cite{Preskill:1982cy}.
On the other hand, the PQ mechanism on its own is far
from constituting a complete theory.
At the conceptional level, it presumes the existence
of an approximate global Abelian symmetry, $U(1)_{\rm PQ}$, which is
spontaneously broken at some energy scale $\Lambda_{\rm PQ}$.
In order to account for the tight experimental upper bound on the QCD vacuum angle,
$\bar{\theta} \lesssim 10^{-10}$~\cite{Baker:2006ts}, this global PQ symmetry needs
to be of excellent quality, which contradicts with general expectations
in the context of quantum gravity~\cite{Kamionkowski:1992mf,Holman:1992us}.
In fact, it is believed that in a consistent quantum theory of gravity all global
symmetries are necessarily broken by gravitational interactions~\cite{Giddings:1988cx}.
At low energies, this is reflected in a series of
higher-dimensional, Planck-suppressed operators
which explicitly introduce large symmetry-breaking effects.
It is, hence, unclear what ensures that the PQ symmetry is actually of such
high quality and not spoiled by gravitational effects at the Planck scale. 
Moreover, the origin of the PQ scale $\Lambda_{\rm PQ}$
remains obscure. 
In typical axion models~\cite{Kim:1979if,Zhitnitsky:1980tq}, the PQ scale usually
takes a value close to the axion decay constant,
$\Lambda_{\rm PQ} \sim \mathcal{O}\left(1\cdots 10\right) f_a$,
where $f_a$ is conventionally constrained (based on astrophysical~\cite{Turner:1989vc}
and cosmological observations~\cite{Preskill:1982cy}) to lie somewhere within
the so-called \textit{axion window},
$10^9 \,\textrm{GeV} \lesssim f_a \lesssim 10^{12} \,\textrm{GeV}$.
The PQ symmetry is therefore likely to be broken at some intermediate scale,
which appears surprising, given that one would naively expect $\Lambda_{\rm PQ}$
to be either tied to the electroweak scale~\cite{Peccei:1977hh,Weinberg:1977ma},
$\Lambda_{\rm PQ} \sim 10^2 \,\textrm{GeV}$, or,
in absence of any other scale, to the string or Planck scale,
$\Lambda_{\rm PQ} \sim \left(10^{17} \cdots 10^{18}\,\textrm{GeV}\right)/
\left(32 \pi^2\right) \sim 10^{15} \cdots 10^{16}\,\textrm{GeV}$~\cite{Choi:1985je}.
The first of these two possibilities has, however, been ruled out experimentally
a long time ago~\cite{Donnelly:1978ty}, while the second one requires
fine-tuning of the initial axion misalignment angle to avoid
the overproduction of axionic DM~\cite{Linde:1987bx}.
This leaves one wondering what other new physics might possibly come into question
as the dynamical origin of the scale $\Lambda_{\rm PQ}$.


The above mentioned problems related to the PQ mechanism appear in a new light
as soon as supersymmetry (SUSY) is brought into play.
Just like the PQ symmetry, SUSY needs to be spontaneously broken at some
intermediate energy scale, $\Lambda_{\rm SUSY}$.
This entails the tempting idea to suppose that
$\Lambda_{\rm PQ}$ and $\Lambda_{\rm SUSY}$ are, in fact, determined
by the same dynamics~\cite{Barger:2004sf,Feldstein:2012bu}.
For instance, we may imagine that SUSY is broken dynamically by
the interactions in some strongly coupled hidden sector.
The virtue of such models of dynamical SUSY breaking (DSB)~\cite{Affleck:1983mk}
is that, in these models, the SUSY breaking scale is generated without any
dimensionful input parameters via the effect of dimensional transmutation, i.e., it ends up
being related to the dynamical scale $\Lambda$ of the
strong interactions, $\Lambda_{\rm SUSY} \sim \Lambda$.
If the same strong dynamics are responsible for the spontaneous breaking
of the PQ symmetry, we then have
\begin{align}
\Lambda_{\rm SUSY} \sim \Lambda \sim \Lambda_{\rm PQ} \sim
10^{11} \cdots 10^{12} \,\textrm{GeV} \,.
\end{align}
Remarkably enough, this estimate coincides with the range of
$\Lambda_{\rm SUSY}$ values that one typically encounters in models of
\textit{pure gravity mediation} (PGM)~\cite{Ibe:2006de,Ibe:2012hu}
(for closely related mediation schemes, see~\cite{Hall:2011jd,Hall:2012zp}).
In this framework for the mediation of SUSY breaking to the visible
sector, sfermions receive large masses of the order of the
gravitino mass $m_{3/2} \sim 10^4 \cdots 10^6 \,\textrm{GeV}$ via
the tree-level scalar potential in supergravity (SUGRA)~\cite{Nilles:1983ge},
while gauginos obtain one loop-suppressed masses via anomaly
mediation~\cite{Dine:1992yw} (see also \cite{D'Eramo:2012qd}).
PGM comes with a number of attractive features
at the cost of a slightly fine-tuned electroweak scale:
It easily accounts for a standard model (SM) Higgs boson mass
of $126\,\textrm{GeV}$~\cite{Aad:2012tfa} thanks to large stop loop
corrections~\cite{Okada:1990vk};
it ensures the unification of the SM gauge couplings,
so that it may be readily embedded into a grand unified theory (GUT);
and it is capable of providing a viable candidate for dark matter either in
the form of winos~\cite{Ibe:2012hu,Ibe:2013jya,Hall:2012zp}, binos~\cite{Harigaya:2013asa}
or higgsinos~\cite{Evans:2014pxa}.
At the same time, it avoids the Polonyi problem usually encountered in
models of gravity mediation~\cite{Coughlan:1983ci}, as SUSY is required
to be broken by a non-singlet field in PGM;
it is free of any cosmological gravitino problem~\cite{Weinberg:1982zq}, as
the gravitino decays way before the onset of big bang nucleosynthesis;
and it is in less tension with constraints on flavor-changing neutral
currents (FCNCs) and $CP$ violation~\cite{Gabbiani:1996hi}, once again because
of the high sfermion mass scale.


\subsection[Unique discrete $R$ symmetry for pure gravity mediation]
{Unique discrete \boldmath{$R$} symmetry for pure gravity mediation}
\label{subsec:Z4RPGM}


Meanwhile, a common idea to protect the PQ symmetry from the dangerous
effects of higher-dimensional operators is to invoke some gauge symmetry
which ensures that the PQ symmetry survives as an approximate
\textit{accidental} symmetry in the low-energy effective
theory~\cite{Holman:1992us,Lazarides:1985bj,Harigaya:2013vja}.
A particularly attractive choice in this context, which we have recently examined
in more detail in~\cite{Harigaya:2013vja}, is to protect
the PQ symmetry by means of a gauged discrete $R$ symmetry, $Z_N^R$.
Such a discrete $R$ symmetry may be the remnant of a spontaneously broken
continuous symmetry in higher dimensions
(for instance, in string theory)~\cite{Imamura:2001es};
and it is an often important
and sometimes even imperative ingredient in SUSY model building:
It prevents too rapid proton decay via perilous dimension-5
operators~\cite{Sakai:1981pk};
it forbids a constant term in the superpotential of the order of
the Planck scale, which would otherwise result in a huge negative
cosmological constant~\cite{Izawa:1997he};
and it may account for the approximate global continuous $R$ symmetry
which is required to realize stable~\cite{Nelson:1993nf} or
meta-stable~\cite{Intriligator:2006dd} SUSY-breaking vacua in a large class
of DSB models.


On top of that, a discrete $R$ symmetry automatically
suppresses the bilinear Higgs mass in the superpotential,
$W\supset\mu\, H_u H_d$, and hence allows for a solution
to the $\mu$ problem~\cite{Kim:1983dt} in the minimal
supersymmetric standard model (MSSM).
In fact, in absence of a bare $\mu$ term, a supersymmetric mass
for the MSSM Higgs fields $H_u$ and $H_d$ needs to be generated in
consequence of $R$ symmetry breaking.
This can, for instance, be done as in the next-to-minimal supersymmetric
standard model (NMSSM)~\cite{Maniatis:2009re}, where one introduces a
SM singlet field $S$ that couples to $H_u H_d$ and which obtains a
vacuum expectation value (VEV) $\left<S\right> = \mu$ during
electroweak symmetry breaking (EWSB).
A more minimal and hence more elegant solution, however, consists in
allowing for a Higgs bilinear term~\cite{Inoue:1991rk}
in the K\"ahler potential (see also~\cite{Dudas:2012hx}),
\begin{align}
K \supset c_H H_u H_d \,,
\label{eq:KGM}
\end{align}
where $c_H$ is a dimensionless constant of $\mathcal{O}(1)$ and which
readily yields a $\mu$ term of the order of the gravitino mass,
$\mu = c_H\, m_{3/2}$, after $R$ symmetry breaking.
This solution to the $\mu$ problem is typically employed in PGM models,
where it renders the $\mu$ and $B_\mu$ parameters in the MSSM Higgs potential 
linearly independent, thereby ensuring the successful occurrence of
radiative EWSB~\cite{Evans:2013lpa}.
On the other hand, the term in Eq.~\eqref{eq:KGM} is only allowed
in the K\"ahler potential as long as
the total $R$ charge of the Higgs bilinear $H_u H_d$ is zero modulo $N$.
In order to assess for which $Z_N^R$ symmetries this requirement
can be fulfilled, it is important to remember that any discrete symmetry
which is supposed to be relevant at low energies should be a good gauge symmetry
even at the quantum level, i.e., it should be free of any gauge
anomalies~\cite{Krauss:1988zc,Ibanez:1991hv}.
This implies in particular as a minimal constraint that the fermions of the MSSM
should equally contribute to the weak and color anomalies
of the discrete $R$ symmetry,%
\footnote{Extra matter fields contributing to these two anomalies
should appear in complete $SU(5)$ multiplets, so as not to disturb
the unification of the SM gauge couplings.
As such, they, too, should equally contribute to the weak and color
anomalies of discrete $R$ symmetry, leaving the difference between these
two anomalies unaffected~\cite{Kurosawa:2001iq}.}
$\mathcal{A}\left[Z_N^R\textrm{--}SU(2)_L\textrm{--}SU(2)_L\right] -
\mathcal{A}\left[Z_N^R\textrm{--}SU(3)_C\textrm{--}SU(3)_C\right] = 0$.
For generation-independent $R$ charges commuting with $SU(5)$
and together with the structure of the MSSM superpotential
(supplemented by heavy neutrino Majorana mass terms in accord with the
seesaw mechanism~\cite{seesaw}), this condition then necessitates
that~\cite{Kurosawa:2001iq} (see also~\cite{Harigaya:2013vja,Evans:2011mf})
\begin{align}
r_{H_u} + r_{H_d} \modulo{N} 4 \,.
\label{eq:rHuHd}
\end{align}
Here, we have introduced $r_X$ as the general symbol to denote the $R$ charge
of the field $X$ and where $\modulo{N}$ serves as a shorthand notation
for equality modulo $N$.
Consequently, only for $N=4$, the $R$ charge of the Higgs bilinear vanishes,%
\footnote{Note that $Z_2^R$ is not an actual $R$ symmetry; instead
it is equivalent to the non-$R$ matter parity of the MSSM~\cite{Dine:2009swa}.\smallskip}
which singles out $Z_4^R$ as the only discrete $R$ symmetry consistent with
PGM and the generation of the
$\mu$ term via the Higgs K\"ahler term in Eq.~\eqref{eq:KGM}.


\subsection{Synopsis: PQ symmetry from dynamical SUSY breaking}


Together, the above observations lead to a remarkably consistent picture
in the context of pure gravity mediation:
\textit{While the magnitude of the PQ scale may be determined by the SUSY-breaking
scale in some appropriate DSB model, the quality of the PQ symmetry
may be safeguarded by the discrete $Z_4^R$ symmetry required to solve the $\mu$ problem.}
In this paper, we shall demonstrate that such a scenario is indeed feasible---we
set out to construct an explicit DSB model featuring an approximate PQ symmetry
that is sufficiently protected by an anomaly-free discrete $Z_4^R$ symmetry.
In doing so, we will restrict ourselves to the arguably simplest case,
i.e., we will content ourselves with
presenting a minimal example based on strong $SU(2)$ dynamics
breaking SUSY \`a la IYIT~\cite{Izawa:1996pk}.
Furthermore, to render the $Z_4^R$ symmetry anomaly-free, we are lead to introduce
a set of new $SU(5)$ multiplets,%
\footnote{We do not consider the possibility of anomaly cancellation
via the Green-Schwarz mechanism in string theory~\cite{Green:1984sg}.
In such a case, the $Z_4^R$ (and several other discrete $R$ symmetries)
could also be rendered anomaly-free solely within the MSSM, i.e., without the need for
an extra matter sector~\cite{Lee:2010gv}.}
which obtain masses via their
coupling to the SUSY-breaking sector.
Here, the phenomenological constraints on our model surprisingly single
out a unique number of extra matter fields: five pairs of $\mathbf{5}$
and $\mathbf{5}^*$ multiplets with supersymmetric masses close to $m_{3/2}$.
Such additional matter states affect the gaugino mass spectrum
in PGM~\cite{Harigaya:2013asa,Nakayama:2013uta} and thus play a crucial
role in determining the composition of the lightest supersymmetric
particle (LSP).
This has, in turn, important consequences for dark matter
and SUSY searches at colliders.
At the same time, for gravitino masses around $100 \,\textrm{TeV}$,
our model predicts an axion decay constant $f_a$
of $\mathcal{O}\left(10^{10}\right)\,\textrm{GeV}$ and an
axionic contribution to the total abundance of dark matter 
of at most $\mathcal{O}\left(10\,\%\right)$.
Meanwhile, the superpartners of the axion, the axino as well as the saxion,
receive masses of the order of the SUSY breaking scale
and hence do not cause any cosmological problems~\cite{Feldstein:2012bu}.
We therefore anticipate our model to offer an appealing solution to the strong
$CP$ problem in the context of dynamical SUSY breaking, which is
consistent with all existing bounds, but which, at the same time, can be readily
probed in a number of terrestrial experiments and astrophysical observations.

The rest of the paper is organized as follows:
In the next section, we will show how to embed the PQ mechanism into the
IYIT model of dynamical SUSY breaking.
Here, some of the technical details regarding the vacuum configuration
of the IYIT model after taking into account $R$ symmetry breaking
have been deferred to Appendix~\ref{app:vacuum}.
In Sec.~\ref{sec:pheno}, we will then comment on the quality of the
PQ symmetry in dependence of the free parameters of our model and discuss
the resulting phenomenological constraints.
Finally, we will conclude and give a brief outlook in Sec.~\ref{sec:conclusions}.


\section{Embedding of the PQ mechanism into the IYIT model}
\label{sec:model}


\subsection{Field content and superpotential of the SUSY-breaking sector}


The IYIT model represents the simplest vector-like model of dynamical SUSY
breaking.
In its most general formulation, it is based on strongly interacting
$Sp(N)$ gauge dynamics, while featuring $2 N_f = 2(N + 1)$ matter fields
$\Psi^i$ in the fundamental representation of $Sp(N)$.
At energies below the dynamical scale $\Lambda$, this theory exhibits a flat
quantum moduli space, which is best described in terms of the $N_f\left(2N_f - 1\right)$
gauge-invariant composite ``meson'' fields
\begin{align}
M^{ij} \simeq \frac{1}{\eta}\frac{1}{\Lambda} \left<\Psi^i \Psi^j\right>
\,, \quad i,j = 1,2,\cdots 2N_f \,.
\label{eq:Mij}
\end{align}
Here, $\eta$ is a dimensionless coefficient which is supposed to ensure
that the meson fields $M^{ij}$ are canonically normalized at low energies.
According to arguments from \textit{naive dimensional analysis}
(NDA)~\cite{Manohar:1983md}, it is expected to be
of $\mathcal{O}\left(4\pi\right)$; but values as small as
$\eta \simeq \pi$ may perhaps also still be admissible.
Similar numerical coefficients are expected to appear in a number of places in
the low-energy description of the IYIT model.
For simplicity, we will, however, ignore the possibility that these factors could
numerically deviate from each other and simply account
for the uncertainties in all dimensionless couplings of our
model by means of a \textit{single} NDA factor $\eta$.


As shown by Seiberg, the meson fields in Eq.~\eqref{eq:Mij} are subject to the following
quantum mechanically deformed moduli constraint~\cite{Seiberg:1994bz},
\begin{align}
\textrm{Pf}\left(M^{ij}\right) \simeq \left(\frac{\Lambda}{\eta}\right)^{N+1} \,,
\label{eq:Pfconstraint}
\end{align}
where the dynamical scale on the right-hand side of this constraint arises in consequence of
nonperturbative effects and where $\textrm{Pf}\left(M\right)$ denotes the
Pfaffian of the antisymmetric meson matrix $M$,
$\left[\textrm{Pf}\left(M\right)\right]^2 = \textrm{det}\left(M\right)$.
A convenient way to implement this constraint when studying the low-energy dynamics
of the IYIT model is to rewrite it in the form of a dynamical superpotential,
\begin{align}
W_{\rm dyn} \simeq \kappa\,\eta\, X \bigg(\frac{\eta}{\Lambda}\bigg)^{N-1}
\left[\textrm{Pf}\left(M^{ij}\right) - \left(\frac{\Lambda}{\eta}\right)^{N+1}\right] \,,
\label{eq:WdynPf}
\end{align}
with the field $X$ acting as a Lagrange multiplier.
Unfortunately, the dynamical K\"ahler potential for $X$ is uncalculable and
hence the physical interpretation of $X$ is ambiguous.
One possibility to account for this ambiguity is to \textit{define}
the dimensionless coupling $\kappa$ such that $X$ is always
canonically normalized, irrespectively of its physical status.
More precisely, if a dynamical K\"ahler potential for $X$ should be generated, 
$X$ would be physical and we would expect $\kappa$ to be some $\mathcal{O}(1)$
constant.
On the other hand, if $X$ should remain unphysical, we could still introduce
$X$ as in Eq.~\eqref{eq:WdynPf}, the only difference being that, now, $\kappa$
would formally blow up to infinity.
In what follows, we will therefore stick to the dynamical superpotential
in Eq.~\eqref{eq:WdynPf}, keeping in mind that there are two sensible regimes
for the coupling $\kappa$:
We should either set $\kappa \sim 1$ or take the limit $\kappa \rightarrow \infty$.


Supersymmetry is broken in the IYIT model by lifting the flat directions
in moduli space by means of appropriate Yukawa couplings.
Let us introduce a singlet field $Z_{ij}$ for each
$M^{ij}$ and couple these singlets to the corresponding ``quark'' bilinears in the tree-level
superpotential,
\begin{align}
W_{\rm tree}^{\rm IYIT} = \frac{1}{2}\lambda_{ij}'\, Z_{ij}\, \Psi^i \Psi^j \,.
\label{eq:Wtree}
\end{align}
At low energies, these Yukawa interactions then give rise to the following effective
superpotential,
\begin{align}
W_{\rm eff}^{\rm IYIT} \simeq
\frac{1}{2} \lambda_{ij}\, \frac{\Lambda}{\eta}\, Z_{ij} \, M^{ij} \,,
\label{eq:Weff}
\end{align}
where we have replaced the high-energies Yukawa couplings $\lambda_{ij}'$ by
their low-energy analogues $\lambda_{ij}$ to account for their RGE running from
energies high above the dynamical scale to energies below the dynamical scale.
For all Yukawa couplings $\lambda_{ij}$ being nonzero, the superpotential in 
Eq.~\eqref{eq:Weff} implies F-term conditions for the meson fields, $M = 0$,
which contradict the deformed moduli constraint in Eq.~\eqref{eq:Pfconstraint},
$\textrm{Pf}\left(M\right) \neq 0$, signaling that SUSY is spontaneously broken.
Another way to put this is to say that the total effective 
superpotential at low energies,
\begin{align}
W_{\rm eff} = W_{\rm dyn} + W_{\rm eff}^{\rm IYIT} \simeq 
\kappa\,\eta\, X \bigg(\frac{\eta}{\Lambda}\bigg)^{N-1}
\left[\textrm{Pf}\left(M^{ij}\right) - \left(\frac{\Lambda}{\eta}\right)^{N+1}\right]
+ \frac{1}{2} \lambda_{ij}\, \frac{\Lambda}{\eta}\, Z_{ij} \, M^{ij} \,,
\label{eq:Wtot}
\end{align}
is of the O’Raifeartaigh type and, hence, SUSY is broken via
the O’Raifeartaigh mechanism~\cite{O'Raifeartaigh:1975pr}.


An interesting feature of the IYIT model, which will be of crucial importance to
us in the following, is that the superpotential in Eq.~\eqref{eq:Wtree} is invariant
under an axial $U(1)_A$ symmetry associated with a global $\Psi^i$ phase rotation,
$\Psi^i \rightarrow \exp\left(i q_i \theta\right)\Psi^i$, which is
anomaly-free under the strongly coupled $Sp(N)$ gauge group.
In \cite{Domcke:2014zqa}, Domcke et al.\ have promoted this $U(1)_A$ to an exact gauge symmetry,
$U(1)_A \rightarrow U(1)_{\rm FI}$, to point out a possibility how to dynamically generate
a field-dependent Fayet-Iliopoulos (FI) D-term in field theory. 
In this paper, we will now identify the same global $U(1)_A$
as the PQ symmetry, $U(1)_A \rightarrow U(1)_{\rm PQ}$, and show that
it may very well serve as a basis for the construction of a viable axion model.


\subsection{SUSY- and PQ symmetry-breaking vacuum at low energies}


In the remainder of this paper, we shall focus on the simplest version of
the IYIT model, i.e., SUSY breaking via strong $Sp(1) \cong SU(2)$ dynamics.
The extension of our construction to the general $Sp(N)$ case is
straightforward; it merely requires a bigger notational effort.
For $N = 1$, we then have to deal with $N_f = 2$ quark flavors, each consisting
of a pair of fundamental quarks fields, $\left(\Psi^1,\Psi^2\right)$ and 
$\left(\Psi^3,\Psi^4\right)$.
Under the $U(1)_A$ flavor symmetry, these fields are charged as follows,
\begin{align}
\left[\Psi^1\right] = \left[\Psi^2\right] = + \frac{1}{2} \,, \quad
\left[\Psi^3\right] = \left[\Psi^4\right] = - \frac{1}{2} \,,
\end{align}
where we have chosen the normalization so that the mesons at low
energies carry integer charges.
In fact, relabeling all meson and singlet fields, $M^{ij}$ and $Z_{ij}$,
according to their $U(1)_A$ charges, the low-energy effective theory ends up consisting
of the following degrees of freedom (DOFs),
\begin{align}
M_+ & = M^{12} \,, & M_- & = M^{34} \,, & M_0^1 & =  M^{13} \,, &
M_0^2 & = M^{14} \,, & M_0^3 & = M^{23} \,, & M_0^4 & = M^{24} \,, \\ \nonumber
Z_- & = Z_{12} \,, & Z_+ & = Z_{34} \,, & Z_0^1 & = Z_{13} \,, &
Z_0^2 & = Z_{14} \,, & Z_0^3 & = Z_{23} \,, & Z_0^4 & = Z_{24} \,.
\end{align}


In terms of these charge eigenstates, the effective superpotential in
Eq.~\eqref{eq:Wtot} now reads
\begin{align}
W_{\rm eff} \simeq \kappa\,\eta\, X
\left[\textrm{Pf}\left(M^{ij}\right) - \left(\frac{\Lambda}{\eta}\right)^2\right]
+ \frac{\Lambda}{\eta} \left(\lambda_+\, M_+\, Z_-
+ \lambda_-\, M_-\, Z_+ + \lambda_0^a\, M_0^a\, Z_0^a\right) \,,
\label{eq:Weffcha}
\end{align}
with $a = 1,2,3,4$, where we have renamed the Yukawa couplings $\lambda_{ij}$
in an obvious manner and where the Pfaffian of the meson matrix can now be expanded
into the following polynomial,
\begin{align}
\textrm{Pf}\left(M^{ij}\right) = M_+ M_- - M_0^1 M_0^4 + M_0^2 M_0^3 \,.
\end{align}
The F-term scalar potential corresponding to this superpotential exhibits
a saddle point at the origin as well as three local minima, in which the
Pfaffian constraint is either approximately satisfied
by the meson bilinear $M_+M_-$, by $M_0^1 M_0^4$ or by $M_0^2 M_0^3$,
with all other meson VEVs vanishing.
The potential energies of these three vacua respectively scale with the products
of the corresponding Yukawa couplings, $\lambda_+\lambda_-$,
$\lambda_0^1\lambda_0^4$, and $\lambda_0^2\lambda_0^3$.
As we intend to identify the $U(1)_A$ flavor symmetry with the PQ
symmetry, we need to make sure that the $U(1)_A$ symmetry is spontaneously
broken at low energies.
This is, however, only achieved once the charged mesons $M_+$ and $M_-$
acquire nonzero VEVs. 
In the following, we shall therefore assume that the product $\lambda_+\lambda_-$
is (at least slightly) smaller than $\lambda_0^1\lambda_0^4$ and $\lambda_0^2\lambda_0^3$,
so that $M_+ M_- \sim \Lambda^2/\eta^2$ and $M_0^a = Z_0^a = 0$
in the true vacuum.
Setting all neutral mesons and singlets to zero in Eq.~\eqref{eq:Weffcha},
we then obtain the effective superpotential describing the fluctuations
of $M_+$ and $M_-$ around the symmetry-breaking vacuum,
\begin{align}
W_{\rm eff} \simeq \kappa\,\eta\, X
\left[M_+ M_- - \left(\frac{\Lambda}{\eta}\right)^2\right]
+ \frac{\Lambda}{\eta} \left(\lambda_+\, M_+\, Z_-
+ \lambda_-\, M_-\, Z_+\right) \,.
\label{eq:WMpm}
\end{align}


In the limit of rigid SUSY, $Z_+$, $Z_-$
and $X$ turn out to be stabilized at zero (see Appendix~\ref{app:vacuum}).
The VEVs of the meson fields $M_+$ and $M_-$ then readily follow from minimizing
the F-term scalar potential corresponding to the above superpotential.
For canonical K\"ahler potential, we obtain
\begin{align}
M_\pm\ = \varepsilon\, \frac{\lambda}{\lambda_\pm}
\frac{\Lambda}{\eta} \,, \quad \lambda = \sqrt{\lambda_+ \lambda_-} \,, \quad
\varepsilon = \left(1 - \zeta\right)^{1/2} \,, \quad
\zeta = \left(\frac{\lambda}{\kappa\,\eta}\right)^2 \,, \label{eq:Mpm}
\end{align}
where $\lambda$ denotes the positive square root of the geometric mean of
$\lambda_+^2$ and $\lambda_-^2$ and with
$\varepsilon$ parameterizing the suppression of $M_\pm$
w.r.t.\ the asymptotic expression in the limit $\kappa \rightarrow \infty$
(or $\zeta \rightarrow 0$), 
\begin{align}
M_\pm = \varepsilon \, M_\pm^0 \,, \quad M_\pm^0 =
\lim_{\zeta\rightarrow0}M_\pm = \frac{\lambda}{\lambda_\pm} \frac{\Lambda}{\eta} \,.
\label{eq:Mpm0}
\end{align}
Here, $\zeta$ can be interpreted as a measure for the coupling strength
of the Yukawa terms in Eq.~\eqref{eq:WMpm}, viz.\ $\lambda$, in comparison to
the coupling strength of the Lagrange term, viz.\ $\kappa\,\eta$, which arises
in consequence of the deformed moduli constraint.
For fixed values of $\kappa$ and $\eta$, unitarity and the fact that $\zeta$
must not exceed $1$ (so that the meson VEVs do not vanish and the PQ symmetry
is, in fact, broken) then restrict the Yukawa coupling $\lambda$ to take a value
between $0$ and $\lambda_{\rm max} \simeq \min\left\{4\pi,\kappa\,\eta\right\}$.
Consequently, the parameters $\lambda$, $\zeta$ and $\varepsilon$ are allowed
to vary within the following ranges,
\begin{align}
0 \leq \lambda \leq \lambda_{\rm max} \,, \quad
0 \leq \zeta \leq \zeta_{\rm max} =
\left(\frac{\lambda_{\rm max}}{\kappa\,\eta}\right)^2 \,, \quad
\left(1 - \zeta_{\rm max}\right)^{1/2}= \varepsilon_{\rm min} \leq \varepsilon \leq 1 \,.
\label{eq:lzerange}
\end{align}
It is important to keep in mind that the above results for the meson VEVs come
with a certain irreducible uncertainty, given the fact that
we are unable to calculate the precise form of the K\"ahler potential
below the dynamical scale.
In this sense, the exact parameter relations in Eqs.~\eqref{eq:Mpm}, \eqref{eq:Mpm0}
and \eqref{eq:lzerange} should be taken with a grain of salt, as they may easily receive
corrections from the noncanonical K\"ahler potential.
On the other hand, we will continue to use the above relations in the following analysis
\textit{for definiteness}.
After all, they allow for a \textit{consistent} treatment of our model in terms
of a well-defined set of parameters with a clear physical interpretation.
In other words, in the following, we will study the IYIT model for the special
case of a canonical K\"ahler potential, guided by the notation that this should represent
an important \textit{benchmark scenario for the more general case}---a benchmark scenario
that we have well under control.
This will, in particular, help us to keep track of the various numerical factors
in our analysis and provide us with better estimates for a handful
of prefactors which one would otherwise simply take to be ``of $\mathcal{O}(1)$''.


Plugging our result for the meson VEVs in Eq.~\eqref{eq:Mpm} back into the
superpotential in Eq.~\eqref{eq:WMpm}, we are able to deduce the SUSY-breaking
F-terms of the fields $Z_+$, $Z_-$ and $X$,
\begin{align}
\left|F_{Z_\pm}\right| = \lambda
\left(1-\zeta\right)^{1/2} \frac{\Lambda^2}{\eta^2} =
\zeta^{1/2} \left(1-\zeta\right)^{1/2} F_0 
\,, \quad \left|F_X\right| = \frac{\lambda^2}{\kappa\,\eta} \frac{\Lambda^2}{\eta^2} =
\zeta \, F_0 \,, \quad F_0 = \kappa\,\eta\, \frac{\Lambda^2}{\eta^2} \,.
\label{eq:FZpmX}
\end{align}
The parameter $\zeta$ can hence also be regarded as a measure
for the size of the F-term of the field $X$,
\begin{align}
\zeta = \frac{\left|F_X\right|}{F_0} \leq 1 \,.
\end{align}
Here, the fact that $F_X$ can be nonzero
in the first place indicates (perhaps somewhat surprisingly) that,
for $\zeta > 0$, the deformed moduli
constraint is actually not exactly satisfied in the true vacuum,
\begin{align}
\left|M_+ M_- - \left(\frac{\Lambda}{\eta}\right)^2\right| =
\zeta \left(\frac{\Lambda}{\eta}\right)^2 \,,
\end{align}
where $\zeta$ serves again as a useful measure to parametrize the deviation from the
situation in the limit $\kappa\rightarrow\infty$ (where the moduli
constraint \textit{is} exactly fulfilled in the true vacuum).
Moreover, given the possibility of nonzero $F_X$, the field $X$ may
turn out to be the most important one among the three
SUSY-breaking fields $Z_+$, $Z_-$ and $X$.
According to Eq.~\eqref{eq:FZpmX}, we namely have
\begin{align}
\left|\frac{F_X}{F_{Z_\pm}}\right| = \left(\frac{\zeta}{1-\zeta}\right)^{1/2} \,,
\label{eq:FXFZ}
\end{align}
so that $\left|F_X\right| < \left|F_{Z_\pm}\right|$ for $\zeta < 1/2$ and
$\left|F_X\right| \geq \left|F_{Z_\pm}\right|$ for $\zeta \geq 1/2$
in a nicely symmetric fashion.
At this point, we should mention that the authors of~\cite{Domcke:2014zqa}
exclusively focused on the regime of large $\kappa$ and hence small $\zeta$,
thereby disregarding the possibility that the meson VEVs in Eq.~\eqref{eq:Mpm0}
could potentially be suppressed.
In this paper, we shall, by contrast, refrain from making any such assumption
regarding the size of $\zeta$ and simply keep $\zeta$ as a free parameter
in our analysis.
This will allow us to consistently account for the possibility of suppressed
meson VEVs, $M_\pm = \varepsilon\, M_\pm^0$, relative to the naive expressions
$M_\pm^0$ which we expect in the limit $\kappa\rightarrow\infty$.
We emphasize that, adapting this procedure, we are not only able to capture
a possible suppression of the meson VEVs due to a nonvanishing F-term for
the field $X$, but---at an effective, phenomenological level---also a possible
suppression due to the uncalculable strong-coupling effects in the K\"ahler potential.
This is one of the main reasons why we decide to stick to our parametrization in
terms of $\zeta$ in the following, despite the uncertainties induced by
the unknown terms in the dynamical K\"ahler potential.


Our results for the singlet F-terms in Eq.~\eqref{eq:FZpmX} now immediately
provide us with an expression for the SUSY breaking scale $\Lambda_{\rm SUSY} \equiv \mu$,
\begin{align}
\Lambda_{\rm SUSY}^2 \equiv \mu^2 =
\left(\left|F_{Z_+}\right|^2 + \left|F_{Z_-}\right|^2 + \left|F_X\right|^2\right)^{1/2} =
\lambda \left(2-\zeta\right)^{1/2} \frac{\Lambda^2}{\eta^2} \,,
\label{eq:LambdaSUSY}
\end{align}
which leads us to yet another interpretation of the parameter $\zeta$.
From Eq.~\eqref{eq:LambdaSUSY}, we infer that $2-\zeta$
counts what may be regarded as the effective number of ``active'' SUSY-breaking
fields $N_{\rm SUSY}^{\rm eff}$, so that $\zeta = 2 - N_{\rm SUSY}^{\rm eff}$.
More precisely, for $\zeta = 0$, the F-term of the field $X$ vanishes and SUSY
is solely broken by the F-terms belonging to the fields $Z_+$ and $Z_-$.
Hence, $N_{\rm SUSY}^{\rm eff} = 2$ in this case.
On the other hand, for $\zeta = 1$, both $F_{Z_+}$ and $F_{Z_-}$ are zero.
SUSY is then solely broken by the F-term of the field $X$ and
$N_{\rm SUSY}^{\rm eff} = 1$.
Correspondingly, intermediate values of $\zeta$ interpolate between
these two extrema of $N_{\rm SUSY}^{\rm eff}$.
Furthermore, we are now able to determine the gravitino mass $m_{3/2}$,
which appears as a constant term, $W_0 = \textrm{const.}$, in the superpotential
upon $R$ symmetry breaking,%
\footnote{Here, we ignore the VEV of the K\"ahler potential, $K_0$, which actually
enters the right-hand side of this relation in form of a factor
$\exp\left(-K_0/M_{\rm Pl}^2/2\right)$.
Since $K_0 \ll M_{\rm Pl}^2$ (see Eq.~\eqref{eq:K0}), this factor is, however,
completely negligible.}
\begin{align}
W \supset W_0 = m_{3/2}\, M_{\rm Pl}^2 \,,
\label{eq:W0}
\end{align}
with $M_{\rm Pl} = \left(8\pi G\right)^{-1/2} \simeq 2.44 \times 10^{18}\,\textrm{GeV}$
denoting the reduced Planck mass.
Requiring that the true vacuum at low energies be a Minkowski vacuum with (almost)
zero cosmological constant, we then have to balance the SUSY-breaking contribution to
the full SUGRA scalar potential, $\Lambda_{\rm SUSY}^4$, against the constant term,
$- 3/M_{\rm Pl}^2\left|W_0\right|^2$, induced by $R$ symmetry breaking.
This gives
\begin{align}
m_{3/2} = \frac{\Lambda_{\rm SUSY}^2}{\sqrt{3}\,M_{\rm Pl}} =
\frac{\lambda \left(2-\zeta\right)^{1/2}}{\sqrt{3}\,\eta^2} \frac{\Lambda^2}{M_{\rm Pl}} \,.
\label{eq:m32}
\end{align}
In conclusion, we find that, as anticipated in the introduction, both the SUSY
breaking scale as well as the gravitino mass turn out to be controlled
by the dynamical scale $\Lambda$.
In order to attain a gravitino mass consistent with PGM, say,
$m_{3/2} \sim 100 \,\textrm{TeV}$, we thus need $\Lambda$ to be of
$\mathcal{O}\left(10^{12}\right)\,\textrm{GeV}$.
In the next section, we shall see how this scale can also be understood as
the scale of PQ symmetry breaking and, in particular, how it is related to
the axion decay constant $f_a$.

\subsection{Identification and decay constant of the axion}


The nonzero VEVs of the charged meson fields $M_+$ and $M_-$ in Eq.~\eqref{eq:Mpm}
spontaneously break the global $U(1)_A$ symmetry of the IYIT superpotential.
We identify this flavor symmetry with the PQ symmetry, which means that the
chiral axion superfield $A$ must correspond to the goldstone multiplet of spontaneous
$U(1)_A$ breaking contained in $M_+$ and $M_-$.
To make this relation manifest, let us expand $M_+$ and $M_-$ around their VEVs
(which satisfy $\lambda_+ \left<M_+\right> = \lambda_- \left<M_-\right>$)
into a flavor-symmetric fluctuation $M$ in the radial direction as well as into a
goldstone phase $\Theta$,
\begin{align}
M_\pm = \frac{1}{\lambda_\pm} \left[\lambda_\pm \left<M_\pm\right>
+ \frac{\lambda_h}{\sqrt{2}}\, M \right] e^{\pm \Theta} \,.
\label{eq:MpmMTheta}
\end{align}
Here, the coupling $\lambda_h$, denoting the positive square root of the harmonic mean
of $\lambda_+^2$ and $\lambda_-^2$,
\begin{align}
\lambda_h = \left[\frac{1}{2}\left(\frac{1}{\lambda_+^2} +
\frac{1}{\lambda_-^2}\right)\right]^{-1/2} \,,
\end{align}
is chosen such that the meson field $M$, i.e., the ``Higgs field'' of PQ symmetry
breaking, is canonically normalized.
Meanwhile, the phase $\Theta$ directly corresponds to the axion multiplet $A$
modulo a proper normalization.
In order to find the precise relation between $\Theta$ and $A$, we have to
examine the canonical K\"ahler potential for $M_+$ and $M_-$ in the
true vacuum, i.e., for $M = 0$,
\begin{align}
K = K_0 \cosh\big(\Theta + \Theta^\dagger\big) +
\Delta\,\sinh\big(\Theta + \Theta^\dagger\big) =
K_0 + \Delta\, \big(\Theta + \Theta^\dagger\big) +
\frac{1}{2} K_0\, \big(\Theta + \Theta^\dagger\big)^2 + \cdots \,,
\label{eq:K0}
\end{align}
where $K_0$ stands for the VEV of the mesonic K\"ahler potential
and $\Delta$ denotes the difference between the two meson VEVs squared,%
\footnote{Note that $\Delta$ would represent a field-dependent and dynamically
generated FI-term for the $U(1)_A$ flavor symmetry, in case this symmetry was
promoted to a gauge symmetry and the meson VEVs were not
identical, $\lambda_+ \neq \lambda_-$~\cite{Domcke:2014zqa}.\smallskip}
\begin{align}
K_0 = \big<\left|M_+\right|^2\big> + \big<\left|M_+\right|^2\big> \,, \quad
\Delta = \big<\left|M_+\right|^2\big> - \big<\left|M_+\right|^2\big> \,.
\end{align}
This leads us to identify the properly normalized axion field $A$ as follows,
\begin{align}
A = K_0^{1/2}\,\Theta \,, \quad
K_0 = \frac{2}{\rho^2}\left(1-\zeta\right)\left(\frac{\Lambda}{\eta}\right)^2 \,, \quad
\rho = \frac{\lambda_h}{\lambda} =
\left[\frac{1}{2}\left(\frac{\lambda_+}{\lambda_-} + \frac{\lambda_-}{\lambda_+}\right)\right]^{-1/2}\,,
\label{eq:ATheta}
\end{align}
Here, the complex scalar $\left(\phi +  i\, a\right)/\sqrt{2}$ contained in
$A = \left\{\phi,a,\tilde{a}\right\}$ now consists of the
axion $a$ and the saxion $\phi$, while the fermionic component of $A$
represents the axino $\tilde{a}$.
Meanwhile, $\rho\in\left[0,1\right]$ is a convenient measure for the magnitude
of the flavor hierarchy in Eq.~\eqref{eq:WMpm}.
For equal Yukawa couplings, $\lambda_+ = \lambda_-$, the superpotential
in Eq.~\eqref{eq:WMpm} is invariant under the exchange of
$\textrm{``$+$''}$ and $\textrm{``$-$''}$ and $\rho = 1$.
On the other hand, as soon as $\lambda_+ \neq \lambda_-$, this symmetry is
broken and $\rho <1$.%
\footnote{At high energies and for generic ``neutral'' Yukawa couplings $\lambda_0^a$,
the exchange symmetry of the superpotential in Eq.~\eqref{eq:WMpm} for
$\lambda_+ = \lambda_-$ cannot be realized at the level of the fundamental
quarks $\Psi^i$.
This renders it an accidental rather than an exact symmetry,
which is expected to be explicitly broken in the K\"ahler potential.
More generally, we stress that, out of the maximal flavor symmetry of the IYIT model,
$U(4)$, we only depend on the PQ symmetry, $U(4) \supset SU(4) \supset U(1)_{\rm PQ}$,
to be a reasonably good symmetry.
All other global symmetries ought to be explicitly broken, in order to avoid
massless particles and/or the formation of topological defects in the early universe.
\label{fn:global}}
The more we increase the flavor hierarchy in the charged meson sector,
the smaller $\rho$ then becomes---until, for $\lambda_+ \gg \lambda_-$
or $\lambda_+ \ll \lambda_-$, it eventually approaches zero.
However, for not too large a hierarchy,
$\left|\log_{10}\left(\lambda_+/\lambda\right)\right| \leq 1/2$,
the parameter $\rho$ always stays rather close to unity,
$\rho \geq (20/101)^{1/2} \simeq 0.44$.
We will therefore ignore the possibility of a parametrically suppressed value of $\rho$
in the following and simply take it to be some $\mathcal{O}(1)$ constant
from now on, i.e., we will work with $\rho \sim 0.3 \cdots 1$.


Next, let us determine the axion decay constant $f_a$.
As we will discuss in Sec.~\ref{subsec:Rsymmetry}, the PQ symmetry
ends up acquiring a color anomaly due to the presence of additional matter states
coupling directly to the SUSY-breaking sector.
This is good news, since a PQ color anomaly is a necessary prerequisite
for any implementation of the PQ mechanism;
it generates the $a\,G\tilde{G}$ term by means of which
the $CP$-violating $\bar{\theta}$ term in the effective QCD Lagrangian
is eventually canceled,
\begin{align}
\mathcal{L}_{\rm eff}^{\rm QCD} \supset
\bar{\theta}\,\frac{\alpha_s}{8\pi}\,
\textrm{Tr}\left[G_{\mu\nu}\,\tilde{G}^{\mu\nu}\right] -
\left|\mathcal{A}_{\rm PQ}\right| \frac{a/\sqrt{2}}{K_0^{1/2}}\,
\frac{\alpha_s}{8\pi}\,
\textrm{Tr}\left[G_{\mu\nu}\,\tilde{G}^{\mu\nu}\right] \,.
\end{align}
Here, $\alpha_s$ is the strong coupling constant,
$G_{\mu\nu}$ and $\tilde{G}^{\mu\nu}$ respectively denote the gluon
field strength tensor and its dual and 
$\mathcal{A}_{\rm PQ}$ stands for the coefficient of the
$U(1)_{\rm PQ}\textrm{--}SU(3)_C\textrm{--}SU(3)_C$ anomaly.
The axion decay constant $f_a$ is now \textit{defined} such that these 
two terms can be combined to yield
\begin{align}
\mathcal{L}_{\rm eff}^{\rm QCD} \supset
\left(\bar{\theta} - \frac{a}{f_a}\right)\frac{\alpha_s}{8\pi}\,
\textrm{Tr}\left[G_{\mu\nu}\,\tilde{G}^{\mu\nu}\right] \,,
\end{align}
which tells us that the decay constant $f_a$ is basically given by the
VEV of the K\"ahler potential,
\begin{align}
f_a = \frac{\sqrt{2}}{\left|\mathcal{A}_{\rm PQ}\right|}\, K_0^{1/2}
= \frac{2}{\left|\mathcal{A}_{\rm PQ}\right|}\frac{1}{\rho}
\left(1-\zeta\right)^{1/2}\, \frac{\Lambda}{\eta}
= \frac{2}{\left|\mathcal{A}_{\rm PQ}\right|}\frac{\varepsilon}{\rho}
\frac{\Lambda}{\eta} \,.
\label{eq:faLambda}
\end{align}


In view of this result, three comments are in order:
(i) Irrespectively of the details of how the PQ mechanism is actually
implemented in a concrete model, naive dimensional analysis leads us to expect
that $f_a$ should be suppressed compared to the PQ breaking scale
by one power of the NDA parameter $\eta$~\cite{Manohar:1983md}.
Our result in Eq.~\eqref{eq:faLambda} obviously complies with this expectation.
(ii) In addition to this, the axion decay constant turns out to be further suppressed
due to various factors beyond the simple NDA estimate, $f_a \sim \Lambda/\eta$.
It is also suppressed by the anomaly coefficient $\left|\mathcal{A}_{\rm PQ}\right|$ as well
as by the suppression factor $\varepsilon$ in the meson VEVs.
Depending on the size of these prefactors, $f_a$ may be smaller
than the dynamical scale $\Lambda$ by one or even more orders of magnitude.
For $\varepsilon \sim 0.1$, $\left|\mathcal{A}_{\rm PQ}\right| \sim 1 \cdots 10$
and $\eta \sim \pi \cdots 4\pi$, for instance, we typically have
$f_a \sim 10^{-2}\,\Lambda$, so that SUSY breaking scales of
$\mathcal{O}\left(10^{12}\right)\,\textrm{GeV}$ result in
axion decay constants of $\mathcal{O}\left(10^{10}\right)\,\textrm{GeV}$.
In fact, as we will see in Secs.~\ref{subsec:constraints} and \ref{subsec:scan},
the $\Lambda$ and $f_a$ values consistent with all phenomenological
constraints will happen to be of exactly these orders of magnitude.
(iii) Similarly to the meson VEVs in Eq.~\eqref{eq:Mpm}, the relation
between $f_a$ and $\Lambda$ in Eq.~\eqref{eq:faLambda} is sensitive to
corrections coming from the dynamically generated K\"ahler potential.
Imagine, for instance, that the mesons $M_\pm$ obtain a quartic K\"ahler
potential due to strong-coupling effects,
\begin{align}
K = M_\pm M_\pm^\dagger + \Delta K \equiv \left(1+c_\pm\right)\,M_\pm M_\pm^\dagger 
\,, \quad
\Delta K  = \pm \,C_\pm \left(\frac{\eta}{\Lambda}\right)^2
\left(M_\pm M_\pm^\dagger\right)^2 \,,
\end{align}
where the size of the coefficients $C_\pm$ is unknown and where
the $c_\pm = \Delta K / \big(M_\pm M_\pm^\dagger\big)$ parametrize the ratio between
the noncanonical corrections and the canonical K\"ahler potential.
Large corrections, $\left|c_\pm\right| \gg 0$, therefore shift the normalization of
the meson fields as well as the VEV of the K\"ahler potential, $K_0$, which in turn
modifies the relation between $f_a$ and $\Lambda$ in Eq.~\eqref{eq:faLambda}.
Such large corrections do, however, not drastically affect our final conclusions, as all
bounds  that we are going to study in Sec.~\ref{subsec:constraints} will actually be bounds
on $\Lambda$.
In our parameter analysis, large corrections to the relation in Eq.~\eqref{eq:faLambda}
would therefore only result in different labels along the $f_a$ axes in our plots;
the functional dependences displayed in these plots would still remain the same.


\subsection{Stabilization of the axino and saxion}
\label{subsec:stabilization}


Next to the axion $a$ itself, the chiral axion multiplet $A$ in Eq.~\eqref{eq:ATheta}
also contains the superpartners of the axion: a two-component Weyl fermion, the axino $\tilde{a}$,
as well as a real scalar, the saxion $\phi$.
These two particles are potentially produced in large numbers in the early
universe by means of various thermal and nonthermal processes.
This might have significant cosmological implications~\cite{Tamvakis:1982mw,Kawasaki:2013ae}.
Axino and saxion decays may, for instance, lead to the overproduction of
dark matter, inject too much entropy into the thermal bath, thereby diluting the
primordial baryon asymmetry, or alter the predictions of big bang nucleosynthesis.
In order to prevent these catastrophic effects from taking place, either both
the axino and saxion abundances need to be adequately suppressed or both
species have to decay sufficiently fast.
Here, the latter solution is, in particular, realized once $\tilde{a}$ and $\phi$
are given sufficiently large masses.
On the other hand, one can show on rather general grounds that, in the supersymmetric
limit, the entire axion multiplet $A$ is necessarily massless~\cite{Kugo:1983ma}.
In the case of unbroken SUSY, the saxion $\phi$ especially represents a flat direction
in the scalar potential of the PQ-breaking sector, so that it is prone to cause
cosmological problems.
A successful stabilization of the PQ-breaking fields ($M_\pm$ in our case) 
can therefore only be achieved as long as $\tilde{a}$ and $\phi$ acquire appropriate
soft masses in the course of spontaneous SUSY breaking.
In the following, we shall show that this is exactly what is happening in our model.


First of all, let us trade the three SUSY-breaking singlet fields $Z_+$, $Z_-$ and $X$
in Eq.~\eqref{eq:WMpm} for the following linear combinations,
\begin{align}
\label{eq:S012}
S_0 = & \: \frac{1}{\left(2-\zeta\right)^{1/2}} \left[\left(1-\zeta\right)^{1/2}\left(Z_+ + Z_-\right)
- \zeta^{1/2} X \right] \,,  \quad
S_1 = \frac{1}{2^{1/2}} \left(Z_+ - Z_-\right) \,, \\\nonumber
S_2 = & \: \frac{1}{\left(2-\zeta\right)^{1/2}} \left[\left(\zeta/2\right)^{1/2}\left(Z_+ + Z_-\right)
+ 2^{1/2}\left(1-\zeta\right)^{1/2} X \right] \,.
\end{align}
As we will see shortly, these fields will end up corresponding to the physical
mass eigenstates in the singlet sector.
In terms of these new fields as well as in terms of the fields $M$ and $\Theta$
in Eq.~\eqref{eq:MpmMTheta}, the effective superpotential
of the IYIT model in Eq.~\eqref{eq:WMpm} can now be rewritten as follows,
\begin{align}
\label{eq:WAMS}
W_{\rm eff} \simeq & \: \left[\mu^2 r^2
+ \frac{m^2}{2\,\mu^2}\left(2\,\textrm{ch}_\Theta\, F_A^2 -
2 \left(1-\,\textrm{ch}_\Theta\right)F_A M
-  M^2\right)\right] S_0 \\ \nonumber
- & \:\, m\, \textrm{sh}_\Theta \left(F_A + M\right) S_1
- m \left[r \left(1-\textrm{ch}_\Theta\right)\left(F_A + M\right)
- \left(\frac{1}{r}
+ \frac{m^2}{2\,r\,\mu^4} F_A M\right)M \right] S_2 \,,
\end{align}
where $\textrm{ch}_\Theta \equiv \cosh\Theta$ and $\textrm{sh}_\Theta \equiv \sinh\Theta$,
where we have introduced $F_A \equiv K_0^{1/2} = 2^{-1/2}\left|\mathcal{A}_{\rm PQ}\right| f_a$
as an alternative symbol for the normalization of the axion multiplet (see Eq.~\eqref{eq:ATheta}),
where $\mu \equiv \Lambda_{\rm SUSY}$ denotes the SUSY breaking scale
(see Eq.~\eqref{eq:LambdaSUSY}) and where the parameters $m$ and $r$ are defined as
\begin{align}
m = \lambda_h\, \frac{\Lambda}{\eta} \,, \quad r = \left(\frac{\zeta}{2-\zeta}\right)^{1/2} \,.
\label{eq:mrdef}
\end{align}


From the superpotential in Eq.~\eqref{eq:WAMS}, we can calculate the scalar potential for the
two scalar DOFs contained in $A = F_A\, \Theta \supset 2^{-1/2}\left(\phi + i a\right)$, i.e.,
for the axion $a$ as well as for the saxion $\phi$.
As discussed in more detail in Appendix~\ref{app:vacuum},
the VEVs of the three singlet fields $S_0$, $S_1$ and $S_2$ vanish
in the rigid SUSY limit.
Therefore, neglecting any SUGRA effects and setting all singlets to zero,
the axion scalar potential induced by the spontaneous breaking of SUSY takes the following form,
\begin{align}
V(\phi,a) = \mu^4 r^2 + m^2 F_A^2 \cosh\left(\sqrt{2}\, F_A^{-1}\phi\right)
 = \mu^4 + \frac{1}{2} m_\phi^2 \, \phi^2\left[1 + \frac{1}{6}\left(\phi/F_A\right)^2
+ \mathcal{O}\left(\left(\phi/F_A\right)^4\right) \right] \,.
\end{align}
Here, we have introduced $m_\phi^2 = 2\, m^2$ to denote the saxion mass and used
the fact that the four parameters $\mu$, $m$, $F_A$ and $r$ are actually not linearly
independent; as one may easily check, they satisfy the relation $\mu^4 r^2 + m^2 F_A^2 = \mu^4$
(see Eqs.~\eqref{eq:LambdaSUSY}, \eqref{eq:ATheta} and \eqref{eq:mrdef}).
The lesson from this scalar potential now is twofold.
First of all, we note that the scalar potential $V(\phi,a)$ does indeed
not depend on $a$, rendering the axion a flat direction.
This is, of course, expected, given that the field $a$ ought to represent the Nambu-Goldstone boson
associated with the spontaneous breaking of the $U(1)_A$ symmetry by construction.
Second, we find that the saxion indeed ends up being stabilized
thanks to the SUSY-breaking dynamics of our model, $\left<\phi\right> = 0$.
Its mass around the origin is controlled by the mass parameter $m \propto \lambda_h\, \Lambda$,
which is closely related to the SUSY breaking scale and which, moreover, goes to zero as soon 
as the SUSY-breaking sector decouples from the dynamics of PQ symmetry breaking (i.e.,
as soon as $\lambda_\pm \rightarrow 0$ for fixed $\kappa\,\eta$ in Eq.~\eqref{eq:WMpm}).


In order to study the interactions of the axion multiplet $A$ in the true vacuum,
it is therefore sufficient to restrict our analysis to the superpotential in Eq.~\eqref{eq:WAMS}
in the limit of small fluctuations of the goldstone phase $\Theta$ around zero,
\begin{align}
\label{eq:Wfinal}
W_{\rm eff} \simeq & \:\, \mu^2 S_0 - m\, A \, S_1  + \frac{m}{r}\, M\,S_2
+ \frac{m^2}{2\,\mu^2} \left(A^2 - M^2\right) S_0 \\ \nonumber
- & \:\, \frac{m}{2\,F_A} \left[2\, M A \, S_1 - \frac{1}{r}\, M^2 S_2
- r\left(A^2-M^2\right) S_2\right] + \cdots \,,
\end{align}
where we have replaced $\Theta$ by $A/F_A$ after expanding in powers of $\Theta$
and where the ellipsis stands for operators of dimension $4$ and higher.
This form of the superpotential provides us with a number of useful physical
insights:
(i) The singlet fields $S_0$, $S_1$ and $S_2$ indeed parametrize the fluctuations
of the physical mass eigenstates around the true vacuum.
Here, $S_1$ turns out to share a Dirac mass $m_A \equiv m$ with the axion field $A$,
while $S_2$ turns out to share a Dirac mass $m_M \equiv m/r$ with the ``radial'' meson field $M$.
Meanwhile, $S_0$ remains massless at tree level.
(ii) As is now evident, the mass parameter $m$ corresponds to the common Dirac mass of
$A$ and $S_1$, while $r$ parametrizes the gap in the mass spectrum, i.e.,
the ratio between the two Dirac masses, $r = m_A / m_M$.
(iii) Among the three singlet fields, $S_0$ is the only one with a nonvanishing F-term.
We can, thus, identify it with the goldstino (or Polonyi) multiplet which is responsible
for the spontaneous breaking of SUSY via its F-term, $\left|F_{S_0}\right| = \mu^2$.
Upon spontaneous SUSY breaking, its fermionic component, the goldstino $\tilde{s}_0$,
is therefore absorbed by the gravitino $\tilde{G}$ (playing the role of its longitudinal DOFs thereafter),
which is why it eventually acquires a mass $m_{\tilde{s}_0} \equiv m_{3/2}$.
At the same time, the scalar component of the goldstino multiplet, the sgoldstino $s_0$,
is a flat direction of the scalar potential at tree level, as it is present in any SUSY-breaking
model of the O'Raifeartaigh type.


At the loop level, the (pseudo)modulus $s_0 \subset S_0$ is lifted via radiative corrections.
The relevant loop diagrams arise from the $A^2S_0$ and $M^2S_0$ Yukawa interactions
in Eq.~\eqref{eq:Wfinal} as well as from the Yukawa interactions of the $X$ component of the
goldstino field, $S_0 = -r\, X +\cdots$, with the neutral mesons in the full effective
superpotential (i.e., from the $M_0^aM_0^b\,X$ terms in Eq.~\eqref{eq:Weffcha}).
This results in the following contribution to the sgoldstino mass~\cite{Chacko:1998si}
(see Appendix~\ref{app:vacuum} for details),%
\footnote{In~\cite{Domcke:2014zqa}, a similar expression for the loop-induced
mass of the pseudoflat direction has been derived (see Eq.~(57) in this paper).
This expression does, however, not feature the weight function $\omega$, as the authors
of~\cite{Domcke:2014zqa} only work in the limit $\kappa\rightarrow\infty$,
where $\omega\rightarrow0$.
In this limit, the deformed moduli constraint is fulfilled exactly, 
the meson field $M$ decouples completely and the $M^2S_0$ interaction
no longer contributes to $m_0^2$.
Similarly, the $M_0^aM_0^bX$ interactions have been neglected in~\cite{Domcke:2014zqa},
so that the terms weighted by $\omega_0$ are missing in this paper.
Meanwhile, the calculation in \cite{Chacko:1998si} does account for the $M_0^aM_0^bX$ interactions.
But, as it is also based on the assumption of an exactly fulfilled moduli constraint
(i.e., on $\kappa\rightarrow\infty$), it, too, misses the contribution coming
from the $M^2S_0$ coupling.}
\begin{align}
m_0^2 = & \: \frac{2\ln2-1}{16\pi^2} \left[1 + \omega(r) +
\frac{2}{\rho^6}\,\bigg(\left(\frac{\lambda_{14}}{\lambda}\right)^2\omega_0\big(\lambda_0^{1,4}\big) +
\left(\frac{\lambda_{23}}{\lambda}\right)^2\omega_0\big(\lambda_0^{2,3}\big)\bigg)\right]
\left(\frac{m}{\mu}\right)^4 m^2 \,, 
\label{eq:sgoldstinom}\\\nonumber
\omega(r) = & \: \frac{1}{2\ln2-1}
\left[\frac{1}{2}\left(1+\frac{1}{r^2}\right)^2\ln\left(1+r^2\right)
- \frac{1}{2}\left(1-\frac{1}{r^2}\right)^2\ln\left(1-r^2\right) - \frac{1}{r^2}\right] \approx r^2 \,.
\end{align}
Here, the function $\omega$, which smoothly interpolates between $\omega(0)=0$ and $\omega(1)=1$,
acts as weight for the relative importance of the $M^2S_0$ interaction compared to
the $A^2S_0$ interaction.
In the case of a degenerate mass spectrum (i.e., for $r = 1$), diagrams with virtual $M$ lines
in the loop yield the same contribution to the sgoldstino mass as diagrams with virtual $A$ lines.
On the other hand, once the meson field $M$ becomes much heavier than the axion field $A$
(i.e., for $r \rightarrow 0$), the sgoldstino mass ceases to receive contributions from
$M$ loops.
In this limit, saxion and axino loops then remain as the only source of
mass generation for the sgoldstino in the charged meson sector.
At the same time, the contributions to $m_0^2$ due to the interaction of $S_0$ with
the neutral mesons are weighted by $\omega_0$,
evaluated as a function of the Yukawa couplings $\lambda_0^{1,4}$ and $\lambda_0^{2,3}$,
respectively.
The full expression for $\omega_0$ is given in Appendix~\ref{app:vacuum}.
For now, we merely remark that, as long as $\lambda_0^a \geq \lambda$ for all $a$,
also this weight smoothly interpolates between $0$ and $1$.
Here, the maximal value, $\omega_0 = 1$, is, in particular, attained
in the flavor-symmetric limit, i.e., for all $\lambda_0^a $ being equal to $\lambda$.
Furthermore, we note that the contributions to $m_0^2$ induced by the neutral meson loops come
with prefactors proportional to $\lambda_{14}^2$ and $\lambda_{23}^2$, which are defined as follows,
\begin{align}
\lambda_{14} = \left[\lambda^4 + \frac{1}{4}\left(\left(\lambda_0^1\right)^2
-\left(\lambda_0^4\right)^2\right)^2\right]^{1/4} \,, \quad
\lambda_{23} = \left[\lambda^4 + \frac{1}{4}\left(\left(\lambda_0^2\right)^2
-\left(\lambda_0^3\right)^2\right)^2\right]^{1/4} \,. \label{eq:lambda1423}
\end{align}
In the limit of equal Yukawa couplings, $\lambda_0^a = \lambda$, the prefactors
$\left(\lambda_{14}/\lambda\right)^2$ and $\left(\lambda_{23}/\lambda\right)^2$
therefore also reduce to unity, so that, in this limit, the sgoldstino mass
in Eq.~\eqref{eq:sgoldstinom} takes the following form,
\begin{align}
\lambda_0^a \equiv \lambda \quad\Rightarrow\quad
m_0^2 = \frac{2\ln2-1}{16\pi^2} \left[1 + \omega(r) +
\frac{4}{\rho^6}\right] \left(\frac{m}{\mu}\right)^4 m^2 \,, \quad
\omega(r)\approx r^2 \,. \label{eq:m02approx}
\end{align}
For simplicity and since we do not expect any large flavor hierarchy
in the IYIT model, we will work with this expression for $m_0^2$
(including the approximation $\omega(r)\approx r^2$) in the following.


In addition, working with an even more precise expression for $m_0^2$ (such as,
for instance, the one in Eq.~\eqref{eq:sgoldstinom}) would not be of much help for another reason:
Unfortunately, next to the perturbative Yukawa interactions encoded in the effective
superpotential, the sgoldstino mass also receives contributions
from the effective K\"ahler potential~\cite{Chacko:1998si}.
The true sgoldstino mass squared, $m_{s_0}^2$, is then given as the sum of
$m_0^2$ and some dynamically generated and \textit{uncalculable} correction,
\begin{align}
m_{s_0}^2 = m_0^2 + \Delta m_{K_{\rm eff}}^2 \,.
\label{eq:ms02}
\end{align}
For large Yukawa couplings, $\lambda \sim \eta$, we expect the uncalculable
correction $\Delta m_{K_{\rm eff}}^2$ to be of similar (but not much greater)
importance as the perturbative result $m_0^2$.
We note that this will be the more relevant case in the context of our phenomenological
study later on (see Sec.~\ref{sec:pheno}).
For smaller Yukawa couplings, $\lambda \ll \eta$, on the other hand,
we have more confidence in the purely perturbative calculation.
The upshot of these considerations is that the true sgoldstino mass
is, most likely, always roughly of the order of the expression in
Eq.~\eqref{eq:m02approx}, $m_{s_0} \sim m_0$.
On top of that, if we further assume $\Delta m_{K_{\rm eff}}^2$ to be positive, $m_0$
represents a lower bound on the actual sgoldstino mass,
\begin{align}
\Delta m_{K_{\rm eff}}^2> 0 \quad\Rightarrow\quad m_{s_0}^2 \geq m_0^2 \,.
\end{align}
This comes in handy, because it allows us to determine a \textit{conservative} upper bound
on the sgoldstino VEV after taking into account the effect of $R$ symmetry breaking
(see Appendix~\ref{app:vacuum}).
Such a conservative upper bound on $\left<S_0\right>$ is useful, since it prevents
us from underestimating the impact of higher dimensional operators
on the quality of the PQ symmetry (see Sec.~\ref{sec:pheno}).


(iv) Finally, a few comments on the masses of the remaining bosonic
and fermionic DOFs contained in $S_0$, $S_1$, $S_2$, $M$, and $A$ are in order.
In the true vacuum and neglecting the effect of SUGRA on the VEV
of the sgoldstino field $S_0$, the scalar masses in our model are given as follows,
\begin{align}
m_{s_0^\pm}^2 \sim m_0^2 \,, \quad
m_\phi^2 = 2\, m^2 \,, \quad m_a^2 = 0 \,, \quad m_{s_1^\pm}^2 = m^2 \,, \quad
m_{m^\pm}^2 = \frac{m^2}{r^2}\left(1 \pm r^2\right) \,, \quad
m_{s_2^\pm}^2 = \frac{m^2}{r^2} \,.
\label{eq:mbos}
\end{align}
Similarly, we obtain for the fermionic masses in the globally supersymmetric limit
\begin{align}
m_{\tilde{s}_0}^2 = 0 \,, \quad m_{\tilde{a}}^2 = m_{\tilde{s}_1}^2 = m^2 \,, \quad
m_{\tilde{m}}^2 = m_{\tilde{s}_2}^2 = \frac{m^2}{r^2} \,.
\label{eq:mfer}
\end{align}
The dependence of these different mass eigenvalues on the Yukawa couplings $\lambda$
and $\kappa$ becomes more transparent, if we rewrite them as functions of
$\zeta = \lambda^2/\left(\kappa^2\,\eta^2\right)$ (see Eqs.~\eqref{eq:Mpm}
and \eqref{eq:mrdef}),
\begin{align}
\frac{m^2}{\rho^2\kappa^2\Lambda^2} = \zeta \,, \quad
\frac{m^2/r^2}{\rho^2\kappa^2\Lambda^2} = 2-\zeta \,, \quad
\frac{m^2/r^2\left(1+r^2\right)}{\rho^2\kappa^2\Lambda^2} = 2\,, \quad
\frac{m^2/r^2\left(1-r^2\right)}{\rho^2\kappa^2\Lambda^2} = 2\left(1-\zeta\right) \,,
\end{align}
These expressions allow us to study the SUSY-breaking and PQ-preserving
limit ($\lambda\rightarrow\kappa\,\eta$) as well as the SUSY-preserving and PQ-breaking
limit ($\lambda\rightarrow0$) of our model in a nice fashion.%
\footnote{As for the sgoldstino, we have
$m_0^2/(\rho^2\kappa^2\Lambda^2)
= (2\ln2-1)(4-2\,\zeta+\rho^6)/\rho^2\lambda^2 /(8\pi^2)\,\zeta \left(2-\zeta\right)^{-2}$,
which turns into $(2\ln2-1)(2+\rho^6)/\rho^2\lambda^2 /(8\pi^2)$
for $\lambda\rightarrow\kappa\,\eta$ ($\simeq 2.3$ for $\lambda = 4\pi$ and $\rho = 1$)
and into $0$ for $\lambda\rightarrow0$.
We also recall that
$m_{3/2}^2/(\rho^2\kappa^2\Lambda^2) =
(\Lambda^2/\eta^2)/(3M_{\rm Pl}^2)/\rho^2\,\zeta\left(2-\zeta\right)$.
This goes to $(\Lambda^2/\eta^2)/(3M_{\rm Pl}^2)/\rho^2$ for
$\lambda\rightarrow\kappa\,\eta$ and to $0$ for $\lambda\rightarrow0$.}
For $\lambda\rightarrow\kappa\,\eta$, the above masses squared (in units of $\rho^2\kappa^2\Lambda^2$)
approach $\left\{1,1,2,0\right\}$, while for $\lambda\rightarrow0$, they turn into
$\left\{0,2,2,2\right\}$.
Here, the massless field in the PQ-preserving limit (the real meson scalar $m_-$) is 
the result of an accidental cancellation in the scalar mass matrix for the special
parameter choice $\lambda = \kappa\,\eta$.
In the limit $\lambda\rightarrow\kappa\,\eta$, the field $m_-$, thus, becomes
the second lightest state in the IYIT model, the only lighter field being the massless axion $a$.
At the same time, the massless fields in the SUSY-preserving limit correspond to
the DOFs contained in $S_0$, $S_1$, and $A$.
We, hence, see once again that it is mandatory to break SUSY in order
to stabilize the axino as well as the saxion.


Next, we note that, according to the above results for the mass eigenvalues in our model,
the physical fields at low energies appear to correspond to four real scalars ($\phi$, $a$, $m_+$, and $m_-$),
three complex scalars ($s_0$, $s_1$, and $s_2$), one Weyl fermion ($\tilde{s}_0$) as well as two 
Dirac fermions ($\left(\tilde{a},\tilde{s}_1\right)$ and $\left(\tilde{m},\tilde{s}_2\right)$).
In fact, all mass degeneracies in Eqs.~\eqref{eq:mbos} and \eqref{eq:mfer} are, however,
lifted through SUGRA effects---see Appendix~\ref{app:vacuum}, where we derive the VEVs of
all singlet fields taking into account the effect of $R$ symmetry breaking and
state the full expressions for all bosonic and fermion masses
given a nonzero value of $\left<S_0\right>$.
The fields in Eqs.~\eqref{eq:mbos} and \eqref{eq:mfer} are therefore only quasi-degenerate,
i.e., they are only degenerate in the rigid SUSY limit.
In the full SUGRA case, we have to deal instead with ten real scalars,
one Weyl fermion and four Majorana fermions.
The mass splittings among the quasi-complex scalars and quasi-Dirac fermions is then
of $\mathcal{O}\left(m_{3/2}\right)$ and therefore quite large.
Last but not least, we mention that, imposing the deformed moduli constraint
exactly, i.e., in the limit $\kappa\rightarrow\infty$, the fields contained in $M$
and $S_2$ become formally infinitely heavy.
In this limit, they are, thus, unphysical and need to be integrated out
(see also the discussion in~\cite{Domcke:2014zqa}).
The effective superpotential of the IYIT model in Eq.~\eqref{eq:Wfinal} then turns into
\begin{align}
W_{\rm eff} \simeq \mu^2 S_0 - m\, A \, S_1 + \frac{m^2}{2\,\mu^2} A^2 S_0 \,.
\label{eq:Wsimple}
\end{align}
which is nothing but the superpotential studied in~\cite{Domcke:2014zqa} (see Eq.~(43) therein). 


\section{Quality of the PQ symmetry and phenomenological constraints}
\label{sec:pheno}


\subsection[Protecting the PQ symmetry by means of an anomaly-free $Z_4^R$ symmetry]
{Protecting the PQ symmetry by means of an anomaly-free \boldmath{$Z_4^R$} symmetry}
\label{subsec:Rsymmetry}


Up to now, we have only discussed the renormalizable interactions
among the fields of the IYIT model.
In the context of SUGRA, we, however, expect gravitational effects at the Planck
scale to induce further, nonrenormalizable interactions among these fields in the
low-energy effective theory.
\textit{A priori}, there is no reason why these additional interactions should happen to
respect the global PQ symmetry that is enjoyed by the IYIT model in the rigid SUSY limit.
Instead, the full effective superpotential as well as the full effective K\"ahler
potential at energies below the dynamical scale are expected to contain
higher-dimensional operators that explicitly break PQ,
\begin{align}
W_{\rm eff}^{\cancel{\rm PQ}} \supset \frac{\Lambda^2}{M_*} M_\pm^2 \,, \:\:
\frac{\Lambda^3}{M_*^3} M_\pm^3 \,, \:\:
\frac{1}{M_*} Z_\pm^4 \,, \:\: \cdots \,, \quad 
K_{\rm eff}^{\cancel{\rm PQ}} \supset \frac{\Lambda^2}{M_*^2} M_\pm^2 \,, \:\:
\frac{\Lambda^3}{M_*^4} M_\pm^3 \,, \:\:
\frac{1}{M_*} Z_\pm^3 \,, \:\: \cdots  \,, \label{eq:WKPQ}
\end{align}
with $M_*$ denoting an appropriate high-energy cut-off scale close to the Planck
scale, $M_* \sim M_{\rm Pl}$.
These operators result in corrections to the ordinary axion potential in QCD,
which causes the axion VEV to shift from its desired value,
$\left<a\right> = f_a \,\bar{\theta}$, to some displaced value,
$\left<a\right> = f_a \left(\bar{\theta} + \Delta\bar{\theta}\right)$,
at which $CP$ is no longer conserved.
In other words, the gravity-induced higher-dimensional operators in the
effective theory re-introduce a nonzero QCD vacuum angle,
$\Delta\bar{\theta}$, (which may easily become very large,
$\Delta\bar{\theta} \gg 10^{-10}$) and, hence, bring us back to the
original strong $CP$ problem.


In order to suppress $\Delta\bar{\theta}$ below the experimental bound,
$\Delta\bar{\theta} \lesssim 10^{-10}$, it is necessary to forbid all
effective operators that explicitly violate PQ up to some high order.
As discussed in the introduction (see also \cite{Harigaya:2013vja}),
this is best done by invoking a protective gauge symmetry that eliminates
all of the relevant dangerous operators from the effective theory.
In this paper, we shall, in particular, rely on a discrete $Z_4^R$ symmetry,
which is well motivated from the perspective of PGM (see our discussion in Sec.~\ref{subsec:Z4RPGM}).
Let us now derive the charge spectrum for such a $Z_4^R$ symmetry in the context of
the IYIT model and assess which PQ-breaking operators in the effective theory it is able for forbid.
The characteristics of general $Z_N^R$ symmetries along with the charge assignment
for the MSSM fields have already been reviewed in~\cite{Harigaya:2013vja},
which is why we will be rather brief in what follows.
First of all, let us group the fields of the MSSM into complete multiplets of $SU(5)$,
$\mathbf{10} = \left(q,u^c,e^c\right)$, $\mathbf{5}^* = \left(d^c,\ell\right)$,
and $\mathbf{1} = \left(n^c\right)$, since we are only interested in charge assignments that
are at least compatible with $SU(5)$ unification.
The MSSM fields are then charged under the discrete $Z_4^R$ symmetry as follows
(for details, see Sec.~2.2.2 and Appendix A in \cite{Harigaya:2013vja}),
\begin{align}
\left(r_{\mathbf{10}},\, r_{\mathbf{5}^*},\, r_{\mathbf{1}},\, r_{H_u},\, r_{H_d}\right) \modulo{4}
\frac{1}{5}\left(1,\, -3,\, 5,\, 8,\, 12\right)
+ \frac{2\,\alpha}{5} \left(1,\, -3,\, 5,\, -2,\, 2\right) \,.
\label{eq:MSSSMcharges}
\end{align}
Here, $\alpha$ is an integer that can take any value between $0$ and $9$.
Also, notice that the row vector multiplied by $2\alpha/5$ on the right-hand side of
Eq.~\eqref{eq:MSSSMcharges} encompasses the charges of the MSSM fields under
the Abelian GUT group $U(1)_X$.
This group, sometimes referred to as ``fiveness'', commutes with $SU(5)$ and
can be represented as a linear combination of the weak hypercharge $Y$ and the
difference between baryon number $B$ and lepton number $L$, i.e., $X = 5\,(B$$-$$L)-4\,Y$.
We note that the MSSM $R$ charges are not uniquely defined, since the MSSM superpotential
(including Majorana mass terms for the singlet neutrino fields $n^c$) happens to be invariant
under $Z_{10} \subset U(1)_X$ transformations.
This ambiguity leaves us with ten different solutions for the MSSM $R$ charges.


Next, we point out that, solely within the MSSM, the $Z_4^R$ symmetry turns out to be
anomalously violated at the quantum level for every possible $R$ charge assignment.
This is illustrated by the fact that the color as well as the weak anomaly coefficient
for the $Z_4^R$ symmetry are always nonzero,
\begin{align}
\mathcal{A}_R^{(C)} = & \:\mathcal{A}\left[Z_4^R\textrm{--}SU(3)_C\textrm{--}SU(3)_C\right] =
6 + N_g \left(3\,r_\mathbf{10} + r_{\mathbf{5}^*} - 4\right) \,,
\label{eq:ARCL}\\ \nonumber
\mathcal{A}_R^{(L)} = & \:\mathcal{A}\left[Z_4^R\textrm{--}SU(2)_L\textrm{--}SU(2)_L\right] =
4 + N_g \left(3\,r_\mathbf{10} + r_{\mathbf{5}^*} - 4\right) + \left(r_{H_u} + r_{H_d} - 2\right)\,,
\end{align}
where $N_g = 3$ stands for the number of SM fermion generations.
To see that $\mathcal{A}_R^{(C)}$ and $\mathcal{A}_R^{(L)}$ are indeed nonzero
(without inserting all possible $R$ charges into Eq.~\eqref{eq:ARCL} by hand),
we have to employ Eq.~\eqref{eq:rHuHd} as well as the following relations between 
the charges $r_\mathbf{10}$, $r_{\mathbf{5}^*}$, $r_{H_u}$, and $r_{H_d}$,
\begin{align}
2\,r_\mathbf{10} + r_{H_u} \modulo{4} 2 \,, \quad 
r_{\mathbf{5}^*} + r_\mathbf{10} + r_{H_d} \modulo{4} 2 \,, 
\end{align}
which directly result from the MSSM Yukawa interactions.
We then find for 
$\mathcal{A}_R^{(C)}$ and $\mathcal{A}_R^{(L)}$,
\begin{align}
\mathcal{A}_R^{(C)} \modulo{4} \mathcal{A}_R^{(L)} \modulo{4} 6 - 4\, N_g \modulo{4} -2 \,.
\end{align}
In order to cancel the MSSM contributions to $\mathcal{A}_R^{(C)}$ and $\mathcal{A}_R^{(L)}$
and, hence, render the $Z_4^R$ symmetry anomaly-free, we are therefore led to introduce new
matter multiplets that transform under $SU(5)$.
Here, the easiest possibility is to simply add a certain number of
$\mathbf{5}$ and $\mathbf{5}^*$ representations, which we shall refer
to as $Q_i$ and $\bar{Q}_i$ in the following.
Given $k \in\mathbb{N}$ of such new ``quark/antiquark'' pairs, the total $R$ charge
of the extra matter fields needs to satisfy
\begin{align}
r_{Q\bar{Q}} = r_{Q} + r_{\bar{Q}} \modulo{4} 2 - \Delta r \,, \quad
\Delta r = - \frac{1}{k} \left(2 + 4\,\ell\right) \,, \quad
\ell = 0,1,\cdots k-1 \,,
\label{eq:rQQ}
\end{align}
so as to make the $Z_4^R$ anomaly coefficients vanish.
We, thus, conclude that our axion model predicts the existence of new SM-charged
fields, without the aid of which we would not able not invoke an anomaly-free
discrete $Z_4^R$ symmetry as a protective gauge symmetry for the PQ symmetry.


Moreover, the new matter fields $Q_i$ and $\bar{Q}_i$ come in handy for another reason.
In order to allow for a successful solution of the strong $CP$ problem, the PQ symmetry
in the IYIT sector needs to exhibit a color anomaly.
This is now easily achieved by coupling $Q_i$ and $\bar{Q}_i$ to the SUSY-breaking
sector.
For instance, and w.l.o.g., we may assume that all $Q\bar{Q}$ pairs couple to $\Psi^1\Psi^2$
via some Planck-suppressed operators.
Above and below the dynamical scale, we then respectively have
\begin{align}
W^Q = \sum_{i=1}^k \frac{C_{Q_i}'}{M_*} \left(Q \bar{Q}\right)_i \Psi^1\Psi^2  \,, \quad
W_{\rm eff}^Q \simeq \sum_{i=1}^k \frac{C_{Q_i}}{M_*} \frac{\Lambda}{\eta} \left(Q \bar{Q}\right)_i M_+ \,,
\label{eq:WQ}
\end{align}
where the dimensionless coefficients $C_{Q_i}'$ and $C_{Q_i}$ are naively expected to be
of $\mathcal{O}(1)$ or at most as large as $4\pi$.
In view of the superpotential terms in Eq.~\eqref{eq:WQ}, three comments are now in order:
(i) Since the meson field $M_+$ carries PQ charge $+1$, each quark/antiquark pair must
carry PQ charge $-1$.
Meanwhile, as no MSSM field couples to the SUSY-breaking sector directly, all MSSM fields
remain uncharged under the PQ symmetry.
The total PQ color anomaly, hence, receives contributions from the new quark fields only,
\begin{align}
\mathcal{A}_{\rm PQ}^{(C)} = k \left(q_Q + q_{\bar{Q}}\right) = -k \,, \quad q_Q + q_{\bar{Q}} = -1 \,.
\label{eq:APQC}
\end{align}
This renders our axion model a special supersymmetric variant of the
KSVZ axion model invented by Kim, Shifman, Vainshtein, and Zakharov a long time ago~\cite{Kim:1979if}. 
(ii) The superpotential couplings in Eq.~\eqref{eq:WQ} also act as mass
terms for the new quark fields.
In fact, upon spontaneous PQ symmetry breaking, each quark pair
acquires a supersymmetric Dirac mass close to the gravitino mass,
\begin{align}
m_{Q_i} = \frac{C_{Q_i}}{M_*} \frac{\Lambda}{\eta} \left<M_+\right> 
= \frac{C_{Q_i}}{\lambda_+} \frac{\sqrt{3}\,M_{\rm Pl}}{M_*}
\left(\frac{1-\zeta}{2-\zeta}\right)^{1/2} m_{3/2}\,.
\label{eq:mQm32}
\end{align}
Therefore, depending on the value of $m_{3/2}$, the extra quark fields may or may not
be light enough to be within the reach of a future multi-TeV collider experiment.
Albeit extremely challenging, the discovery of $\mathcal{O}(k)$ new $SU(5)$ multiplets in the vicinity
of the gravitino mass would then, of course, be a smoking-gun signal of our axion model.


(iii) Last but not least, we note that the form of the superpotential in Eq.~\eqref{eq:WQ}
together with the IYIT superpotential in Eq.~\eqref{eq:Weffcha}
suffices to fix the $R$ charges of all fields in the IYIT sector,
\begin{align}
r_{M_{\pm}} = \pm\Delta r \,, \quad r_{Z_{\pm}} = 2 \pm\Delta r \,, \quad
r_{M_0^a} = 0 \,, \quad r_{Z_0^a} = r_X = 2 \,.
\label{eq:rIYIT}
\end{align}
Here, we have required that the $Z_4^R$ symmetry be anomaly-free under the
strongly coupled $SU(2)$.
Simply by itself, this implies that the charges of all IYIT quarks must sum
to zero: $r_{M_+} + r_{M_-} = 0$ and, thus, $r_X = 2$.
Notice that, if this was not the case, the dynamically generated superpotential
(see Eq.~\eqref{eq:WdynPf}) would explicitly break $R$ symmetry,
$Z_4^R \rightarrow \emptyset$, so that we could no longer rely on $R$ symmetry
as a tool to constrain the low-energy effective theory.
In particular, we would loose control over the dynamically generated
terms in the superpotential and K\"ahler potential, which might
lead to too large a gravitino mass or other unwanted effects.
Moreover, we point out that our result in Eq.~\eqref{eq:rIYIT} reveals an
interesting relation between the charges of the fields in the IYIT sector
under the local $Z_4^R$ symmetry and the charges of the same fields
under the continuous global $R$ symmetry of the IYIT model in the rigid SUSY limit.
Under the latter, all meson fields are uncharged, while all singlet fields carry
charge $+2$.
If we denote these global $R$ charges by $r^0$ for the individual fields
(and if we denote the corresponding PQ charges by $q$), we arrive at
\begin{align}
r = r^0 + q\,\Delta r \,,
\label{eq:rrq}
\end{align}
which holds for every field in the SUSY-breaking sector.
This is to say that, invoking an anomaly-free $Z_4^R$ symmetry
in the IYIT sector, we are actually doing nothing else but gauging a discrete
subgroup of the global $U(1)_R \times U(1)_A$ symmetry of the IYIT model.
As we shall demonstrate in the next section, this discrete gauge symmetry 
then allows us to eliminate dangerous higher-dimensional operators in
the effective theory.
At the same time, it also fixes the structure of the renormalizable
interactions in the IYIT model.
Without invoking any further symmetry, the renormalizable superpotential
and K\"ahler potential could also contain terms such as 
\begin{align}
W_{\rm eff}^{\cancel{\rm PQ}} \supset  Z_\pm^2 \,, \:\: Z_\pm^3 \,, \:\: \cdots\,, \quad 
K_{\rm eff}^{\cancel{\rm PQ}} \supset Z_\pm \,, \:\: Z_\pm^2 \,, \:\: \cdots \,.
\end{align}
So far, we have simply ignored this issue;
and now we see that, in general (i.e., for most values of the two
integers $k$ and $\ell$), such terms are automatically forbidden
by the $Z_4^R$ symmetry.


Finally, we mention that, as a result of the relation in Eq.~\eqref{eq:rrq},
the SUSY-breaking sector on its own turns out not to break the
$Z_4^R$ symmetry---even though the charged meson fields $M_\pm$
carry nonzero $R$ charge and obtain large VEVs.
Here, the point is that we can always rotate away the charges of the charged
meson fields by means of a global PQ phase transformation, such
that $r\rightarrow r'\equiv r^0$.
The strong dynamics of the IYIT sector therefore only break SUSY as
well as the global PQ symmetry, but leave the gauged $Z_4^R$ intact, though,
$Z_4^R \times U(1)_{\rm PQ} \rightarrow Z_4^{R\prime}$.
This remnant $Z_4^{R\prime}$ symmetry is then only broken,
$Z_4^{R\prime}\rightarrow Z_2^R$, by the constant term in the superpotential,%
\footnote{If the constant term in the superpotential is generated at very
high energies (for instance, via gaugino condensation~\cite{Veneziano:1982ah}),
we do not have to worry about any cosmological consequences
of $R$ symmetry breaking.
In such a case, all dangerous topological defects created during
$R$ symmetry breaking will simply be inflated away.\smallskip}
$W_0 = m_{3/2}\,M_{\rm Pl}^2$, as well as by higher-dimensional
operators in the effective theory (such as the quark mass term in Eq.~\eqref{eq:WQ}).
Here, it is interesting to observe that the $Z_2^R$ parity that
we are eventually left with can be identified with the $R$ parity of the MSSM.
Our model therefore automatically accounts for the origin of $R$ parity
in the MSSM.
That is, in contrast to many other models, it does not require any extension by,
say, a gauged $B$$-$$L$ symmetry to do so~\cite{Martin:1992mq}.


\subsection{Constraints on the axion decay constant}
\label{subsec:constraints}


We have not yet uniquely specified all properties of the
$Z_4^R$ symmetry. 
Our construction still exhibits two free parameters:
$k$, the number of extra quark pairs, as well as $\Delta r$,
the shift in the global $R$ charges $r^0$ (see Eqs.~\eqref{eq:rQQ} and \eqref{eq:rrq}).
In the following, we shall now examine for which values of these
parameters we have a chance of arriving at a viable phenomenology
as well as how the other parameters of our model (the axion decay constant $f_a$,
the gravitino mass $m_{3/2}$, the Yukawa coupling $\lambda$, etc.) are respectively 
constrained in these different scenarios.


First of all, we note that, in order to forbid as many PQ-breaking operators
as possible, it turns out advantageous to choose the integer $\ell$ in
Eq.~\eqref{eq:rQQ} such that $\Delta r$ ends up being a fraction
and not an integer, $\Delta r \not\in \mathbb{N}$.
This already rules out scenarios with only one or two extra quark pairs
from the start%
\footnote{For $k=1$, we are unable to forbid tadpole terms for the singlet fields
$Z_\pm$ in the renormalizable K\"ahler potential, $K \supset Z_\pm$, while for $k=2$,
we are unable to forbid supersymmetric mass terms for the same fields in the
renormalizable superpotential, $W \supset Z_\pm^2$.
These scenarios are, therefore, unfeasible from the very beginning.}
and implies that only the $k$-th powers of the fields $M_\pm$ and $Z_\pm$
can appear in the effective superpotential as well as the effective
K\"ahler potential.
The crucial point here is that, for $\Delta r \not\in \mathbb{N}$,
the smallest integer multiple of $\Delta r$ is nothing but
$k$ times $\Delta r$,
\begin{align}
k\, \Delta r \modulo{4} 2 \,.
\end{align}
For even and odd values of $k$, the lowest-dimensional
PQ-breaking  operators in $W_{\rm eff}^{\cancel{\rm PQ}}$ and
$K_{\rm eff}^{\cancel{\rm PQ}}$ are then respectively given as follows,
\begin{align}
W_{\rm eff}^{\cancel{\rm PQ}} \supset
\begin{cases}
M_\pm^k \,, \:\: Z_\pm^k  & ;\:\: \textrm{$k$ even} \\
M_\pm^k \,, \:\: m_{3/2}\,Z_\pm^k & ;\:\: \textrm{$k$ odd} 
\end{cases} \,, \quad \label{eq:WKk}
K_{\rm eff}^{\cancel{\rm PQ}} \supset 
\begin{cases}
m_{3/2}\,M_\pm^k \,, \:\: m_{3/2}\,Z_\pm^k  & ;\:\: \textrm{$k$ even} \\
m_{3/2}\,M_\pm^k \,, \:\: Z_\pm^k & ;\:\: \textrm{$k$ odd} 
\end{cases} \,.
\end{align}
Recall that the mesons $M_\pm$ acquire VEVs of $\mathcal{O}\left(\Lambda\right)$
(see Eq.~\eqref{eq:Mpm}), while, in the context of SUGRA, the singlets $Z_+$ and $Z_-$
obtain VEVs of $\mathcal{O}\left(m_{3/2}\right)$ (see Eq.~\eqref{eq:ZXVEVs}).
Together with $m_{3/2}\sim\Lambda^2/M_{\rm Pl}$ (see Eq.~\eqref{eq:m32}),
these estimates allow us to assess the order of magnitude of the respectively
most important corrections to the axion
scalar potential, $\Delta V_a$, induced by these PQ-breaking operators,
\begin{align}
W_{\rm eff}^{\cancel{\rm PQ}} \:\:\rightarrow\:\: \Delta V_a  \sim
\left(\frac{m_{3/2}}{M_{\rm Pl}}\right)^{k+1-c} M_{\rm Pl}^4 \,, \quad
K_{\rm eff}^{\cancel{\rm PQ}} \:\:\rightarrow\:\: \Delta V_a  \sim
\left(\frac{m_{3/2}}{M_{\rm Pl}}\right)^{k+2+c} M_{\rm Pl}^4 \,,
\label{eq:WKDeltaV}
\end{align}
where $c=1$ for even $k$ and $c=0$ for odd $k$.
Here, notice that the meson operators require a different power counting than
the singlet operators (see also Eq.~\eqref{eq:WKPQ}).
As the meson fields are, in fact, composite fields, $M^{ij} \sim \Psi^i\Psi^j / \Lambda$
(see Eq.~\eqref{eq:Mij}), each meson field is actually
accompanied by one power of the dynamical scale,
so that each meson VEV is bound to come with a suppression
factor of $\mathcal{O}\left(\Lambda/M_{\rm Pl}\right)$.
Therefore, despite the hierarchy between the actual VEVs,
$\left<M_\pm\right> \gg \left<Z_\pm\right>$, the effect of the
respective meson and singlet operators ends up being comparable,%
\footnote{This different power counting in the case of the meson operators
represents a distinctive feature of our \textit{dynamical} axion model, which
distinguishes it from our earlier axion models presented in \cite{Harigaya:2013vja}.
In this earlier work, the PQ-breaking fields are taken to be elementary fields,
$M_\pm\rightarrow P,\bar{P}$, so that their VEVs do not end up being suppressed
by a factor of $\mathcal{O}\left(\Lambda/M_{\rm Pl}\right)$.
This allows, \textit{inter alia}, for the possibility of extra quark fields
as heavy as the dynamical scale, $m_{Q_i} \sim \Lambda$, and increases the magnitude
of the PQ-breaking terms in the axion potential.
In the present paper, the mass scale of the new quark fields is, by contrast,
tied to the gravitino mass, $m_{Q_i} \sim \Lambda^2/M_{\rm Pl} \sim m_{3/2}$,
and the PQ-breaking terms in the axion potential are
generally more strongly suppressed.
We emphasize that it is these differences that explain why we cannot
simply use the results for general $Z_N^R$ symmetries obtained in \cite{Harigaya:2013vja}
and apply them to the present scenario in the special case of a $Z_4^R$ symmetry.
Instead, a new and dedicated study is necessary.}
\begin{align}
\frac{\Lambda}{M_{\rm Pl}}\left<M_{\pm}\right> \sim \frac{\Lambda^2}{M_{\rm Pl}} \sim m_{3/2}
\sim \left<Z_{\pm}\right> \,.
\end{align}


The main lesson from Eq.~\eqref{eq:WKDeltaV} is that all PQ-breaking effects induced
by the effective K\"ahler potential in Eq.~\eqref{eq:WKk} are suppressed compared to
the corresponding effects induced by the effective superpotential by at least one power
of the ratio $m_{3/2}/M_{\rm Pl}$.
This is perhaps not much of a surprise, since the K\"ahler potential in Eq.~\eqref{eq:WKk}
is holomorphic in the fields $M_\pm$ and $Z_\pm$, so that it can only contribute to the
total scalar potential via pure SUGRA terms.
By comparison, the lowest-dimensional \textit{nonholomorphic} terms
in $K_{\rm eff}^{\cancel{\rm PQ}}$ are obtained by multiplying
the terms in Eq.~\eqref{eq:WKk} by the $R$-invariant field products
$M_\pm M_\pm^*$ and $Z_\pm Z_\pm^*$, respectively.
These higher-dimensional terms then yield corrections to the axion scalar potential
which are of the same order of magnitude as the corrections induced by the holomorphic
terms in $K_{\rm eff}^{\cancel{\rm PQ}}$.
In summary, we therefore find that the PQ-breaking effects stemming from the
K\"ahler potential are always suppressed and that it suffices to focus on the PQ-breaking
operators contained in the superpotential in the following.


Let us now be a bit more specific and write down the operators in
$W_{\rm eff}^{\cancel{\rm PQ}}$ in Eq.~\eqref{eq:WKPQ}
including all prefactors, powers of the dynamical scale $\Lambda$,
powers of the cut-off scale $M_*$, etc.,
\begin{align}
W_{\cancel{\rm PQ}}^{\rm eff} \simeq 
\frac{C_{Z_\pm}}{k!}\left(1 \textrm{ or } \frac{m_{3/2}}{M_*}\right)
\frac{Z_\pm^k}{M_*^{k-3}} +
\frac{C_{M_\pm}}{\left(k!\right)^2}\frac{1}{\eta^2}
\left(\frac{\eta\,\Lambda}{M_*}\right)^k\hspace{-0.15cm}\frac{M_\pm^k}{M_*^{k-3}}\,,
\label{eq:WPQeff}
\end{align}
with the coefficients $C_{Z_\pm}$ and $C_{M_\pm}$
denoting some unknown constants of $\mathcal{O}(1)$
and where the prefactor of $Z_\pm^k$ is determined by whether the integer
$k$ is chosen to be even or odd (see Eq.~\eqref{eq:WKPQ}).
The most dangerous corrections to the axion scalar potential resulting
from this superpotential
(deriving from F-term contributions well as from A-term contributions in SUGRA)
are the following,
\begin{align}
\Delta V_a = & \:\,
\frac{C_{Z_\pm}}{(k-1)!}\,\lambda_\pm
\left(1 \textrm{ or } \frac{m_{3/2}}{M_*}\right)
\frac{\Lambda}{\eta} \frac{M_\pm^* Z_\mp^{k-1}}{M_*^{k-3}}
\label{eq:DeltaVa}\\\nonumber
+ & \:\, \frac{C_{M_\pm}}{\left(k!\right)^2} \frac{1}{\eta^2}
\left(\frac{\eta\,\Lambda}{M_*}\right)^k
\left(k\,\kappa\, \eta\, X^* M_\mp^* + k\,\lambda_\pm\, \frac{\Lambda}{\eta}\, Z_\mp^* +
(k-3)\, m_{3/2}\, M_\pm \right)\frac{M_\pm^{k-1}}{M_*^{k-3}} + \textrm{h.c.} \,,
\end{align}
where all chiral fields are understood to represent their scalar components.
In order to make the dependence of these terms on the axion field value $a$ manifest,
we need to expand the charged fields $M_\pm$ and $Z_\pm$ around their VEVs
(see Eq.~\eqref{eq:MpmMTheta}).
Taking into account the fact that $\left<Z_\pm\right> \neq 0$ in SUGRA
(see Appendix~\ref{app:vacuum}), we then have for the complex scalars
contained in $M_\pm$ and $Z_\pm$,
\begin{align}
M_\pm = \left<M_\pm\right>\,\exp\left(\pm \frac{i\,a}{\sqrt{2}\,F_A}\right) \,, \quad
Z_\pm = \left<Z_\pm\right>\,\exp\left(\pm \frac{i\,a}{\sqrt{2}\,F_A}\right) \,,
\label{eq:MpmZpm}
\end{align}
where we have set all further scalar DOFs contained in $M_\pm$ and $Z_\pm$
to zero.
In passing, we also mention that, in SUGRA, the scale $F_A$ also receives
contributions from the singlet fields $Z_\pm$,
\begin{align}
F_A = K_0^{1/2} \,, \quad
K_0 = \big<\left|M_+\right|^2\big> + \big<\left|M_-\right|^2\big> + 
\big<\left|Z_+\right|^2\big> + \big<\left|Z_-\right|^2\big> \,.
\label{eq:FASUGRA}
\end{align}
However, since the VEVs of the singlets are much smaller than the meson VEVs,
this is only a small correction compared to the globally supersymmetric case.
In the following, we shall therefore neglect the SUGRA corrections to $F_A$
in Eq.~\eqref{eq:FASUGRA} and simply work with the expression in Eq.~\eqref{eq:ATheta}.


Plugging the expressions in Eq.~\eqref{eq:MpmZpm} into the scalar potential
in Eq.~\eqref{eq:DeltaVa}, we find that all dangerous operators in
the axion scalar potential can be brought into the following form,
\begin{align}
\Delta V_a \supset \frac{1}{2}\,v^4 \left[\exp\left(\pm i\,\frac{k\,a}{\sqrt{2}\,F_A}\right)
+ \textrm{h.c.}\right] = v^4 \cos\left(\frac{k\,a}{\sqrt{2}\,F_A}\right) \,, \quad
v \sim \left(\frac{m_{3/2}}{M_{\rm Pl}}\right)^{(k+1-c)/4} M_{\rm Pl} \,,
\label{eq:v4cos}
\end{align}
for some appropriate mass scale $v$ that differs from operator to operator.
This correction to the scalar potential needs to be compared with the
instanton-induced scalar axion potential in QCD,
\begin{align}
V_a^{(0)} \simeq m_a^2\, f_a^2
\left[1-\cos\left(\bar{\theta}-\frac{a}{f_a}\right)\right] \,, \quad
m_a = \frac{z^{1/2}}{1+z} \frac{m_\pi f_\pi}{f_a} \simeq
600\,\textrm{\textmu eV} \left(\frac{10^{10}\,\textrm{GeV}}{f_a}\right) \,,
\end{align}
with $m_a$ denoting the axion mass in QCD, which is determined
by the $\pi^0$ mass $m_{\pi^0} \simeq 135\,\textrm{MeV}$, the $\pi^0$
decay constant $f_{\pi^0} \simeq 92 \,\textrm{MeV}$,
the ratio of the up and the down quark mass,
$z=m_u / m_d \simeq 0.56$, as well as by the axion decay constant
$f_a$~\cite{Agashe:2014kda}.
The sum of $V_a^{(0)}$ and $\Delta V_a$
is then no longer minimized at the CP-conserving field value $\left<a\right> = f_a\,\bar{\theta}$,
but rather at $\left<a\right> = f_a\left(\bar{\theta} + \Delta\bar{\theta}\right)$, where
\begin{align}
\Delta\bar{\theta} = \Delta\bar{\theta}_0\,
\sin\left(\frac{k}{\left|\mathcal{A}_{\rm PQ}\right|}\,\bar{\theta}\right) 
= \Delta\bar{\theta}_0\, \sin\bar{\theta} \,, \quad
\Delta\bar{\theta}_0 = \frac{k}{\left|\mathcal{A}_{\rm PQ}\right|} \frac{v^4}{m_a^2\, f_a^2}
= \frac{v^2}{m_a^2\, f_a^2} \,,
\label{eq:deltath}
\end{align}
up to corrections of $\mathcal{O}\left(\Delta\bar{\theta}_0^{\,2}\right)$
and where we have used that $\left|\mathcal{A}_{\rm PQ}\right| = k$ (see Eq.~\eqref{eq:APQC}).
According to the experimental bound on the QCD angle,
$\Delta\bar{\theta}_0$ must be smaller than $10^{-10}$, which leads us to
\begin{align}
v \lesssim 10^{-2.5}\,\Lambda_a \simeq 240\,\textrm{keV} \,, \quad 
\Lambda_a = \left(m_a\, f_a\right)^{1/2} \simeq 77\,\textrm{MeV} \,.
\label{eq:vbound}
\end{align}
The energy scale of the PQ-breaking operators in the axion potential therefore needs
to be extremely suppressed, i.e., it should be even smaller than half the electron mass!
Given our estimate of the energy scale $v$ in Eq.~\eqref{eq:v4cos}
and taking the gravitino mass to be of $\mathcal{O}(100)\,\textrm{TeV}$,
this implies that scenarios with only $k=3$ or $k=4$ pairs of extra quarks can
be safely ruled out,
\begin{align}
\left(\frac{m_{3/2}}{M_{\rm Pl}}\right)^{(k+1-c)/4} M_{\rm Pl} \lesssim 240\,\textrm{keV}
\,,\quad m_{3/2} \sim 100\,\textrm{TeV} \,, \quad
\quad\Rightarrow\quad k \geq k_{\rm min} \sim 5
\label{eq:kmin}
\end{align}
Whether or not $k=5$ extra quark pairs are phenomenologically viable is hard to
tell in view of this rather simplified estimate.
The case $k=5$, thus, requires a more careful analysis.
In fact, as we shall see in the following, it turns out that $k=5$ new quark pairs
are not only viable, but also the \textit{unique} number of new quark pairs that
will allow us to satisfy all bounds at the same time.


In order to constrain scenarios with $k \geq 5$ extra quark pairs more precisely,
we need to know the exact expressions for the energy scale $v$ belonging
to the respective terms in $\Delta V_a$ in Eq.~\eqref{eq:DeltaVa}.
These expressions simply follow from substituting
all fields in Eq.~\eqref{eq:DeltaVa} with their VEVs
(see Eq.~\eqref{eq:Mpm} as well as Eq.~\eqref{eq:ZXVEVs} in Appendix~\ref{app:vacuum}
for our results for $\left<M_\pm\right>$, $\left<Z_\pm\right>$, 
and $\left<\left|X\right|\right>$, respectively),
\begin{align}
v_{Z_\pm}^4 = & \: \frac{2\,C_{Z_\pm}}{(k-1)!}\,\lambda_\pm
\left(1 \textrm{ or } \frac{m_{3/2}}{M_*}\right)
\frac{\Lambda}{\eta} \frac{\left<M_\pm\right> \left<Z_\mp\right>^{k-1}}{M_*^{k-3}} \,,
\label{eq:vZM}\\ \nonumber
v_{M_\pm}^4 = & \: \frac{2\,C_{M_\pm}}{\left(k!\right)^2} \frac{1}{\eta^2}
\left(\frac{\eta\,\Lambda}{M_*}\right)^k
\left(k\,\kappa\, \eta\,\left<\left|X\right|\right> \left<M_\mp\right>
+ k\,\lambda_\pm\, \frac{\Lambda}{\eta}\, \left<Z_\mp\right> +
(k-3)\, m_{3/2}\, \left<M_\pm\right> \right)\frac{\left<M_\pm\right>^{k-1}}{M_*^{k-3}} \,,
\end{align}
where the additional factors of $2$ cancel with the factor $1/2$ in Eq.~\eqref{eq:v4cos}.
Imposing the requirement that these scales be sufficiently
suppressed compared to the ``axion scale'' (see Eqs.~\eqref{eq:deltath} and \eqref{eq:vbound}), 
\begin{align}
v_{Z_+}^4 + v_{Z_-}^4 \lesssim \Delta\bar{\theta}_0^{\rm max}\, \Lambda_a^4 \,, \quad
v_{M_+}^4 + v_{M_-}^4 \lesssim \Delta\bar{\theta}_0^{\rm max}\, \Lambda_a^4 \,, \quad
\Delta\bar{\theta}_0^{\rm max} = 10^{-10} \,,
\end{align}
we are then able to derive two $k$-dependent upper bounds
on the axion decay constant $f_a$,
\begin{align}
Z_\pm^k \:\:\rightarrow\:\: & f_a \lesssim f_Z^{(k)}  = A_Z^{(k)}\left(\zeta,\rho,\kappa,\eta\right)
F_Z^{(k)} \,, &
F_Z^{(k)} = & \: \left(M_{\rm Pl}\, M_*\right)^{1/2} \left(\frac{\Delta\bar{\theta}_0^{\rm max}\,
\Lambda_a^4}{C_Z\, M_{\rm Pl}\, M_*^3}\right)^{1/(2(k+1-c))} \hspace{-0.1cm}, 
\label{eq:faZM} \\ \nonumber
M_\pm^k \:\:\rightarrow\:\: & f_a \lesssim f_M^{(k)} = A_M^{(k)}\left(\zeta,\rho,\kappa,\eta\right)
F_M^{(k)} \,, &
F_M^{(k)} = & \: M_* \left(\frac{\Delta\bar{\theta}_0^{\rm max}\,
\Lambda_a^4\, M_{\rm Pl}}{C_M\, M_*^5}\right)^{1/(2(k+1))} \,,
\end{align}
Here, $A_Z^{(k)}$ and $A_M^{(k)}$ represent two
dimensionless prefactors, the precise values of which depend on four crucial parameters
of our model: $\zeta$ (i.e., the Yukawa coupling $\lambda$ in the IYIT superpotential,
see Eq.~\eqref{eq:Mpm}), $\rho$ (i.e., the flavor hierarchy in the IYIT sector,
see Eq.~\eqref{eq:ATheta}), $\kappa$ (i.e., the physical status of the Lagrange multiplier
field $X$, see Eq.~\eqref{eq:WdynPf}), and $\eta$ (i.e., the numerical uncertainty of
all coupling constants in the effective theory induced by strong-coupling effects,
see Eq.~\eqref{eq:Mij}),
\begin{align}
\textrm{$k$ even:} \quad & A_Z^{(k)}\left(\zeta,\rho,\kappa,\eta\right) = 
\left[\frac{3^{(k-1)/2}\,k!\,\left(\kappa\eta\right)^{2k-3}}
{\left(2\pi\right)^{2(k-1)}\,k^{2k+1}\,\rho^{2k}}
\frac{\zeta^{k-3/2} \left(1-\zeta\right)^{k/2}}{\left(2-\zeta\right)^{5(k-1)/2}}\,
B^{k-1}\right]^{1/(2k)} \,, \label{eq:AZM}\\ \nonumber
\textrm{$k$ odd:} \quad & A_Z^{(k)}\left(\zeta,\rho,\kappa,\eta\right) =
\left[\frac{4\times3^{k/2}\,k!\,\left(\kappa\eta\right)^{2(k-2)}}
{\left(2\pi\right)^{2(k-1)}\,k^{2k+3}\,\rho^{2(k+1)}}
\frac{\zeta^{k-2} \left(1-\zeta\right)^{k/2+1}}{\left(2-\zeta\right)^{5k/2-2}} \,
B^{k-1}\right]^{1/(2k+2)} \,, \\ \nonumber
& A_M^{(k)}\left(\zeta,\rho,\kappa,\eta\right) =
\left[\frac{2^{2k+1}\,3^{1/2}\,k!^2\,\kappa\eta\,C^{k/2}}{k^{2(k+1)}\,\eta^{2(k-1)}\,\rho^{k+2}}
\frac{\zeta^{1/2}\left(1-\zeta\right)^{k/2+1}}{\left(2-\zeta\right)^{1/2}}
\frac{B}{D}\right]^{1/(2k+2)} \,,
\end{align}
where we have introduced the symbols $B$, $C$, and $D$ for the ease of notation,
\begin{align}
B = & \: \left(2\ln2-1\right)\left(4-2\,\zeta+\rho^6\right) \,, \\ \nonumber
C = & \: 1 + \left(1-\rho^4\right)^{1/2} \,, \\ \nonumber
D = & \: 32 \pi^2\,k\, (2-\zeta)^2 \left[\rho^{2k-2} C + \rho^{-2} \left(2-C\right) C^k\right] +
(k-3)\,\left(\kappa\eta\right)^2\,\zeta\,B \left(\rho^{2k}+C^k\right) \,.
\end{align}


It is illustrative to evaluate the two bounds in Eq.~\eqref{eq:faZM}
for a few representative parameter values.
For $k = 5$ and $k=6$, for instance, and setting $C_Z = C_M = 1$ as well as
$M_* = M_{\rm Pl}$, the two energy scales $F_Z^{(k)}$ and $F_Z^{(k)}$
in Eq.~\eqref{eq:faZM} take the following values,
\begin{align}
F_Z^{(5)} = F_Z^{(6)} = F_M^{(5)} \simeq 1.1 \times 10^{11} \,\textrm{GeV} \,, \quad
F_M^{(6)} \simeq 1.3 \times 10^{12} \,\textrm{GeV} \,,
\end{align}
which is well above the lower astrophysical bound on the axion decay constant,
$f_a \gtrsim 10^9\,\textrm{GeV}$ (see Sec.~\ref{subsec:commonorigin}).
At the same time, for $\kappa=1$, $\eta=4\pi$ and assuming identical Yukawa couplings
in the IYIT superpotential (i.e., $\rho=1$),
$A_Z^{(5,6)}$ and $A_M^{(5,6)}$ are all of $\mathcal{O}(0.1)$ for almost all values of $\zeta$,
\begin{align}
A_Z^{(5)} \sim A_Z^{(6)} \sim A_M^{(5)} \sim A_M^{(6)} \sim 0.1 \,.
\label{eq:AZM56}
\end{align}
Thus, for both scenarios, $k=5$ and $k=6$, we find that the axion decay constant is typically
constrained to be at most of $\mathcal{O}\left(10^{10}\right)\,\textrm{GeV}$.
Given this result, it is worthwhile to to recall that, in order to realize
gravitino masses of $\mathcal{O}\left(100\right)\,\textrm{TeV}$, we anticipate
$f_a$ to actually take a value close to $10^{10}\,\textrm{GeV}$
(see the discussion below Eq.~\eqref{eq:faLambda}).
Our above estimate of the upper bound on $f_a$ is, hence, consistent
with this expectation---albeit it seems as if $f_a$ should be rather close to
its upper bound in order to allow for the possibility of
gravitino masses of $\mathcal{O}\left(100\right)\,\textrm{TeV}$.
We will specify these statements in the next section, where we will
finally present our bounds on $f_a$ along with the corresponding values of
$m_{3/2}$.
Before we are able to do so, there is, however, one more issue which
we need to address.
In addition to the upper bounds on $f_a$ derived above, the requirement of
perturbative gauge coupling unification at the GUT scale also results
in \textit{lower} bounds on $f_a$.


The new quark flavors affect the running of the SM gauge coupling constants.
As we take the new quark fields to transform in complete $SU(5)$ multiplets, i.e.,
$\mathbf{5}$ and $\mathbf{5}^*$, the gauge couplings still unify at the same energy
scale as in the MSSM, $\Lambda_{\rm GUT} \simeq 2\times10^{16}\,\textrm{GeV}$.
Between the new quark mass scale, $m_Q$, and the GUT scale, the new quarks, however,
contribute to the beta functions of the SM gauge coupling constants, which results
in a faster running and, hence, a larger value of the GUT gauge coupling,
$\alpha_{\rm GUT} = g_{\rm GUT}^2/\left(4\pi\right)$
than in the MSSM, $\alpha_{\rm GUT}^{\rm MSSM} \simeq 1/25$.
Thus, in order for our model to be consistent all the way up to the GUT scale, we 
must require that the SM gauge couplings still unify at some perturbative value, i.e.,
that $\alpha_{\rm GUT}$ does not exceed unity.
The actual value of $\alpha_{\rm GUT}$ in our model depends on the details of the MSSM
mass spectrum as well as on the number and the mass scale of the new quark flavors
(i.e., $k$ and $m_Q$).
In particular, it increases with $k$ and decreases with $m_Q$.
For a given MSSM mass spectrum and a fixed value of $k$, the requirement that $\alpha_{\rm GUT}$
must remain perturbative can then be translated into a lower bound on the
mass scale of the new quark flavors,%
\footnote{While in principle this is a correct statement, in practice,
we need to be a bit more careful:
Because of the uncertainty in the low-energy input
parameters, the uncertainties in the MSSM mass spectrum, etc., the SM gauge
couplings do not always unify \textit{exactly} at $\Lambda_{\rm GUT}$.
Instead, the electroweak gauge couplings often unify, $\alpha_1 = \alpha_2$,
before they actually reach the strong gauge coupling $\alpha_3$.
At the technical level, we therefore have to impose the condition that $\alpha_3$
(and not ``$\alpha_{\rm GUT}$'') must remain perturbative, i.e., we require
$\alpha_3 \leq 1$ at the scale where $\alpha_1$ and $\alpha_2$ unify.}
\begin{align}
\alpha_{\rm GUT} = \alpha_{\rm GUT}\left(m_{\rm MSSM}; k,m_Q\right) \,, \quad
\alpha_{\rm GUT} \leq 1 \:\:\Rightarrow\:\:
m_Q \geq m_Q^{\rm min}\left(m_{\rm MSSM};k\right) \,.
\end{align}
In the context of PGM, the MSSM spectrum is basically characterized by two scales:
(i) the gravitino mass $m_{3/2}$, which determines the masses of all sfermions
as well as of the higgsinos,
and (ii) the gaugino mass scale $m_{1/2}$, which is related to the gravitino mass
via a loop factor in PGM, $m_{1/2} \sim m_{3/2}/\left(16\pi^2\right)$, and which
determines the masses of the MSSM gauginos.
Motivated by the perspective of neutralino dark matter, we shall take the gaugino mass
scale to be of $\mathcal{O}(1)\,\textrm{TeV}$ and treat $m_{3/2}$ as a free
parameter in the following.
Consequently, the lower bound on the new quark mass scale, $m_Q^{\rm min}$, 
then becomes a function of $m_{3/2}$ and $k$ only.


\begin{figure}
\begin{minipage}{0.46\textwidth}
\includegraphics[width=\textwidth]{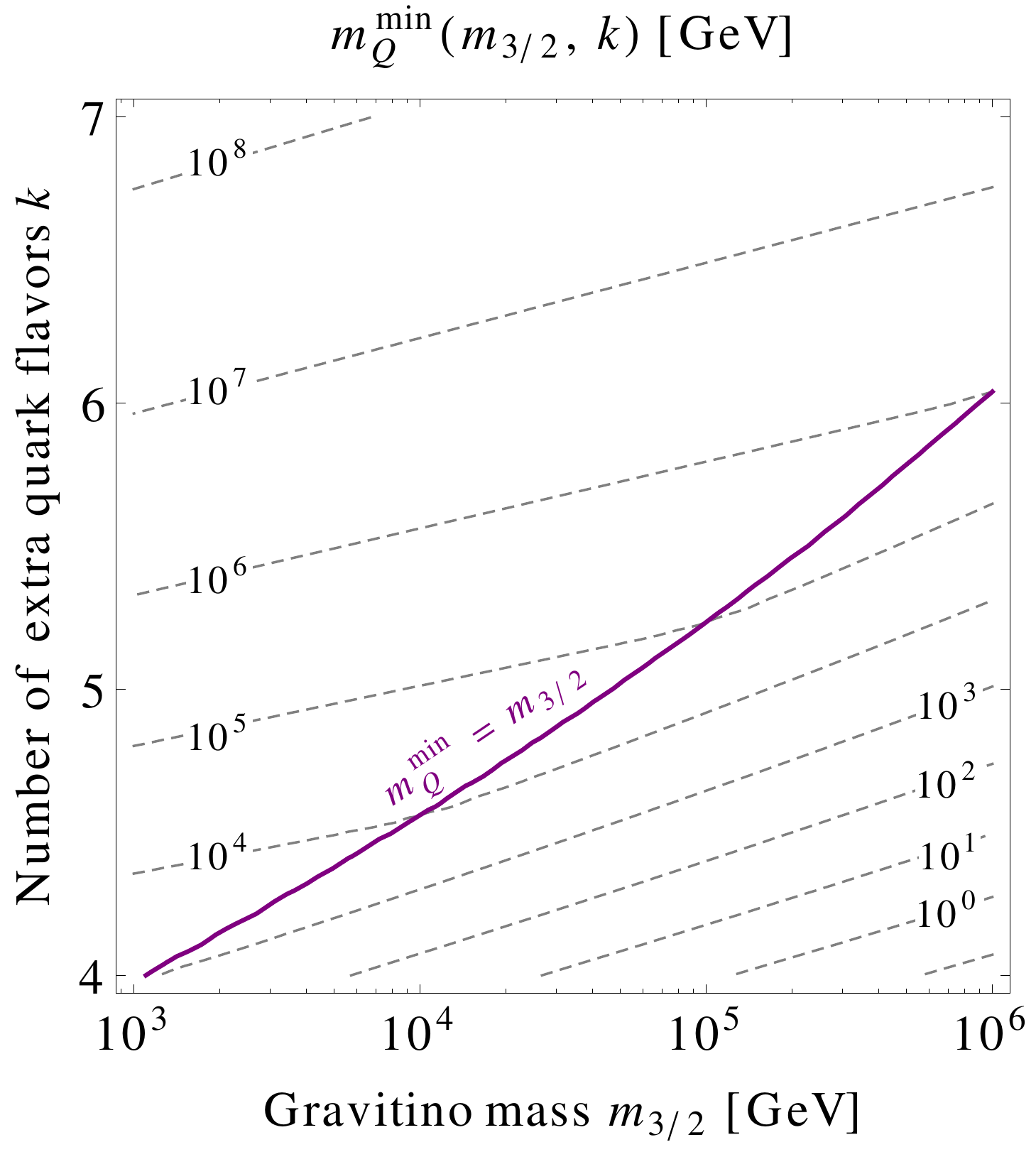}
\end{minipage}
\hfill
\begin{minipage}{0.49\textwidth}
\includegraphics[width=\textwidth]{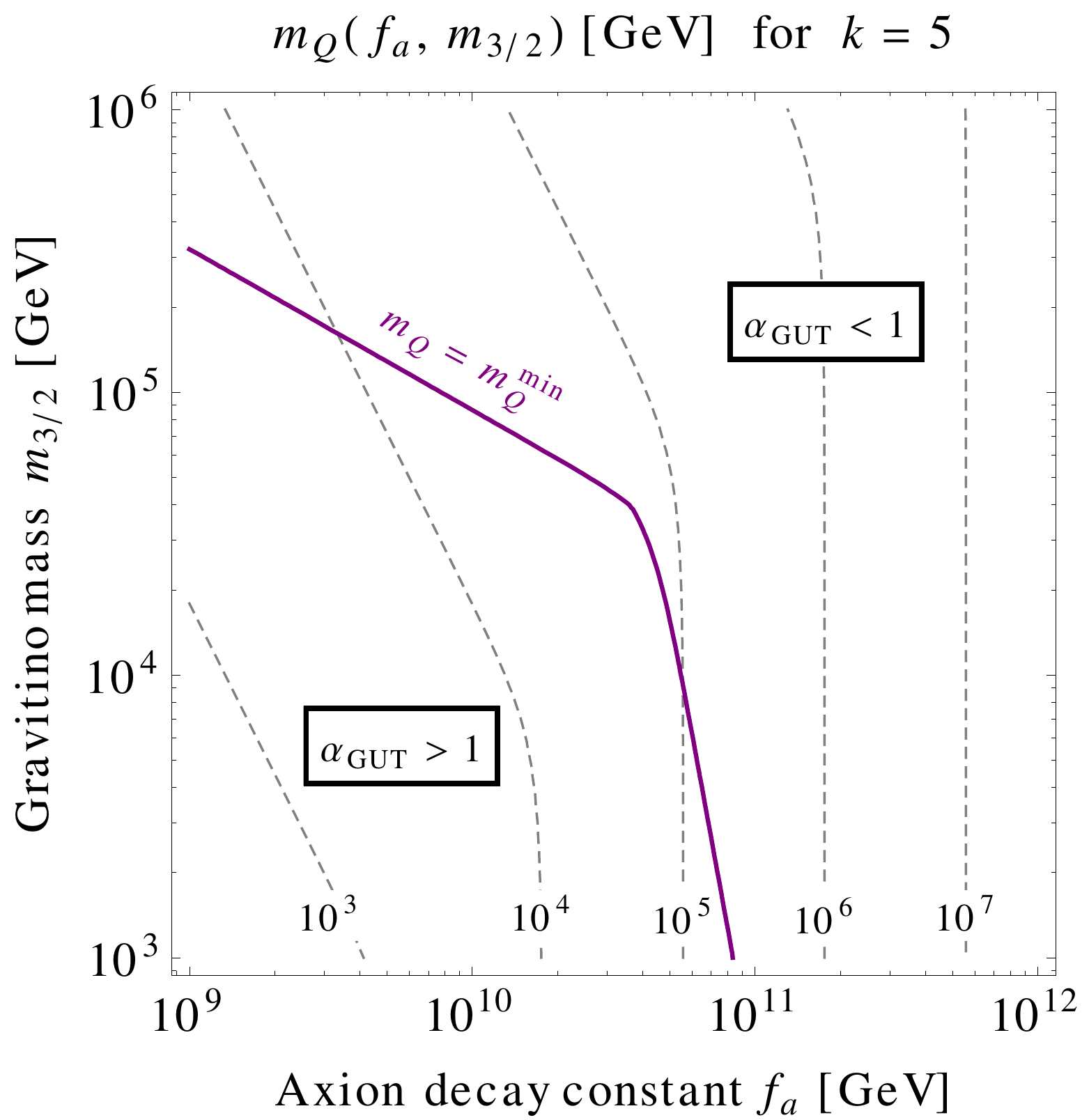}
\end{minipage}
\caption{\textbf{Left panel:} Lower bound on the masses of the new quark fields, $m_Q^{\rm min}$,
as function of the gravitino mass and the number of extra quark flavors
(see Eqs.~\eqref{eq:mQmin} and \eqref{eq:pq}).
\textbf{Right panel:} New quark mass scale, $m_Q$, as well as
lower bound on the axion decay constant as function of the gravitino mass,
$f_Q^{(k)}$, according to the requirement of perturbative gauge coupling unification
(i.e., according to the requirement $m_Q \geq m_Q^{\rm min}$, see Eqs.~\eqref{eq:faQm32}
and \eqref{eq:mQfam32}) for $k=5$.
Here, all other relevant parameters have been chosen as follows:
$\rho=1$, $\kappa=1$, $\eta=4\pi$, $C_{Q_i} = 4\pi$, $M_* = M_{\rm Pl}$.}
\label{fig:mQ}
\end{figure}


In order to determine $m_Q^{\rm min}$ as a function of these two parameters,
we have to solve the renormalization group equations (RGEs) for the SM gauge
couplings. 
We do so numerically and accounting for, in total,
three different mass thresholds:
We set all gaugino masses to $1\,\textrm{TeV}$, take the masses of
all other MSSM sparticles to be equal to $m_{3/2}$, and assume all
new quark flavors to have a common mass equal to $m_Q$.
Between the $Z$ pole and $\textrm{max}\left\{m_{3/2},m_Q\right\}$, we
simply use the ordinary MSSM one-loop beta functions (including the contributions
from the new quark pairs), while for energies between
$\textrm{max}\left\{m_{3/2},m_Q\right\}$ and $\Lambda_{\rm GUT}$, we
perform a two-loop calculation in the $\overline{\textrm{\footnotesize DR}}$ scheme.
%
The idea behind this procedure is that the SM gauge couplings become more
sensitive to small changes in the beta functions, the larger they are.  
We should therefore be a bit more careful in tracking the running of SM
gauge couplings at larger energies than at lower energies, which is
why we switch from a one-loop analysis to a two-loop analysis beyond
the last mass threshold.
The result of our calculation is shown in the left panel of Fig.~\ref{fig:mQ},
which displays $m_Q^{\rm min}$ as a function of $m_{3/2}$ and $k$.
Note that, here, $k$ is treated as a continuous parameter, although, of course,
only integer values of $k$ are physically sensible.
For $k=5$ and $k=6$, for instance, and a gravitino mass of $100\,\textrm{TeV}$, we
respectively find (setting all other parameters to the same values as in Fig.~\ref{fig:mQ}),
\begin{align}
k=5 \,: \quad m_Q^{\rm min} \simeq 1.8 \times 10^4 \,\textrm{GeV} \,, \quad
k=6 \,: \quad m_Q^{\rm min} \simeq 2.1 \times 10^6 \,\textrm{GeV} \,.
\end{align}
The purple line in the left panel of Fig.~\ref{fig:mQ} indicates the boundary between two
different hierarchy schemes that are possible within our set-up.
Above the purple line, the new quarks are required to be heavier than
the MSSM sfermions; below the purple line, they can also be lighter
than the MSSM sfermions.
In passing, we also mention that our numerical result for $m_Q^{\rm min}$ is nicely
fit by the following analytical expression,
\begin{align}
m_Q^{\rm min} \simeq 10^p\,\textrm{GeV} \left(\frac{m_{3/2}}{100\,\textrm{TeV}}\right)^q \,,
\label{eq:mQmin}
\end{align}
where the powers $p$ and $q$ can be expanded into polynomials in
$\Delta k = k - 5.2$,
\begin{align}
p = \begin{cases}
5.0 + 2.9\,\Delta k - 0.85\, \Delta k^2 \\
5.0 + 1.8\,\Delta k - 0.20\, \Delta k^2
\end{cases} \,, \quad
q = \begin{cases}
-1.2 + 0.40\,\Delta k + 0.11\, \Delta k^2 & ;\:\: m_Q^{\rm min} \lesssim m_{3/2} \\
-0.37 - 0.02\,\Delta k + 0.03\, \Delta k^2 & ;\:\: m_Q^{\rm min} \gtrsim m_{3/2}
\end{cases} \,.
\label{eq:pq}
\end{align}
Here, $k \simeq 5.2$ corresponds to the $k$ value for which
$m_Q^{\rm min} \simeq m_{3/2} \simeq 10^5\,\textrm{GeV}$.
We emphasize that this result for $m_Q^{\rm min}$ holds independently of all
other details of our axion model.
In fact, it represents nothing but the universal lower bounds on the masses
of $k$ pairs of $\mathbf{5}$ and $\mathbf{5}^*$ multiplets
imposed by the requirement of perturbative gauge coupling unification
for a specific PGM-inspired MSSM mass spectrum.
For this reason, we believe that it may also be useful in the context of
other scenarios, where the MSSM particle content is supplemented by further
$SU(5)$ representations.


For given $k$ and $m_{3/2}$, the constraint on the new quark mass scale
in Eq.~\eqref{eq:mQmin} now implies a lower bound on $f_a$.
To see this, let us rewrite $m_Q$ in Eq.~\eqref{eq:mQm32} as a function 
of $f_a$ and $\zeta$.
Eqs.~\eqref{eq:m32} and \eqref{eq:faLambda} allow us to write the gravitino mass
as a function of $f_a$ and $\zeta$ first, which leads us to
\begin{align}
m_{3/2} = 
\frac{\kappa\,\eta\,k^2\,\rho^2\,\zeta^{1/2}\left(2-\zeta\right)^{1/2}}
{4\left(1-\zeta\right)^{1/2}} \frac{f_a^2}{\sqrt{3}\,M_{\rm Pl}} \,, \quad
m_Q = \frac{C_Q\,k^2\,\rho^2}{4\left(1-\zeta\right)^{1/2}}\,\frac{f_a^2}{M_*} \,.
\label{eq:mQfazeta}
\end{align}
Requiring $m_Q$ to be larger than $m_Q^{\rm min}$ then provides us with the
following lower bound on $f_a$,
\begin{align}
f_a \gtrsim f_Q^{(k)} = \frac{2\left(1-\zeta\right)^{1/4}}{C_Q^{1/2}\,k\,\rho}\,
\left[m_Q^{\rm min}\left(m_{3/2},k\right) M_*\right]^{1/2} \,, \quad 
m_{3/2} = m_{3/2}\big(f_Q^{(k)},\zeta\big) \,,
\label{eq:faQ}
\end{align}
with $m_Q^{\rm min}\left(m_{3/2},k\right)$ being given in Eqs.~\eqref{eq:mQmin}
and \eqref{eq:pq} and with $m_{3/2}\left(f_a,\zeta\right)$ being given in Eq.~\eqref{eq:mQfazeta}.
Notice that Eq.~\eqref{eq:faQ} only represents an \textit{implicit} definition of
our lower bound on the axion decay constant, as $f_Q^{(k)}$ still appears in the
argument of the gravitino mass on the right-hand side.
In order to evaluate our lower bound on the axion decay constant numerically,
it is therefore still necessary, for any given set of input parameter values,
to solve Eq.~\eqref{eq:faQ} self-consistently for $f_Q^{(k)}$.
Alternatively, we may also trade the $\zeta$ dependence of $f_Q^{(k)}$
for a dependence on the gravitino mass.
To do so, we simply have to use the following relation, which immediately follows
from Eq.~\eqref{eq:mQfazeta},
\begin{align}
\zeta = 1 - \left[1 + \left(\frac{4\sqrt{3}\,m_{3/2}\,M_{\rm Pl}}
{\kappa\,\eta\,k^2\,\rho^2\,f_a^2}\right)^2\right]^{-1/2} \,.
\label{eq:zetam32}
\end{align}
Plugging this relation into Eq.~\eqref{eq:faQ} and solving for $f_Q^{(k)}$,
we obtain the following \textit{explicit} expression,
\begin{align}
f_a \gtrsim f_Q^{(k)} = 
\left(\frac{4\sqrt{3}\,m_{3/2}\,M_{\rm Pl}}{\sqrt{2}\,\kappa\,\eta\,k^2\,\rho^2}\right)^{1/2}
\left[\left(1+\left[
\frac{\sqrt{2}\,\kappa\,\eta}{C_Q}
\frac{m_Q^{\rm min}\left(m_{3/2},k\right)\,M_*}
{\sqrt{3}\,m_{3/2}\,M_{\rm Pl}}\right]^4\right)^{1/2}-1\right]^{1/4} \,.
\label{eq:faQm32}
\end{align}
For $k=5$ and $k=6$, for instance, and taking $m_{3/2}$ to be $100\,\textrm{TeV}$,
this bounds evaluates to (again setting all other parameters to the same
values as in Fig.~\ref{fig:mQ}),
\begin{align}
k=5 \,: \quad f_Q^{(k)} \simeq 7.8 \times 10^9 \,\textrm{GeV} \,, \quad
k=6 \,: \quad f_Q^{(k)} \simeq 2.1 \times 10^{11} \,\textrm{GeV} \,. 
\end{align}
At the same time, Eq.~\eqref{eq:zetam32} also allows us to rewrite $m_Q$ as
a function of $f_a$ and $m_{3/2}$,
\begin{align}
m_Q = \frac{C_Q\,k^2\,\rho^2}{4}\,\frac{f_a^2}{M_*}
\left[1 + \left(\frac{4\sqrt{3}\,m_{3/2}\,M_{\rm Pl}}
{\kappa\,\eta\,k^2\,\rho^2\,f_a^2}\right)^2\right]^{1/4} \,.
\label{eq:mQfam32}
\end{align}
We plot the expressions for $f_Q^{(k)}$ and $m_Q$ in Eqs.~\eqref{eq:faQm32} and \eqref{eq:mQfam32}
in the right panel of Fig.~\ref{fig:mQ} for the special case of $k=5$ extra quark pairs.
For $k=5$ and the values of the gravitino mass that we are most interested in,
$m_{3/2} \sim 100\,\textrm{TeV}$, we  again find a bound of
$\mathcal{O}\left(10^{10}\right)\,\textrm{GeV}$ on the axion
decay constant---which this time is a lower bound and not an upper bound on $f_a$.
In summary, it therefore seems as if, for $k=5$, the axion decay must indeed be of 
$\mathcal{O}\left(10^{10}\right)\,\textrm{GeV}$, i.e., it must neither be much smaller
nor much larger than $10^{10}\,\textrm{GeV}$ in order to satisfy all phenomenological
constraints at the same time (see also our remarks below
Eqs.~\eqref{eq:faLambda} and \eqref{eq:AZM56}, respectively).
In the next section, we are now going to specify these statements in a bit more detail.


\subsection{Final results: viable region in parameter space}
\label{subsec:scan}


The axion decay constant is bounded
from above as well as from below (see Eqs.~\eqref{eq:faZM} and \eqref{eq:faQ}).
Thus, in order to asses the viability of our model, we have to search for
regions in parameter space where \textit{not all} possible values of $f_a$
are ruled out, but which still allow for a viable range for $f_a$,
\begin{align}
f_Q^{(k)} \lesssim f_a \lesssim f_{\cancel{\rm PQ}}^{(k)} \,, \quad
f_{\cancel{\rm PQ}}^{(k)} = \textrm{min}\left\{f_Z^{(k)},f_M^{(k)}\right\} \,.
\end{align}
Here, the bounds $f_Q^{(k)}$ and $f_{\cancel{\rm PQ}}^{(k)}$
are functions of, in total, nine different parameters, which provides us with a lot
of freedom when it comes to picking a concrete realization of our axion model.
Due to this large parametric freedom, the bounds $f_Q^{(k)}$ and $f_{\cancel{\rm PQ}}^{(k)}$
can in principle vary over many orders of magnitude, so that it turns out
impossible to derive a single \textit{unique} range of values
that the axion decay constant is confined to.
Also, a systematic scan of the nine-dimensional parameter scan appears to
be difficult (and perhaps also not very revealing).
Therefore, we will simply focus on certain representative parameter choices
in the following, trying to assess what is achievable in our model.
To do so, let us first recall which nine
parameters $f_Q^{(k)}$ and $f_{\cancel{\rm PQ}}^{(k)}$ actually depend on:
\begin{itemize}
\item The number of extra quark pairs, $k$ (see Eq.~\eqref{eq:rQQ}). 
According to our considerations in Sec.~\ref{subsec:constraints},
the integer $k$ must be $k=5$ or larger (see Eq.~\eqref{eq:kmin}).
At the same time, increasing the value of $k$ implies an increase
in all three bounds on $f_a$.
For too many extra quark pairs, the lower bound $f_Q^{(k)}$ will therefore
begin to exceed the upper boundary of the phenomenologically viable window
for the axion decay constant, $f_Q^{(k)} \gtrsim 10^{12}\,\textrm{GeV}$.
For this reason, we will restrict ourselves to scenarios
with $k=5$, $k=6$ or $k=7$ extra quark pairs in the following.
\item The Yukawa coupling in the IYIT superpotential, $\lambda$, or alternatively the
parameter $\zeta \in \left[0,1\right]$, which parametrizes the suppression of the meson
VEVs $\left<M_\pm\right>$ (see Eqs.~\eqref{eq:Weffcha} and \eqref{eq:Mpm}).
Note that $\zeta$ can also always be traded for the gravitino mass $m_{3/2}$ via the relation
in Eq.~\eqref{eq:zetam32}.
In the following, we will mainly be interested in those values of $\lambda$ (or $\zeta$)
that yield a gravitino mass of $100\,\textrm{TeV}$.
This then eliminates the coupling $\lambda$ as a free parameter from our analysis.
\item The parameter $\rho \in \left[0,1\right]$, which represents a measure of the
hierarchy among the Yukawa couplings $\lambda_+$ and $\lambda_-$ in the IYIT sector
(see Eq.~\eqref{eq:ATheta}).
As evident from Eqs.~\eqref{eq:faZM} and \eqref{eq:faQm32}, all bounds on $f_a$ increase
when going to smaller values of $\rho$.
Here, the lower bound $f_Q^{(k)}$ increases, in particular, faster than the
upper bound $f_M^{(k)}$.
In order to maximize the allowed region in parameter space, we should therefore choose
the parameter $\rho$ as large as possible, $\rho = 1$.
Interestingly enough, this coincides with the flavor-symmetric limit, $\lambda_+ = \lambda_-$,
in the IYIT sector and, hence, might be regarded as a sensible and well motivated choice.
\item The parameter $\kappa$, which indicates the physical status of the Lagrange
multiplier field $X$.
As noted below Eq.~\eqref{eq:Mij}, $\kappa$ should be either treated as an
$\mathcal{O}(1)$ coupling or sent to infinity.
In the former case (i.e., when the field $X$ is assumed to be physical),
larger $\kappa$ values turn out to be more advantageous
for our purposes, $\kappa \gtrsim 1$, because going to larger values of $\kappa$
relaxes both the bounds $f_Q^{(k)}$ and $f_Z^{(k)}$ (the bound $f_M^{(k)}$ is rather
insensitive to $\kappa$).
We will therefore distinguish between three different cases in the following:
$\kappa = 1$, $\kappa = 4$, and $\kappa\rightarrow\infty$.
\item The NDA parameter $\eta$, which captures the numerical uncertainty of
all coupling constants in the low-energy effective theory induced by
strong-coupling effects (see Eq.~\eqref{eq:Mij}).
Larger $\eta$ implies a stronger bound on $f_a$ coming from
the $M_\pm^k$ meson operators in the superpotential (see Eq.~\eqref{eq:vZM}
and \eqref{eq:AZM}), which is why we should actually choose $\eta$ as small
as possible, $\eta \simeq \pi$.
On the other hand, $\eta$ is naively expected to be of
$\mathcal{O}\left(4\pi\right)$, which is why we will consider two cases
in the following: $\eta = \pi$ and $\eta = 4\pi$.
\item The high-energy cut-off scale $M_*$ in the PQ-breaking operators in
$W_{\cancel{\rm PQ}}^{\rm eff}$ and $K_{\cancel{\rm PQ}}^{\rm eff}$ (see Eq.~\eqref{eq:WKPQ}).
As we take these operators to be generated via gravitational interactions, $M_*$
is expected to be close to the Planck scale.
For now, we will therefore simply set $M_* = M_{\rm Pl}$.
\item The three dimensionless coefficients $C_Q$, $C_{Z_\pm}$, and $C_{M_\pm}$
(see Eqs.~\eqref{eq:WQ} and \eqref{eq:WPQeff}), which we expect to take values
somewhere between $1$ and $4\pi$.
To maximize the allowed region in parameter space, the coefficients $C_{Z_\pm}$ and $C_{M_\pm}$
should be chosen as small as possible (see Eq.~\eqref{eq:faZM}), while 
the coefficient $C_Q$ should be chosen as large as possible (see Eq.~\eqref{eq:faQ}).
We will therefore set $C_{Z_\pm} = C_{M_\pm} = 1$ and $C_Q = 4\pi$ in the following.
\end{itemize}


\begin{figure}
\centering
\includegraphics[width=0.48\textwidth]{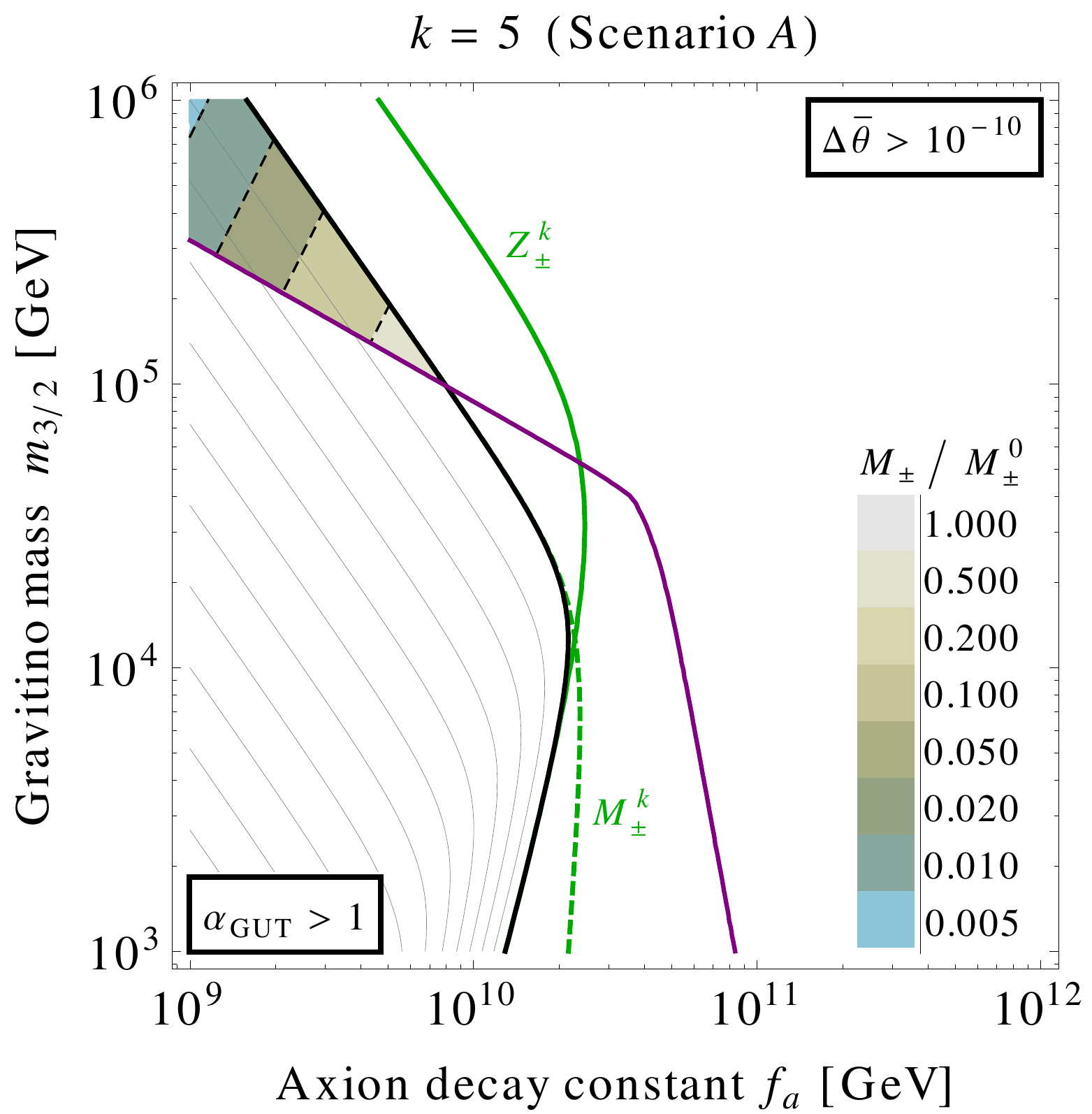}
\hfill\includegraphics[width=0.48\textwidth]{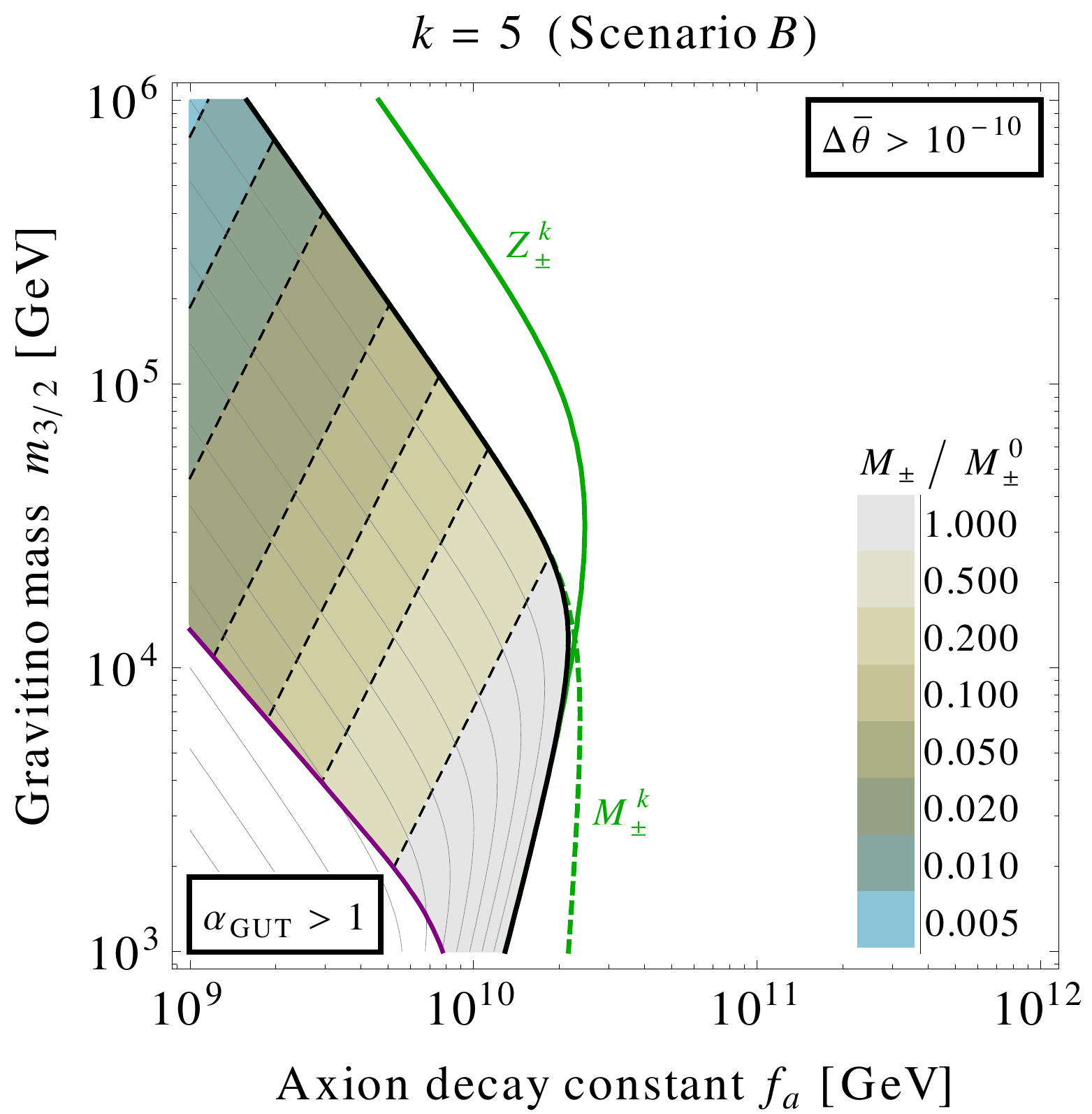}

\bigskip\bigskip
\includegraphics[width=0.48\textwidth]{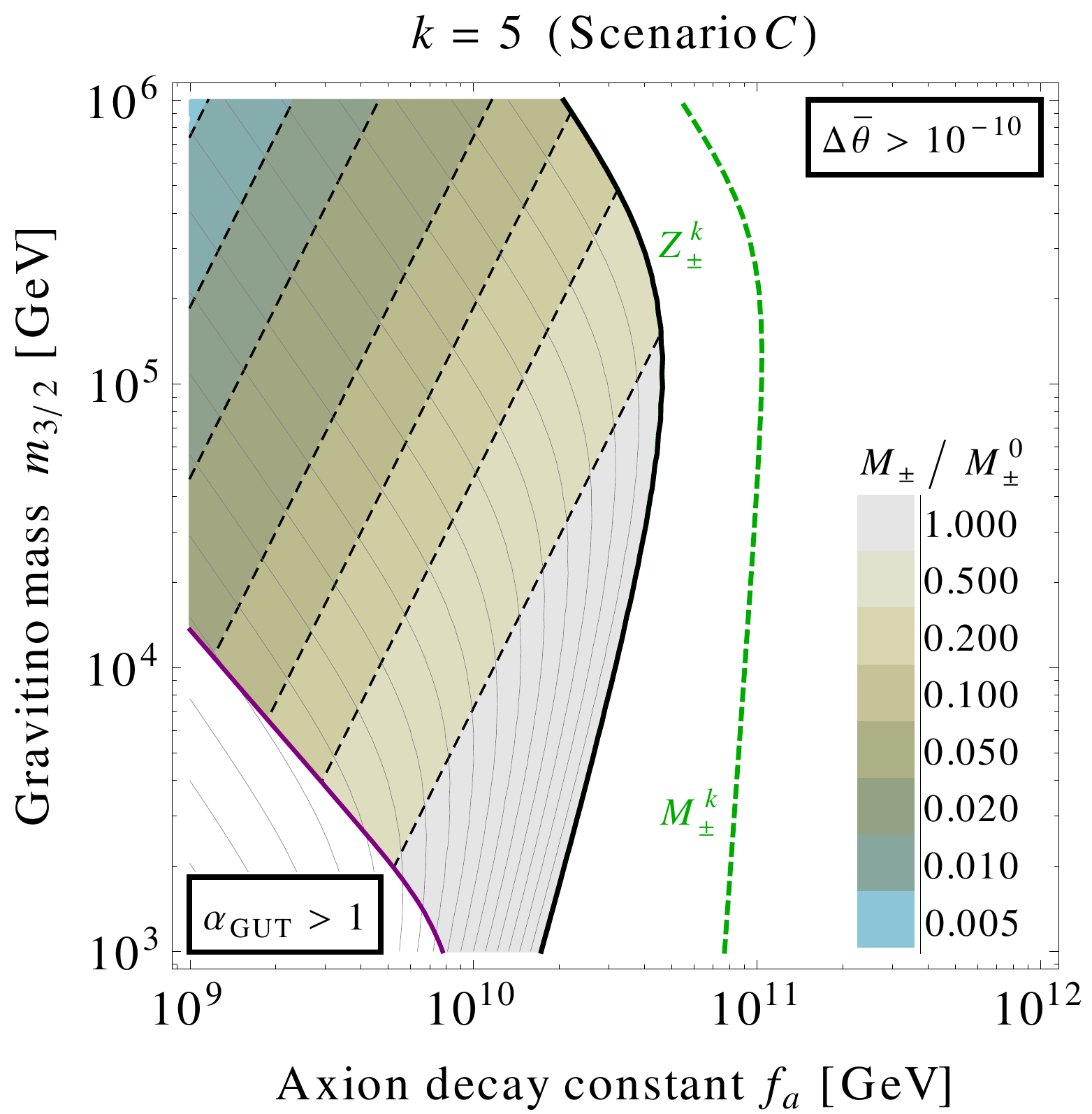}
\hfill\includegraphics[width=0.48\textwidth]{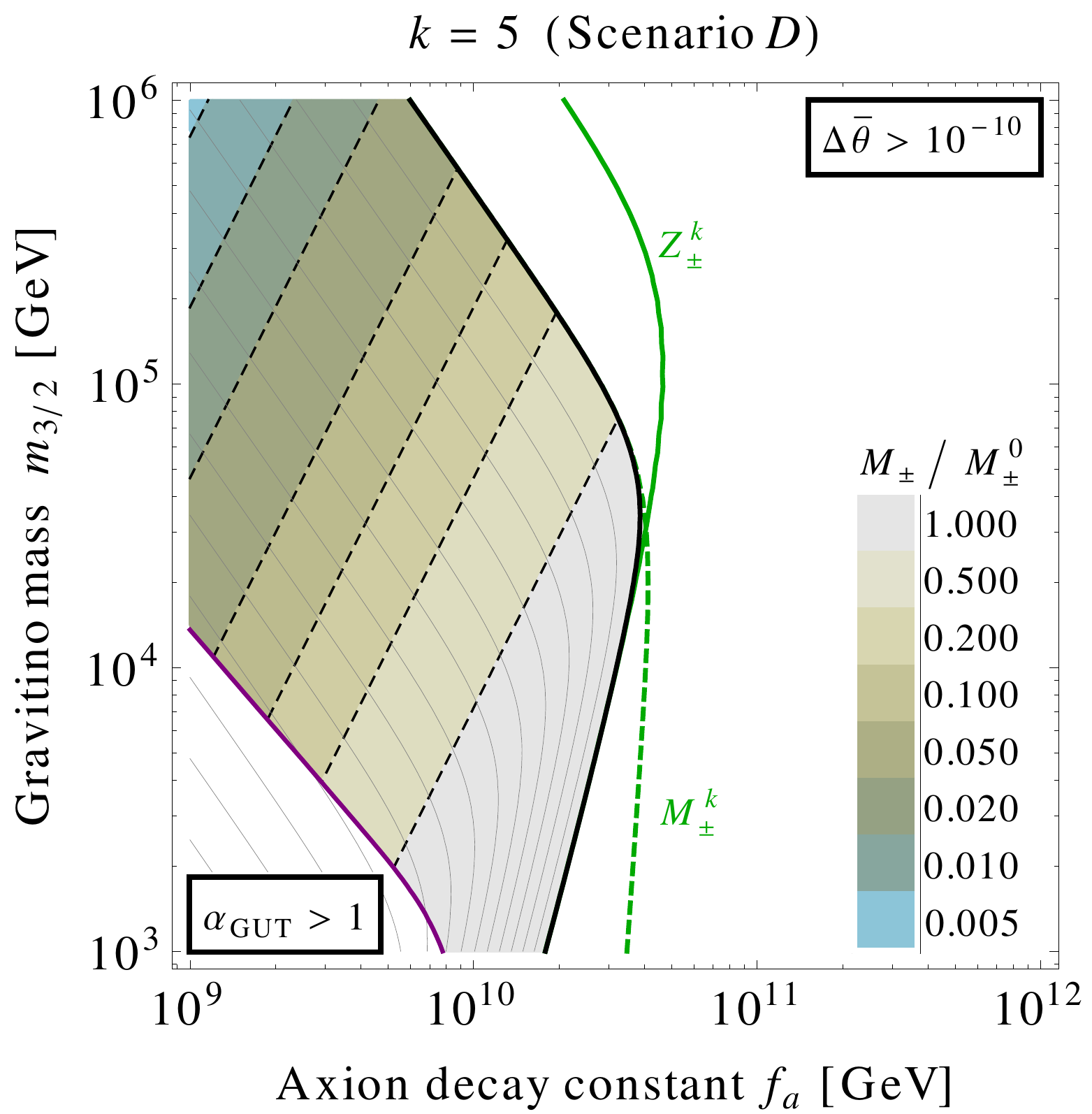}
\caption{Constraints on the axion decay constant $f_a$ and the gravitino mass $m_{3/2}$
for the four different parameter scenarios specified in
Eqs.~\eqref{eq:scenarioA}, \eqref{eq:scenarioB}, and \eqref{eq:scenarioCD}.
The thick solid and dashed green lines represent the upper bounds $f_Z^{(k)}$ and $f_M^{(k)}$
(see Eq.~\eqref{eq:faZM}), respectively, while the thick purple lines show
the lower bound $f_Q^{(k)}$ (see Eq.~\eqref{eq:faQ}).
The color code and the dashed black lines indicate the suppression of the meson VEVs
compared to the asymptotic expression $M_\pm^0$ (see Eq.~\eqref{eq:Mpm0}), i.e.,
the value of the parameter $\varepsilon = \left(1 -\zeta\right)^{1/2}$,
where $\zeta = \left(\lambda/\kappa/\eta\right)^2$ (see Eq.~\eqref{eq:Mpm}).
Here, large suppression corresponds to a large Yukawa coupling
$\lambda \simeq \kappa\,\eta$, while small suppression corresponds to a small
Yukawa coupling $\lambda\ll \kappa\,\eta$.
The thin solid gray lines indicate the value of $\Delta\bar{\theta}$
in integer steps on a logarithmic scale, $\Delta\bar{\theta} = 10^{-11},\,10^{-12},\,\cdots$.
The thick black solid line marks the values of $f_a$ and $m_{3/2}$ for which 
$\Delta\bar{\theta} = 10^{-10}$.}
\label{fig:k5}
\end{figure}


Let us now study all combinations of the parameters $k$, $\kappa$, and $\eta$
according to the above list of restrictions.
Remarkably enough, it turns out that there is actually only \textit{one} combination,
which happens to allow for a viable range of $f_a$ values for a gravitino
mass of $100\,\textrm{TeV}$!
\begin{align}
\textrm{Scenario $A$:} \qquad k = 5 \,, \quad \kappa = 4 \,, \quad \eta = \pi 
\quad\Rightarrow\quad
m_{3/2} = 100\,\textrm{TeV} \,, \quad f_a \simeq 8 \times 10^9\,\textrm{GeV}  \,.
\label{eq:scenarioA}
\end{align}
In this scenario (referred to as Scenario $A$ in the following), the axion decay constant
ends up being tightly constrained to a value close to $f_a \simeq 10^{10}\,\textrm{GeV}$
(as expected).
In addition, we now see that our axion model turns out to yield a
\textit{unique} prediction for the number of extra quark pairs:
We have to introduce exactly $k=5$ pairs of $\mathbf{5}$ and $\mathbf{5}^*$
multiplets---no more, no less.
Furthermore, we find that the parameter $\kappa$ is required to take a finite value.
This means that the Lagrange multiplier field $X$ \textit{must} correspond
to a dynamical field.
Sending $\kappa$ to infinity (and hence assuming the Lagrange multiplier $X$ to be unphysical)
is not an option in Scenario $A$.
Finally, if we allow the gravitino mass to vary, also
our constraints on $f_a$ begin to change.
This is shown in the upper left panel of Fig.~\ref{fig:k5},
which displays the constraints
on $f_a$ and $m_{3/2}$ in the case of Scenario $A$.
As illustrated by this plot, the meson VEVs are always suppressed
in Scenario $A$, $M_\pm / M_\pm^0 \lesssim 0.1$.
This indicates that, in Scenario $A$, the Yukawa coupling $\lambda$ is required
to be rather large.


While $f_a$ and $m_{3/2}$ are found to be tightly constrained in the minimal version of our
model (i.e, in Scenario $A$), there are several (almost trivial) possibilities
to modify our model, so as to relax the bounds on parameter space.
For instance, we may assume a different mechanism to generate the masses of the new quark
pairs than in Eq.~\eqref{eq:WQ}.
So far, we have taken the new quark pairs to be coupled to the SUSY-breaking
sector via gravitational interactions, i.e., we have taken the high-energy
cut-off scale in Eq.~\eqref{eq:WQ} to be the scale $M_* \sim M_{\rm Pl}$.
This, however, does not necessarily need to be the case.
The new quark pairs might also couple to the SUSY-breaking sector via the exchange of
GUT messenger fields $\Psi'$ and $\bar{\Psi}'$ with masses of
$\mathcal{O}\left(\Lambda_{\rm GUT}\right)$ and transforming as
fundamentals of both the strongly coupled $SU(2)$ as well as of $SU(5)$,
\begin{align}
W^Q \supset Q_i \bar{\Psi}' \Psi^1 + \bar{Q}_i\Psi'\Psi^2 + M' \Psi' \bar{\Psi}' 
\quad\Rightarrow\quad
W_{\rm eff}^Q \supset \frac{1}{M'} \frac{\Lambda}{\eta} \left(Q \bar{Q}\right)_i M_+ \,,
\quad M' \sim \Lambda_{\rm GUT} \,.
\end{align}
In this case, the cut-off scale in the effective quark superpotential $W_{\rm eff}^Q$
is no longer of $\mathcal{O}\left(M_{\rm Pl}\right)$, but rather of
$\mathcal{O}\left(\Lambda_{\rm GUT}\right)$.
Effectively, such a situation can be accounted for in our analysis by increasing
the coefficient $C_Q$ in Eq.~\eqref{eq:WQ} by a factor
$M_{\rm Pl} / \Lambda_{\rm GUT} \sim 100$.
Setting $C_Q$ to $C_Q = 100 \times 4\pi$ then significantly widens the allowed 
region in parameter space (see the upper right panel of Fig.~\eqref{fig:k5}).
We shall refer to this scenario as Scenario $B$,
\begin{align}
\textrm{Scenario $B$:} \qquad k = 5 \,, \quad \kappa = 4 \,, \quad \eta = \pi \,, \quad
C_Q = 100 \times 4\pi \,.
\label{eq:scenarioB}
\end{align}
For $m_{3/2} = 100\,\textrm{TeV}$, increasing $C_Q$ to such a large value
basically removes the lower bound on $f_a$ coming from the requirement of perturbative
gauge coupling unification.
The axion decay constant then ends up being constrained by the astrophysical bound
$f_a \gtrsim 10^9\,\textrm{GeV}$ as well as by $f_{\cancel{\rm PQ}}^{(k)}$,
\begin{align}
m_{3/2} = 100\,\textrm{TeV} \quad\Rightarrow\quad
10^9\,\textrm{GeV} \lesssim f_a \lesssim 8\times 10^9 \,\textrm{GeV} \,.
\end{align}
Here, smaller values of $f_a$ require the meson VEVs $\left<M_\pm\right>$ to be
increasingly suppressed compared to the asymptotic expression $M_\pm^0$
(see Eq.~\eqref{eq:Mpm}).%
\footnote{Recall that, in our formal calculation (employing a canonical K\"ahler potential),
this is achieved by fine-tuning the Yukawa coupling $\lambda$, so that it
increasingly approaches its maximal value
$\lambda_{\rm max} = \kappa\,\eta = 4\pi$ (see Eq.~\eqref{eq:Mpm}).\smallskip}
Too strong a suppression, however, appears implausible, both from the standpoint
of our explicit calculation as well as according to our general expectation regarding
the behavior of the strongly coupled IYIT sector at low energies.
We therefore believe that the axion decay constant has, in general, a
tendency of being as large as possible, so as to reduce the suppression
of the meson VEVs.
As for Scenario $B$, this means that, despite the significant relaxation
of the lower bound $f_Q^{(k)}$, we actually still expect $f_a$ to be of
$\mathcal{O}\left(10^{10}\right)\,\textrm{GeV}$.


Another trivial possibility to relax the bounds on parameter space is to
increase the cut-off scale $M_*$ by some factor of $\mathcal{O}(1\cdots4\pi)$.%
\footnote{This may also be desirable from the perspective of flavor-changing
neutral currents, which may still be a little bit too large for
$m_{3/2} = 100\,\textrm{TeV}$.
Slightly increasing the cut-off scale $M_*$ can then help to
\textit{fully} solve the FCNC problem also for a gravitino mass of $100\,\textrm{TeV}$, 
i.e., without the need for going to $m_{3/2}$ as large as, say,
$1000\,\textrm{TeV}$~\cite{Gabbiani:1996hi,Bhattacherjee:2012ed}.}
If we do so starting with Scenario $A$, it becomes difficult to realize gravitino masses
of $100\,\textrm{TeV}$ because $f_Q^{(k)}$ increases too drastically.
On the other hand, combining Scenario $B$ with a larger cut-off scale
does provide us with a viable scenario that also admits a gravitino mass of $100\,\textrm{TeV}$.
In this case, also $\kappa$ and $\eta$ can again be set to different values,
\begin{align}
\textrm{Scenario $C$:} \qquad k = &\: 5 \,, & \kappa = &\: 4 \,, & \eta = &\: \pi \,,
& C_Q = & \: 100 \times \left(4\pi\right)^2 \,, & M_* = &\: 4\pi\, M_{\rm Pl} \,, \\ \nonumber
\textrm{Scenario $D$:} \qquad k = &\: 5 \,, & \kappa = &\: 1 \,, & \eta = &\: 4\pi \,,
& C_Q = & \: 100 \times \left(4\pi\right)^2 \,, & M_* = &\: 4\pi\, M_{\rm Pl} \,. 
\end{align}
Here, we have multiplied $C_Q$ by another factor of $4\pi$ to keep the 
ratio $C_Q / M_*$ fixed at the same value as in Scenario $B$.
The bounds on the $f_a$--$m_{3/2}$ parameter space for these two scenarios
are shown in the two lower panels of Fig.~\ref{fig:k5}.
For $m_{3/2} = 100\,\textrm{TeV}$, the axion decay constant is again bounded
by the lower astrophysical bound, $f_a \gtrsim 10^9\,\textrm{GeV}$, in these two scenarios.
At the same time, the upper bound on $f_a$
now increases by roughly half an order of magnitude,
\begin{align}
\textrm{Scenario $C$:} \quad f_a \lesssim 5 \times 10^{10}\,\textrm{GeV} \,, \quad
\textrm{Scenario $D$:} \quad f_a \lesssim 3 \times 10^{10}\,\textrm{GeV} \,.
\label{eq:scenarioCD}
\end{align}
Guided by the notion that too strong a suppression of the meson VEVs tends to be
unrealistic, we suppose that also in Scenarios $C$ and $D$ the axion decay
constant most likely takes a value close to the upper end of the allowed range.
That is, once again, we expect $f_a \sim 10^{10} \,\textrm{GeV}$.


Finally, we mention that, relaxing our restrictions on $C_Q$ and $M_*$ similarly
as in Eq.~\eqref{eq:scenarioCD}, i.e., assuming the new quark masses to be generated
at the GUT scale and slightly raising the cut-off scale $M_*$ above the reduced Planck
mass $M_{\rm Pl}$, a number of further interesting scenarios become available.
For instance, scenarios with $k>5$ extra quark pairs now become viable, such as
\begin{align}
\textrm{Scenario $E$:} \qquad k = &\: 6 \,, & \kappa = &\: 1 \,, & \eta = &\: 4\pi \,,
& C_Q = & \: 100 \times \left(4\pi\right)^2 \,, & M_* = &\: 4\pi\, M_{\rm Pl} \,,
\end{align}
in the case of which the axion decay constant is again required to take a value
of $\mathcal{O}\left(10^{10}\right)\,\textrm{GeV}$,
\begin{align}
m_{3/2} = 100\,\textrm{TeV} \quad\Rightarrow\quad
7 \times 10^9 \,\textrm{GeV} \lesssim f_a \simeq 5 \times 10^{10}\,\textrm{GeV} \,.
\end{align}
Furthermore, we may now also assume that the field $X$ is unphysical and
send $\kappa$ to infinity,
\begin{align}
\textrm{Scenario $F$:} \qquad k = &\: 6 \,, & \kappa \rightarrow &\: \infty \,, & \eta = &\: 4\pi \,,
& C_Q = & \: 100 \times \left(4\pi\right)^2 \,, & M_* = &\: 4\pi\, M_{\rm Pl} \,. 
\end{align}
In this scenario, the axion decay constant is then more or
less constrained to a certain value,
\begin{align}
m_{3/2} = 100\,\textrm{TeV} \quad\Rightarrow\quad
f_a \simeq 5 \times 10^{10}\,\textrm{GeV} \,.
\end{align}
A more systematic study of these (and possibly other) scenarios
is left for future work.
For now, we merely conclude by observing that our model indeed appears
to be compatible with all bounds in large parts of parameter space.
Without any further assumptions, the number of extra pairs is fixed to be $k=5$,
while the decay constant $f_a$ is generally expected to take a value of
$\mathcal{O}\left(10^{10}\right)\,\textrm{GeV}$.


\section{Conclusions and Outlook}
\label{sec:conclusions}


In this paper, we have demonstrated how
the PQ solution to the strong $CP$ problem might be inherently connected
to the dynamics of spontaneous SUSY breaking.
To give a concrete example of our idea, we have embedded the PQ mechanism
into the IYIT model of dynamical SUSY breaking (i.e.\ into a strongly coupled $SU(2)$
gauge theory with four matter and six singlet fields), which has led us to
a particular supersymmetric variant of the KSVZ axion model.
As a direct consequence of this embedding, we found that the scale
of PQ symmetry breaking, $\Lambda_{\rm PQ}$, is no longer an arbitrary
(and somewhat mysterious) input parameter, but rather directly tied to
the dynamical scale $\Lambda$ of the strong interactions in the SUSY-breaking
sector, $\Lambda_{\rm PQ} \sim \Lambda$.
As the same dynamical scale also determines the scale of SUSY breaking
in the IYIT model, $\Lambda_{\rm SUSY} \sim \Lambda$, a PQ scale of
$\mathcal{O}\left(10^{11}\cdots10^{12}\right)\,\textrm{GeV}$ then
implies a large SUSY breaking scale and, hence, a large gravitino mass,
$m_{3/2} \sim 100\,\textrm{TeV}$.
The proposed connection between the dynamics of PQ symmetry and SUSY
breaking therefore turns out to go very well with the idea of pure
gravity mediation.


Besides that, the notion of pure gravity mediation is also crucial
to our axion model for another reason:
In order to protect the PQ symmetry from the dangerous effect
of higher-dimensional operators induced by gravitational interactions around
the Planck scale, one has to invoke a protective
gauge symmetry---for instance, as proposed in \cite{Harigaya:2013vja},
a discrete $R$ symmetry.
Among all possible $Z_N^R$ symmetries, pure gravity mediation
singles out the special case of a $Z_4^R$ symmetry, which is the
only discrete $R$ symmetry that allows to generate the
MSSM $\mu$ term via a Higgs bilinear term in the K\"ahler potential.
As we were able to show, such a discrete $Z_4^R$ symmetry
then manages to suppress all PQ-breaking operators in the superpotential
and K\"ahler potential up to a high order,
thereby ensuring that the PQ symmetry is of sufficiently good quality.
Solely within the MSSM, however, a discrete $Z_4^R$ symmetry does not represent
a good symmetry, as it is anomalously violated at the quantum level
by $SU(2)_L$ and $SU(3)_C$ instanton effects.
Therefore, in order to render the $Z_4^R$ symmetry anomaly-free, the presence
of further SM-charged is required.%
\footnote{By contrast, solely within the IYIT sector, the $Z_4^R$ \textit{is}
anomaly-free and even preserved in the true vacuum.
It is, therefore, neither broken explicitly nor spontaneously by the strong
interactions, which retains $R$ symmetry as a useful tool to study the
low-energy dynamics of the SUSY-breaking sector (see our discussion
related to Eq.~\eqref{eq:rIYIT}).}
For this reason, we have assumed the existence of $k$ new quark/antiquark
pairs $\left(Q_i,\bar{Q}_i\right) \sim \left(\mathbf{5},\mathbf{5}^*\right)$
in this paper, which obtain masses of the order of the gravitino mass
from coupling to the SUSY-breaking sector.
Provided an appropriate $R$ charge, these new quark fields then cancel the
$Z_4^R$ anomalies, which puts us in the position to employ the $Z_4^R$ symmetry
as a protective gauge symmetry after all.


The axion model presented in this paper comes with a number of attractive
conceptional and phenomenological implications.
The extra matter fields, for instance, contribute to the running
of the SM gauge couplings, which increases the value of the GUT gauge coupling
constant.
Requiring that the SM gauge couplings should unify at a perturbative value
therefore puts a lower bound on the mass scale of the new quark fields.
For one thing, this constrains the parameter space of our model
(i.e., it provides us with a lower bound on the axion decay constant).
For another, we note that the effect of several new $SU(5)$ multiplets with masses
$m_Q \sim m_{3/2}$ on the running of the SM gauge couplings might also
play the role of a selection criterion in the landscape of string vacua.
Of course, such an assertion is highly speculative; but we have the feeling
that it is worth being pointed out nonetheless.
Imagine, for instance, that the SM gauge couplings are bound to
unify at some $\mathcal{O}(1)$ value at the GUT scale.
The fact that the new quark fields obtain their masses via couplings to
the SUSY-breaking sector may then potentially bias the distribution of
different values of the SUSY breaking scale---maybe the SUSY breaking scale
happens to be very large, so that $m_{3/2} \sim 100\,\textrm{TeV}$, simply because
otherwise the SM gauge couplings would run over too long a distance
between the GUT scale and the new quark mass threshold.
This would then alter the ratios of the SM gauge couplings at the
electroweak scale and, for one reason for another (in the context of nuclear
and/or atomic physics), maybe exclude the possibility of habitable universes.


Apart from this perhaps far-fetched speculation, our model also makes
a number of predictions which are testable in present-day or near-future experiments.
First of all, the new quark fields may, for instance, be directly detectable
in a multi-TeV collider experiment.
In this context, it is interesting to remark that our model
(at least in its simplest form) surprisingly singles out a unique number of extra quark
pairs: We have to add exactly five pairs of new matter fields.
Remarkably enough, this leads to a situation where the $R$ charges of
the new quark fields, the $R$ charges of the fields in the SUSY-breaking sector
as well as  the $R$ charges of the MSSM fields all look very similar. 
For $k=5$ (and only for $k=5$), the $R$ charges of all fields in our model
turn out to be multiples of $1/5$.
Whether or not this points at something deep remains to be seen;
but it is certainly an interesting observation.
Moreover, we find that a sufficient suppression of all PQ-breaking
effects typically requires the axion decay constant to take a value
not much larger than $f_a \sim 10^{10} \,\textrm{GeV}$.
This has several interesting implications for cosmology.
To begin with, let us remark that, in our model, the PQ symmetry
should be broken before the end of inflation, i.e., the PQ scale
needs to exceed the inflationary Hubble scale,
$\Lambda_{\rm PQ}\gtrsim H_{\rm inf}$.
If this was not the case, dangerous axion domain walls (with domain wall
number $N_{\rm DW} = \left|\mathcal{A}_{\rm PQ}\right| = k  > 1$)
would form during the QCD phase transition, dominating the energy
density of the universe soon after their production~\cite{Sikivie:1982qv}.
However, if the PQ symmetry is already broken during inflation,
we have to pay attention that the isocurvature perturbations induced
by the axion fluctuation during inflation, $\delta \bar{\theta} \simeq H_{\rm inf}/(2\pi)$,
do not violate any of the stringent bounds derived from the precise observations of
the cosmic microwave background (CMB)~\cite{Kawasaki:2013ae}.
For $f_a \sim 10^{10} \,\textrm{GeV}$ and an initial axion misalignment
angle $\bar{\theta}$ of $\mathcal{O}(1)$, this constrains the
Hubble rate during inflation to a rather small value,
$H_{\rm inf} \lesssim 10^8\,\textrm{GeV}$.
Our axion model is therefore only compatible with small-field models
of inflation.
Or put differently, from the perspective of our model, we are led to
expect that upcoming CMB experiments will unfortunately not be able to see
any signs of tensor perturbations in the CMB.
That is, if the inflationary Hubble rate should indeed be as small as
$10^8\,\textrm{GeV}$, or even smaller, the tensor-to-ratio
is at most of $\mathcal{O}(10^{-13})$,
which is unfortunately out of reach for any planned CMB experiment.
Besides that, an axion decay constant of $\mathcal{O}(10^{10})\,\textrm{GeV}$
(in combination with $\bar{\theta} \sim 1$ and $\delta\bar{\theta}\ll 1$),
results in an axionic contribution to the relic density of dark matter
of about $\mathcal{O}(10\,\%)$~\cite{Turner:1985si}. 
The remaining DM density is then accounted for by weakly interacting
massive particles (WIMPs) in the form of MSSM neutralinos in our model.
For this reason, we are confident that both axion as well as WIMP dark matter
searches may, in principle, be able to find positive signals.
Moreover, for $f_a \sim 10^{10}\,\textrm{GeV}$, the axion mass lies in the meV range.
Such relatively heavy axions could, for instance, be searched for in
fifth-force experiments searching for axion-mediated long range forces~\cite{Arvanitaki:2014dfa}
or in experiments aiming at measuring the proton electric dipole
moment~\cite{Anastassopoulos:2015ura}.
Given the fact that $\Delta\bar{\theta}$ may easily take a value only
slightly below the upper bound $\Delta\bar{\theta}^{\rm max}$ in our model,
$\Delta\bar{\theta} \sim 10^{-11} \cdots 10^{-12}$,
(see the upper left panel of Fig.~\ref{fig:k5}) such experiments look
indeed promising.


In summary, we therefore conclude that our axion model not only appears to provide an
interesting link between dynamical SUSY breaking and the PQ mechanism,
it also gives rise to a rich phenomenology
that is going to be tested in current and upcoming experiments.
This is exciting and hopefully only a first step towards a better understanding
of supersymmetry, dark matter and the new physics lurking behind the strong $CP$
problem---which, as we believe, should certainly star some kind of
\textit{axion field of dynamical origin} as the main protagonist.


\subsubsection*{Acknowledgements}

This work has been supported in part by Grants-in-Aid for Scientific Research
from the Ministry of Education, Culture, Sports, Science, and Technology (MEXT), Japan,
No.\ 24740151 and No.\ 25105011 (M.\,I.) as well as No.\ 26104009 (T.\,T.\,Y.);
Grant-in-Aid No.\ 26287039 (M.\,I.\ and T.\,T.\,Y.) from the Japan
Society for the Promotion of Science (JSPS); and by
the World Premier International Research Center Initiative (WPI), MEXT, Japan.
K.\,H.\ has been supported in part by a JSPS Research
Fellowship for Young Scientists.


\appendix


\section{Exact vacuum of the IYIT model for canonical K\"ahler potential}
\label{app:vacuum}


In this appendix, we compute the exact VEVs of the SUSY-breaking singlet fields
in the IYIT model, $Z_+$, $Z_-$, and $X$, under the simplifying assumption of
a canonical K\"ahler potential for all relevant fields.
Of course, this assumption can never hold true \textit{exactly}, as,
in general, strong-coupling effects will always induce
higher-dimensional terms in the effective K\"ahler potential.
Still, we deem a calculation based on a canonical K\"ahler potential useful
for several reasons.
First of all, we expect it to represent an important benchmark scenario for the more
general case---a benchmark scenario that we have
well under control and that allows us to obtain a better understanding of the
various parameter dependences in our model.
Moreover, since we lack the ability to calculate the dynamical corrections
to the K\"ahler potential, there are, in fact, not many alternatives
to assuming a canonical K\"ahler potential, if we are interested in more
than just some rough order-of-magnitude estimates.
Starting from a canonical K\"ahler potential, we are able
to derive a consistent set of expressions in terms of a number of well-defined 
parameters.
This would, by contrast, not be possible, if we also intended to
account for the uncertainties related to the effective K\"ahler potential.
On top of that, in case the dynamically generated corrections to the K\"ahler
potential become smaller and smaller, the expressions that we are going to derive
in the following become increasingly accurate, approximating the true
results in the IYIT model with arbitrary precision.
All in all, we therefore believe that the simplifying assumption
of a canonical K\"ahler potential---while not exactly reflecting the actual
situation in the IYIT model---still captures many of the aspects that we are
interested in and that it is, hence, worth a closer examination.


The starting point of our analysis is the effective superpotential in Eq.~\eqref{eq:Weffcha}
in combination with a canonical K\"ahler potential for all DOFs in the low-energy
effective theory,
\begin{align}
W_{\rm eff} \simeq & \: \kappa\,\eta\, X
\left[\textrm{Pf}\left(M^{ij}\right) - \left(\frac{\Lambda}{\eta}\right)^2\right]
+ \frac{\Lambda}{\eta} \left(\lambda_+\, M_+\, Z_-
+ \lambda_-\, M_-\, Z_+ + \lambda_0^a\, M_0^a\, Z_0^a\right) \,, \\\nonumber
K_{\rm eff} \simeq & \: \left|X\right|^2 + \left|Z_+\right|^2 + \left|Z_-\right|^2
+ \left|M_+\right|^2 + \left|M_-\right|^2 + \sum_{a=1}^4 \left|Z_0^a\right|^2
+ \sum_{a=1}^4 \left|M_0^a\right|^2 \,.
\end{align}
In the true vacuum of the scalar potential corresponding to these input functions
(assuming $\lambda_+\lambda_-$ to be the smallest among the three products
$\lambda_+\lambda_-$, $\lambda_0^1\lambda_0^4$, and $\lambda_0^2\lambda_0^3$),
SUSY is broken by the F-term of the following linear combination of the fields
$Z_+$, $Z_-$, and $X$ (see Eqs.~\eqref{eq:LambdaSUSY} and \eqref{eq:S012}),
\begin{align}
S_0 =  \frac{1}{\left(2-\zeta\right)^{1/2}} \left[\left(1-\zeta\right)^{1/2}\left(Z_+ + Z_-\right)
- \zeta^{1/2} X \right] \,,\quad
\left|F_{S_0}\right| = \mu^2 = \lambda \left(2-\zeta\right)^{1/2} \frac{\Lambda^2}{\eta^2} \,.
\end{align}
Taking into account the spontaneous breaking of $R$ symmetry in the context of SUGRA,
this linear combination turns out to acquire a nonzero VEV of $\mathcal{O}\left(m_{3/2}\right)$.
To see this, first of all note that $R$ symmetry breaking induces a constant term $W_0$
in the superpotential (see Eq.~\eqref{eq:W0}),
\begin{align}
W \supset W_0 = m_{3/2}\, M_{\rm Pl}^2 \,.
\end{align}
Together with the SUSY-breaking tadpole term for the goldstino field $S_0$
in the effective superpotential, $W_{\rm eff} \supset \mu^2 S_0$ (see Eq.~\eqref{eq:Wfinal}),
and together with the loop-induced mass for the complex sgoldstino $s_0 \subset S_0$ in the effective
scalar potential, $V_{\rm eff} \supset m_{s_0}^2 \left|s_0\right|^2$
(see Eq.~\eqref{eq:m02approx}), this constant superpotential gives rise to
the following total scalar potential for the complex scalar $s_0$,
\begin{align}
V_{\rm eff} = m_{s_0}^2 \left|s_0\right|^2 - 2\, m_{3/2}\, \mu^2 \left(s_0 + s_0^*\right) \,.
\end{align}
We, thus, find that the interplay between the constant superpotential $W_0$ and the
tadpole term $\mu^2 S_0$ breaks the rotational invariance in the complex $s_0$ plane. 
That is, while the imaginary part of $s_0$ remains stabilized at $0$ thanks to the
loop-induced mass $m_{s_0}$, the real component of $s_0$ obtains a linear potential
proportional to $m_{3/2}\,\mu^2$,
which shifts its VEV from $0$ to some value of $\mathcal{O}\left(m_{3/2}\right)$,
\begin{align}
\left<\textrm{Re}\left\{s_0\right\}\right> = \frac{2\,\mu^2}{m_{s_0}^2} \, m_{3/2} \,, \quad
\left<\textrm{Im}\left\{s_0\right\}\right> = 0 \quad\Rightarrow\quad
\left<S_0\right> = \frac{2\,\mu^2}{m_{s_0}^2} \, m_{3/2} \,.
\label{eq:S0VEV}
\end{align}
This result readily translates into expressions for
$\left<Z_+\right>$, $\left<Z_-\right>$, and $\left<X\right>$.
All we need to know is the inverse of the transformation between the two
field bases $\left(Z_+,Z_-,X\right)$ and $\left(S_0,S_1,S_2\right)$ in Eq.~\eqref{eq:S012},
\begin{align}
Z_\pm = & \: \frac{1}{\left(2-\zeta\right)^{1/2}} \left[\left(1-\zeta\right)^{1/2} S_0
\pm 2^{-1/2}\left(2-\zeta\right)^{1/2} S_1 + \left(\zeta/2\right)^{1/2} S_2\right] \,,
\\\nonumber
X = & \: \frac{1}{\left(2-\zeta\right)^{1/2}} \left[-\zeta^{1/2} S_0
+ 2^{1/2}\left(1-\zeta\right)^{1/2} S_2 \right] \,.
\end{align}
Taking into account that the singlets $S_1$ and $S_2$ do \textit{not} obtain a nonzero VEV,
this leads us to
\begin{align}
\left<Z_\pm\right> = \left(\frac{1-\zeta}{2-\zeta}\right)^{1/2} \left<S_0\right> 
= \frac{1}{\sqrt{2}}\left(1-r^2\right)^{1/2} \left<S_0\right> \,, \quad
\left<X\right> = - \left(\frac{\zeta}{2-\zeta}\right)^{1/2} \left<S_0\right> 
= - r \left<S_0\right> \,,
\label{eq:ZXVEV}
\end{align}
where we have used Eq.~\eqref{eq:mrdef} to rewrite the $\zeta$-dependent
coefficients in terms of the parameter $r$.
From Eqs.~\eqref{eq:S0VEV} and \eqref{eq:ZXVEV}, we now see that all
singlet VEVs crucially depend on the loop-induced sgoldstino
mass $m_{s_0}$.
In order to obtain usable expressions for $\left<Z_+\right>$, $\left<Z_-\right>$,
and $\left<X\right>$, we therefore need to determine this mass parameter
as precisely as possible.
This is what we shall do next.


The sgoldstino mass $m_{s_0}$ follows from the one-loop effective
Coleman-Weinberg potential,
\begin{align}
m_{s_0}^2 = \left.\frac{\partial^2\, V_{\rm CW}}{\partial s_0\, \partial s_0^*}\right|_{s_0 = 0} \,, \quad
V_{\rm CW} = \frac{1}{64\pi^2}\, \textrm{STr}\left[M^4 \left(\ln\left(\frac{M^2}{Q^2}\right) + c\right)\right] \,,
\label{eq:VCW}
\end{align}
where $M^2$ stands for the total mass matrix of the IYIT model squared, $Q$ denotes an appropriate
renormalization scale for the low-energy effective theory and $c$ is a constant that is sometimes
introduced for cosmetic reasons, but which may as well also be simply absorbed into the scale $Q$.
In order to evaluate $V_{\rm CW}$ and determine $m_{s_0}$, we therefore need to compute
the entire mass spectrum of the IYIT model for a nonzero value of the goldstino field $S_0$.
In the charged meson sector (which includes the chiral superfields $M_\pm$, $Z_\pm$, and $X$),
the physical mass eigenstates correspond to ten real scalars, one Weyl fermion and four Majorana
fermions (see also our discussion at the end of Sec.~\ref{subsec:stabilization}).
Here, the bosonic DOFs consist of the axion $a$, the saxion $\phi$
as well as the real and imaginary parts of the complex scalars contained in the goldstino field,
$s_0^\pm$, the singlet field $S_1$ (which shares a Dirac mass with the axion field), $s_1^\pm$,
the radial meson field, $m^\pm$, and the singlet field $S_2$ (which shares a Dirac
mass with the radial meson field), $s_2^\pm$.
Meanwhile, the fermionic DOFs consist of the goldstino $\tilde{s}_0$ as well
as four Majorana fermions forming two pairs of quasi-Dirac fermions,
$(\tilde{a},\tilde{s}_1)$ and $(\tilde{m},\tilde{s}_2)$,
where $\tilde{a}$ stands for the axino.
A straightforward calculation of the bosonic mass matrix in global SUSY and at tree
level, accounting for nonzero $S_0$, then yields
\begin{align}
\label{eq:specchabos}
m_\phi^2 = & \: m^2 \left[\frac{3}{2} + \frac{m^2}{2\,\mu^4}\left|S_0\right|^2 +
\frac{1}{2}\left(1 + 6\, \frac{m^2}{\mu^4}\left|S_0\right|^2
+ \frac{m^4}{\mu^8} \left|S_0\right|^4\right)^{1/2}\right] \,, \quad
m_a^2 = m_{s_0^\pm}^2 = 0 \,, \\ \nonumber
m_{s_1^-}^2 = & \: m^2 \left[\frac{3}{2} + \frac{m^2}{2\,\mu^4}\left|S_0\right|^2 -
\frac{1}{2}\left(1 + 6\, \frac{m^2}{\mu^4}\left|S_0\right|^2
+ \frac{m^4}{\mu^8} \left|S_0\right|^4\right)^{1/2}\right] \,, \quad
m_{s_1^+}^2 = m^2 + \frac{m^2}{\mu^4}\left|S_0\right|^2 \,, \\ \nonumber
m_{m^\pm}^2 = & \: \frac{m^2}{r^2} \left[1 \pm \frac{r^2}{2} + \frac{r^2m^2}{2\,\mu^4}\left|S_0\right|^2 \pm
\frac{r^2}{2}\left(1 + \frac{2}{r^2}\left(2\pm r^2\right)\, \frac{m^2}{\mu^4}\left|S_0\right|^2
+ \frac{m^4}{\mu^8} \left|S_0\right|^4\right)^{1/2}\right] \,, \\ \nonumber
m_{s_2^\pm}^2 = & \: \frac{m^2}{r^2} \left[1 \pm \frac{r^2}{2} + \frac{r^2m^2}{2\,\mu^4}\left|S_0\right|^2 \mp
\frac{r^2}{2}\left(1 + \frac{2}{r^2}\left(2\pm r^2\right)\, \frac{m^2}{\mu^4}\left|S_0\right|^2
+ \frac{m^4}{\mu^8} \left|S_0\right|^4\right)^{1/2}\right] \,,
\end{align}
while a similar calculation of the fermionic mass matrix provides us with
\begin{align}
\label{eq:specchafer}
m_{(\tilde{a},\tilde{s}_1)}^2 = & \: m^2 \left[1 + \frac{m^2}{2\,\mu^4}\left|S_0\right|^2
\pm \left(\frac{m^2}{\mu^4}\left|S_0\right|^2 +
\frac{m^4}{4\,\mu^8}\left|S_0\right|^4 \right)^{1/2}\right] \,,
\quad m_{\tilde{s}_0}^2 = 0 \,, \\ \nonumber
m_{(\tilde{m},\tilde{s}_2)}^2 = & \: \frac{m^2}{r^2}\left[1 + \frac{r^2m^2}{2\,\mu^4}\left|S_0\right|^2
\pm r^2\left(\frac{1}{r^2}\frac{m^2}{\mu^4}\left|S_0\right|^2 +
\frac{m^4}{4\,\mu^8}\left|S_0\right|^4 \right)^{1/2}\right] \,.
\end{align}


At the same time, the neutral meson sector (which includes the chiral superfields $M_0^a$
and $Z_0^a$, where $a = 1,2,3,4$) also features $S_0$-dependent mass eigenvalues.
The reason for this is the coupling of the field $X = - r\, S_0 + \cdots$ to the Pfaffian
of the complete meson matrix in Eq.~\eqref{eq:Weffcha}.
In fact, the total effective superpotential for the neutral meson fields takes
the following form,
\begin{align}
W_{\rm eff} \supset \kappa\,\eta\left[r\,S_0 - \left(1 - r^2\right)^{1/2} S_2\right]
\left(M_0^1 M_0^4 - M_0^2 M_0^3 \right) + \lambda_0^a\, M_0^a\, Z_0^a \,,
\end{align}
which gives rise to eight complex scalars as well as to four Dirac fermions.
Here, the masses of the four complex scalars, $m_{14}^\pm$ and $m_{23}^\pm$,
contained in the neutral meson fields $M_0^a$ are gives as
\begin{align}
\label{eq:specneubosmes}
m_{m_{14}^\pm}^2 = & \: \left(\sigma_{14}^2 \pm \lambda_{14}^2\right)
\left(\frac{\Lambda}{\eta}\right)^2 \left[1 \pm \frac{\kappa^2\,\eta^2\,r^2}
{\lambda_{14}^2\,\Lambda^2/\eta^2} \left|S_0\right|^2 +
\mathcal{O}\left(\left|S_0\right|^4\right)\right] \,,
\\\nonumber
m_{m_{23}^\pm}^2 = & \: \left(\sigma_{23}^2 \pm \lambda_{23}^2\right)
\left(\frac{\Lambda}{\eta}\right)^2 \left[1 \pm \frac{\kappa^2\,\eta^2\,r^2}
{\lambda_{23}^2\,\Lambda^2/\eta^2} \left|S_0\right|^2 +
\mathcal{O}\left(\left|S_0\right|^4\right)\right] \,,
\end{align}
where we have introduced the symbols $\sigma_{14}$, $\lambda_{14}$, $\sigma_{23}$,
and $\lambda_{23}$, for the ease of notation (see Eq.~\eqref{eq:lambda1423}),
\begin{align}
\sigma_{14} = \left[\frac{1}{2}\left(\left(\lambda_0^1\right)^2 +
\left(\lambda_0^4\right)^2\right)\right]^{1/2}\,,\quad
\lambda_{14} = \left(\lambda^4 + \delta_{14}^4\right)^{1/4} \,, \quad
\delta_{14} = \left[\frac{1}{2}\left(\left(\lambda_0^1\right)^2 -
\left(\lambda_0^4\right)^2\right)\right]^{1/2} \,,
\label{eq:sigmalambda14}
\end{align}
and similarly for $\sigma_{23}$, $\lambda_{23}$, and $\delta_{23}$.
By contrast, the masses of the complex scalars $z_0^a$ contained in the neutral
singlet fields $Z_0^a$ turn out to be independent of $S_0$ up to
corrections of $\mathcal{O}\big(\left|S_0\right|^4\big)$,
\begin{align}
\label{eq:specneubossin}
m_{z_0^a}^2 = \left(\lambda_0^a\right)^2 \left(\frac{\Lambda}{\eta}\right)^2
+ \mathcal{O}\left(\left|S_0\right|^4\right) \,, \quad a = 1,2,3,4 \,.
\end{align}
Last but not least, the masses of the four Dirac fermions in the neutral meson sector
are given as
\begin{align}
\label{eq:specneufer}
m_{(\tilde{m}_0^1,\tilde{z}_0^1)}^2 = & \: \left(\lambda_0^1\right)^2 \left(\frac{\Lambda}{\eta}\right)^2
\left[1 + \frac{1}{2}\frac{\kappa^2\,\eta^2\,r^2}
{\delta_{14}^2\,\Lambda^2/\eta^2} \left|S_0\right|^2 +
\mathcal{O}\left(\left|S_0\right|^4\right)\right] \,, \\ \nonumber
m_{(\tilde{m}_0^2,\tilde{z}_0^2)}^2 = & \: \left(\lambda_0^2\right)^2 \left(\frac{\Lambda}{\eta}\right)^2
\left[1 + \frac{1}{2}\frac{\kappa^2\,\eta^2\,r^2}
{\delta_{23}^2\,\Lambda^2/\eta^2} \left|S_0\right|^2 +
\mathcal{O}\left(\left|S_0\right|^4\right)\right] \,, \\ \nonumber
m_{(\tilde{m}_0^3,\tilde{z}_0^3)}^2 = & \: \left(\lambda_0^3\right)^2 \left(\frac{\Lambda}{\eta}\right)^2
\left[1 - \frac{1}{2}\frac{\kappa^2\,\eta^2\,r^2}
{\delta_{23}^2\,\Lambda^2/\eta^2} \left|S_0\right|^2 +
\mathcal{O}\left(\left|S_0\right|^4\right)\right] \,, \\ \nonumber
m_{(\tilde{m}_0^4,\tilde{z}_0^4)}^2 = & \: \left(\lambda_0^4\right)^2 \left(\frac{\Lambda}{\eta}\right)^2
\left[1 - \frac{1}{2}\frac{\kappa^2\,\eta^2\,r^2}
{\delta_{14}^2\,\Lambda^2/\eta^2} \left|S_0\right|^2 +
\mathcal{O}\left(\left|S_0\right|^4\right)\right] \,.
\end{align}


With the expressions in Eqs.~\eqref{eq:specchabos}, \eqref{eq:specchafer},
\eqref{eq:specneubosmes}, \eqref{eq:specneubossin}, and \eqref{eq:specneufer} at our
disposal, we now know the entire mass spectrum of the IYIT model up to
$\mathcal{O}\big(\left|S_0\right|^2\big)$.
This allows us to evaluate the
Coleman-Weinberg potential and, hence, determine the sgoldstino mass.
Differentiating $V_{\rm CW}$ w.r.t.\ to $s_0$ and $s_0^*$ 
finally leads us to the following result for $m_{s_0}$ (see Eq.~\eqref{eq:VCW}),%
\footnote{Recall that, for the purposes of this appendix,
we are assuming the K\"ahler potential to be canonical.
However, in a more realistic context, we would also expect the presence of uncalculable
higher-dimensional terms in the K\"ahler potential, which would yield further
contributions to the sgoldstino mass (see our discussion related to Eq.~\eqref{eq:ms02}).}
\begin{align}
m_{s_0}^2 = & \: \frac{2\ln2-1}{16\pi^2} \left[1 + \omega(r) +
\frac{2}{\rho^6}\,\bigg(\left(\frac{\lambda_{14}}{\lambda}\right)^2\omega_0\big(s_{14},t_{14}\big) +
\left(\frac{\lambda_{23}}{\lambda}\right)^2\omega_0\big(s_{23},t_{23}\big)\bigg)\right]
\frac{m^6}{\mu^4} \,, \label{eq:ms02full}\\\nonumber
\omega(r) = & \: \frac{1}{2\ln2-1}\left[f(r) - \frac{1}{r^2}\right] \,, \quad
\omega_0\left(s,t\right) = \frac{1}{2\ln2-1}
\left[f\left(s\right) - \frac{1}{s^2} \, t^2f\left(t\right)\right] \,, 
\end{align}
where the function $f$ stands for the following combination of logarithms,
\begin{align}
f(x) = \frac{1}{2}\left(1+\frac{1}{x^2}\right)^2\ln\left(1+x^2\right)
- \frac{1}{2}\left(1-\frac{1}{x^2}\right)^2\ln\left(1-x^2\right) \,,
\end{align}
and where the parameters $s_{14}$, $t_{14}$, $s_{23}$, and $t_{23}$ are defined as follows
(see Eq.~\eqref{eq:sigmalambda14}),
\begin{align}
s_{14} = \frac{\lambda_{14}}{\sigma_{14}} \,, \quad 
t_{14} = \frac{\delta_{14}}{\sigma_{14}} \,, \quad 
s_{23} = \frac{\lambda_{14}}{\sigma_{23}} \,, \quad 
t_{23} = \frac{\delta_{14}}{\sigma_{23}} \,.
\label{eq:stdef}
\end{align}
The weight function $\omega$ in Eq.~\eqref{eq:ms02full} accounts for the
relative importance of loop diagrams with internal $M$ lines compared to
loop diagrams with internal $A$ lines.
Similarly, $\omega_0$ is a measure for the relative importance of loop diagrams
involving neutral meson fields.
It reduces to $\omega$ in the flavor-symmetric limit, which is characterized by
the parameter $t$ going to zero, i.e., $\omega_0(s,0) = \omega(s)$.
Both weight functions are normalized such that they smoothly interpolate between $0$ and $1$,
\begin{align}
\omega(0) = 0 \,, \quad \omega(1) = 1 \,, \quad \omega_0(s,t=s) = 0 \,, \quad \omega_0(1,0) = 1 \,.
\label{eq:omega01}
\end{align}
Here, note that $t$ can never exceed $s$ by definition, $t \leq s$ (see Eqs.~\eqref{eq:sigmalambda14}
and \eqref{eq:stdef}).
The two functions $\omega$ and $\omega_0$ can, moreover, be conveniently approximated by the
following polynomials,
\begin{align}
\omega(r) \approx r^2 \,, \quad \omega_0(s,t) \approx s^2 - \left(\frac{t}{s}\right)^2 t^2 \,,
\end{align}
which nicely reproduce the exact identities in Eq.~\eqref{eq:omega01} as well as the
fact that $\omega_0(s,0) = \omega(s)$.


Next, let us evaluate the sgoldstino mass in the flavor-symmetric limit, i.e.,
for all $\lambda_0^a$ being equal to $\lambda$.
This will provide us with a much simpler expression for $m_{s_0}$ that
approximates the full result in Eq.~\eqref{eq:ms02full} reasonably well as long as
there is no large hierarchy among the Yukawa couplings. 
In the flavor-symmetric limit, we are then allowed to perform the following
simplifications,%
\footnote{We only perform these simplifications in order to obtain a more practical
(approximate) expression for the sgoldstino mass, i.e., we do not assume that
any larger global flavor symmetry is \textit{actually} realized, since this would lead,
for instance, to problems involving massless particles
(see Eq.~\eqref{eq:specneubosmes} as well as the discussion in Footnote~\ref{fn:global}).}
\begin{align}
\delta_{14} = \delta_{23} = 0 \,, \quad
\lambda_{14} = \lambda_{23} = \sigma_{14} = \sigma_{23} = \lambda \,, \quad
s_{14} = s_{23} = 1 \, \quad t_{14} = t_{23} = 0 \,.
\end{align}
The weight function $\omega_0$ therefore simply evaluates
twice to unity, so that $m_{s_0}$ turns into,
\begin{align}
m_{s_0}^2 = \frac{2\ln2-1}{16\pi^2} \left[1 + \omega(r) + \frac{4}{\rho^6}\right]
\frac{m^6}{\mu^4} \approx
\frac{2\ln2-1}{16\pi^2} \left[1 + r^2 + \frac{4}{\rho^6}\right] \frac{m^6}{\mu^4} \,.
\end{align}
This is our final result for the sgoldstino mass.
Plugging it into Eq.~\eqref{eq:S0VEV}, we find for $\left<S_0\right>$
\begin{align}
\left<S_0\right> 
= \frac{2\,(2-\zeta)^{3/2}}{\left(2\ln2-1\right)\left[1 + \omega(r) + 4/\rho^6\right]\rho^6}
\frac{16\pi^2}{\lambda^3} \, m_{3/2} \approx
\frac{(2-\zeta)^{5/2}}{\left(2\ln2-1\right)\left(4-2\,\zeta + \rho^6\right)}
\frac{16\pi^2}{\lambda^3} \, m_{3/2} \,.
\label{eq:S0final}
\end{align}
We stress that, allowing for the possibility of a noncanonical K\"ahler potential,
this result receives corrections due to dynamically generated contributions
to the sgoldstino mass.
If we assume these contributions to be positive, the above expression for
the goldstino VEV can also be understood as a conservative upper limit.
On the other hand, for negative mass corrections coming from the
dynamical K\"ahler potential, the actual value for $\left<S_0\right>$
ends up being larger than the expression in Eq.~\eqref{eq:S0final}.
In the main body of this paper, we shall assume that the former of these two
possibilities is realized.
In this case, working with our result in Eq.~\eqref{eq:S0final} will then
correspond to a conservative treatment of the effect of higher-dimensional
operators on the quality of the PQ symmetry.


Finally, we are ready to compute the VEVs of the singlet fields $Z_+$, $Z_-$, and $X$.
Making use of the relations in Eq.~\eqref{eq:ZXVEV} as well as of our result
for $\left<S_0\right>$ in Eq.~\eqref{eq:S0final}, we eventually find
\begin{align}
\left<Z_\pm\right> \approx \frac{(1-\zeta)^{1/2}\,(2-\zeta)^2}
{\left(2\ln2-1\right)\left(4-2\,\zeta + \rho^6\right)}
\frac{16\pi^2}{\lambda^3} \, m_{3/2} \,, \:\:
\left<X\right> \approx \frac{-\zeta^{1/2}\,(2-\zeta)^2}{\left(2\ln2-1\right)
\left(4-2\,\zeta + \rho^6\right)} \frac{16\pi^2}{\lambda^3} \, m_{3/2} \,.
\label{eq:ZXVEVs}
\end{align}
It is instructive to consider the behavior of these expressions for certain
extreme parameter choices.
For $\zeta\rightarrow0$, for instance, all three VEVs become arbitrarily large,
$\left<Z_\pm\right>$, $\left<\left|X\right|\right> \gg m_{3/2}$.
The reason for this is simply the inverse cubic power of $\lambda$
appearing in all of the above VEVs.
The two prefactors multiplying $16\pi^3/\lambda^3 \, m_{3/2}$ in Eq.~\eqref{eq:ZXVEVs}
stay, by contrast, also finite in the limit $\zeta \rightarrow 0$.
For $\zeta\rightarrow 1$, on the other hand, $\left<Z_\pm\right>$ approaches $0$,
while $\left<\left|X\right|\right>$ is bounded from below,
\begin{align}
\left<\left|X\right|\right> \geq \frac{16\pi^2/\lambda_{\rm max}^3}
{\left(2\ln2-1\right)\left(2+\rho^6\right)}\, m_{3/2}
\simeq 0.07 \left(\frac{3}{2+\rho^6}\right)
\left(\frac{4\pi}{\lambda_{\rm max}}\right)^3 m_{3/2} \,.
\end{align}
Interestingly enough, we also find that
all three VEVs typically turn out to be smaller than $m_{3/2}$ for
most values of $\zeta$.
More precisely, for $\kappa\,\eta = 4\pi$, $\rho =1$ and in terms
of the coupling $\lambda$, we obtain
\begin{align}
\kappa\,\eta = 4\pi \quad\Rightarrow\quad \left<Z_\pm\right> \leq m_{3/2}
\:\:\:\textrm{for}\:\:\: \lambda \gtrsim 2.0\,\pi \,, \quad
\left<\left|X\right|\right> \leq m_{3/2}
\:\:\:\textrm{for}\:\:\: \lambda \gtrsim 1.5\,\pi \,.
\end{align}
This is advantageous from the perspective of our axion model,
as it indicates that dangerous higher-dimensional operators involving powers of the fields
$Z_+$, $Z_-$ and/or $X$ may not have as strong an effect on the quality
of the PQ symmetry as one may naively expect (i.e., if one simply estimated
$\left<Z_\pm\right>$ and $\left<\left|X\right|\right>$ to be some values of
$\mathcal{O}(m_{3/2})$).
Lastly, we mention that, in the limit of the deformed moduli constraint
being exactly fulfilled, i.e., for $\kappa \rightarrow \infty$, the VEV
of the singlet field $X$ vanishes completely, $\left<\left|X\right|\right> = 0$,
while $\left<Z_\pm\right>$ turns into a simple function of $\lambda$,
\begin{align}
\kappa\rightarrow\infty \quad\Rightarrow\quad 
\left<Z_\pm\right> = \frac{4}{\left(2\ln2-1\right)\left(4+\rho^6\right)}
\frac{16\pi^2}{\lambda^3} \, m_{3/2}
\simeq 0.16 \left(\frac{5}{4+\rho^6}\right)
\left(\frac{4\pi}{\lambda}\right)^3 m_{3/2} \,.
\end{align}
Here, the fact that $X$ vanishes is consistent
with the observation that $X$ becomes infinitely heavy for
$\kappa\rightarrow\infty$, indicating that, in this limit,
it is an unphysical field that needs to be integrated out.



\end{document}